\documentstyle[12pt,epsf]{article}

\newcommand{\wb}{\bar}
\newcommand{\wh}{\widehat}
\newcommand{\wt}{\widetilde}
\newcommand{\MM}{{\cal M}}

\newcommand{\AAA}{{\cal A}}
\newcommand{\II}{{\cal I}}
\newcommand{\DD}{{\cal D}}
\newcommand{\HH}{{\cal H}}
\newcommand{\DDA}{{\cal D}}
\newcommand{\BB}{{\cal B}}
\newcommand{\TH}{{\cal T}}
\newcommand{\KK}{{\cal K}}
\newcommand{\p}{\partial}

\begin{document}

\newcommand{\be}{\begin{equation}}
\newcommand{\ee}{\end{equation}}
\newcommand{\ben}{\begin{eqnarray}\displaystyle}
\newcommand{\een}{\end{eqnarray}}
\newcommand{\refb}[1]{(\ref{#1})}
\newcommand{\sectiono}[1]{\section{#1}\setcounter{equation}{0}}
\renewcommand{\theequation}{\thesection.\arabic{equation}}

\begin{titlepage}

{}~ \hfill\vbox{\hbox{hep-th/9802051}
}\break

\vskip 3.5cm

\centerline{\huge\bf An Introduction to}
\medskip
\centerline{\huge\bf Non-perturbative String 
Theory\footnote{Based on lectures given at the Isaac Newton Institute 
and the Department of Applied Mathematics
and Theoretical Physics, University of Cambridge, UK.}}

\vspace*{6.0ex}

\centerline{\large \rm Ashoke Sen\footnote{E-mail: sen@mri.ernet.in}}

\vspace*{1.5ex}

\centerline{\large \it Mehta Research Institute of Mathematics}
 \centerline{\large \it and Mathematical Physics}

\centerline{\large \it  Chhatnag Road, Jhoosi,
Allahabad 211019, INDIA}

\vspace*{4.5ex}

\centerline {\bf Abstract}

In this review I discuss some basic aspects of non-perturbative
string theory.
The topics include test of duality
symmetries based on the analysis of the low energy
effective action and the 
spectrum of BPS states, relationship between different duality
symmetries, an introduction to M- and F-theories,
black hole entropy in string theory, and Matrix theory.

\end{titlepage}

\tableofcontents

\newpage

\baselineskip=18pt

\setcounter{section}{-1}

\sectiono{Introduction} \label{s0}

During the last few years, our understanding of string theory has
undergone a dramatic change. The key to this development is the 
discovery of duality symmetries, which relate
the strong and weak coupling limits of apparently different
string theories. These symmetries not only relate
apparently different string theories,
but give us a way to compute certain strong coupling results in
one string theory by mapping it to a weak coupling result in a
dual string theory. In this review I shall try to
give an introduction to this exciting subject. However, instead
of surveying all the important developments in this
subject I shall try to explain the basic ideas
with the help of a few simple examples. 
I apologise for the inherent bias in
the choice of examples and the topics, this is solely due to
the varied degree of familiarity that I have
with this vast subject.
I have also not
attempted to give a complete list of references. Instead I have
only included those references whose results have been directly
used or mentioned in this article. A complete list of references
may be obtained by looking at the citations to some of the
original papers in spires. There are also many other reviews in
this subject where more references can be 
found\cite{REVB}-\cite{REVE}.
I hope that this review will serve the
limited purpose of initiating a person  with a knowledge of
perturbative string theory into this area. (For an introduction
to perturbative string theory, see \cite{GSW}). 

The review will be divided into ten main sections as described
below.
\begin{enumerate}
\item 
A brief review of perturbative string theory:
In this section I shall very briefly recollect some of the
results of perturbative string theory which will be useful to us
in the rest of this article. This will in no way constitute an
introduction to this subject; at best it will serve as a reminder
to a reader who is already familiar with this subject.
\item
Notion of duality symmetry: In
this section I shall describe the notion of
duality symmetry in string theory, a few
examples of duality conjectures in
string theory, and the general procedure for testing these
duality conjectures.
\item
Analysis of the low energy effective action: In this section I shall
describe how one arrives at various duality conjectures by
analyzing the low energy effective action of string theory.
\item
Precision test of duality based on the spectrum of BPS states: In
this section I shall discuss how one can device precision
tests of various duality conjectures based on the analysis of the
spectrum of a certain class of supersymmetric states in string
theory.
\item
Interrelation between various dualities: In this section I shall
try to relate the various duality conjectures introduced in the
sections \ref{s1} - \ref{s3} by `deriving' 
them from a basic set of duality
conjectures. I shall also discuss what we mean by
relating different dualities and try to
formulate the rules that must be followed during such a
derivation.
\item Duality in theories with $<16$ supersymmetries: The
discussion in sections \ref{s2}-\ref{s4} is focussed on string
theories with at least 16 supersymmetry generators. In this
section I consider theories with less number of supersymmetries.
Specifically we shall focus our attention on theories with eight
supercharges, which correspond to N=2 supersymmetry in four
dimensions. 
\item
M-theory: In this section I discuss the emergence of a new
theory in eleven dimensions  $-$now known as M-theory $-$
from the strong coupling limit of type IIA string theory. 
I also discuss how compactification of M-theory gives rise to
new theories that cannot be regarded as perturbative
compactification of a string theory.
\item
F-theory: In this section I shall discuss yet another novel
way of generating non-perturbative compactification of string
theory based on a construction known as F-theory. This class of
compactification is non-perturbative in the sense that the string
coupling necessarily becomes strong in some regions of the
internal
compact manifold, unlike conventional compactification where the
string coupling can be kept small everywhere on the internal
manifold.
\item
Microscopic derivation of the black hole entropy: In this 
section I shall discuss how many of the techniques and ideas that
were used to test various duality conjectures in string theory
can be used to give a microscopic derivation of the
Bekenstein-Hawking entropy and Hawking radiation from black
holes.
\item 
Matrix theory: In this final section I shall discuss a proposal
for a non-perturbative definition of M-theory and various other
string theories in terms of quantum mechanics of $N\times N$
matrices in the large $N$ limit.

\end{enumerate}

Throughout this article I shall work in units where $\hbar=1$
and $c=1$.

\sectiono{A Brief Review of Perturbative String Theory}
\label{spert}

String theory is based on the simple idea that elementary
particles, which appear as point-like objects to the present day
experimentalists, are actually different vibrational modes of
strings. 
The energy per unit length of the string, known as string
tension, is parametrized as $(2\pi\alpha')^{-1}$, where $\alpha'$
has the dimension of (length)$^2$. As we shall describe later,
this theory automatically contains gravitational interaction
between elmentary particles, but in order to correctly reproduce
the strength of this interaction, we need to choose
$\sqrt{\alpha'}$ to be of the order of $10^{-33}cm$. Since
$\sqrt{\alpha'}$ is the only length parameter in the theory, the
typical size of a string is of the order of $\sqrt{\alpha'}\sim
10^{-33}cm$
$-$ a distance that cannot be resolved by present
day experiments. Thus there is no direct way of testing string
theory, and its appeal lies in its theoretical consistency.
\begin{figure}[!ht] 
\begin{center}
\leavevmode
\epsfbox{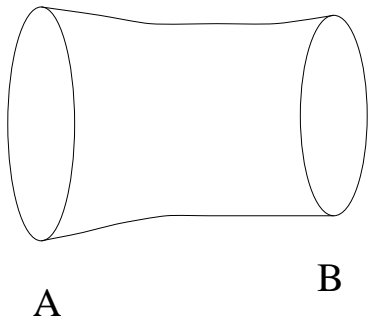}
\end{center}
\caption[]{\small Propagation of a closed string.} 
\label{fp4}
\end{figure}
\begin{figure}[!ht] 
\begin{center}
\leavevmode
\epsfbox{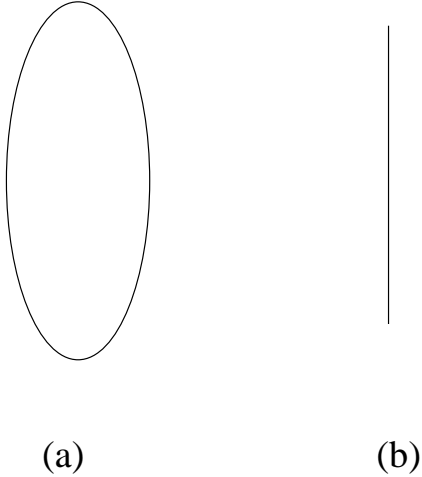}
\end{center}
\caption[]{\small (a) A closed string, and (b) an open string.} 
\label{fp1}
\end{figure}

The basic principle behind constructing a quantum theory of
relativistic string is quite simple. Consider propagation of a
string from a space-time configuration A to a space-time
configuration B. During this motion the string sweeps out a two
dimensional surface in space-time, known as the string
world-sheet (see Fig.\ref{fp4}). The amplitude for the
propagation of the string from the space-time position A to space-time
position B is given by the weighted sum over all world-sheet
bounded by the initial and the final locations of the string. The
weight factor is given by $e^{-S}$ where $S$ is the product of
the string tension and the area of the world-sheet. It turns out
that this procedure by itself does not give rise to a fully
consistent string theory. In order to get a fully consistent
string theory we need to add some internal fermionic degrees of
freedom to the string and generalize the notion of area by adding
new terms involving these fermionic degrees of freedom. The leads
to five (apparently) different consistent string theories in
(9+1) dimensional space-time, as we shall describe.

In the first quantized formalism, the dynamics of a point
particle
is described by quantum mechanics. Generalizing this we see that
the first quantized description of a string will involve a (1+1)
dimensional quantum field theory. However unlike a conventional
quantum field theory where the spatial directions have infinite
extent, here the spatial direction, which labels the coordinate
on the string, has finite extent. It represents a compact circle if
the string is closed (Fig.\ref{fp1}(a)) and a finite line interval
if the string is open (Fig.\ref{fp1}(b)). This (1+1) dimensional
field theory is known as the world-sheet theory.
The fields in this (1+1) dimensional
quantum field theory and the boundary conditions on
these fields vary in different string theories. 
Since the spatial direction of the
world-sheet theory has finite extent, each world-sheet field can
be regarded as a collection of infinite number of harmonic
oscillators labelled by the quantized momentum along this spatial
direction. Different states of the string are obtained by acting
on the Fock vacuum by these oscillators. This gives an infinite
tower of states.  Typically each string theory
contains a set of massless states and an infinite tower of
massive states. 
The massive string states typically have mass of the
order of $(10^{-33}cm)^{-1}\sim 10^{19}GeV$ and are far beyond
the reach of the present day accelerators. Thus the interesting
part of the theory is the one involving the massless states. We
shall now briefly describe the spectrum and interaction in
various string theories and their compactifications.

\subsection{The spectrum} \label{sspec}

There are five known
fully consistent string theories in ten dimensions. They are
known as
type IIA, type IIB, type I, $E_8\times E_8$ heterotic and
SO(32) heterotic string theories respectively. 
Here we give
a brief description of the degrees of freedom and the
spectrum of massless states in each of these
theories. We shall give the description in the so called
light-cone gauge which has the advantage that all states in the
spectrum are physical states. 
\begin{enumerate}
\item{Type II string theories}: In this case
the world-sheet theory is a free
field theory containing eight scalar fields and eight Majorana
fermions. These eight scalar fields are in fact common to all
five string theories, and represent the eight transverse
coordinates of a string moving in a nine dimensional space. It is
useful to regard the
eight Majorana fermions as sixteen Majorana-Weyl fermions, eight
of them having left-handed chirality and the other eight having
right-handed chirality. We shall refer to these as left- and
right-moving fermions respectively.
Both the type II string 
theories contain only closed strings; hence the spatial
component of the world-sheet is a circle. The eight
scalar fields satisfy
periodic boundary condition as we go around the circle. The
fermions have a choice of having periodic or anti-periodic
boundary conditions. It is customary to refer to periodic 
boundary condition as Ramond (R) boundary condition\cite{RAMOND}
and 
anti-periodic boundary condition as Neveu-Schwarz (NS) boundary
condition\cite{NEVSCH}. It turns out that in order to get a
consistent string theory we need to include in our theory different
classes of string states, some of which have periodic and some of
which have anti-periodic boundary condition on the fermions. 
In all there are four classes of states which need to be
included in the spectrum: 
\begin{itemize}
\item NS-NS where we put anti-periodic boundary conditions on
both the left- and the right-moving fermions,
\item NS-R where we put anti-periodic boundary condition on
the left-moving fermions and periodic boundary condition on the
right-moving fermions,
\item R-NS where we put periodic boundary condition on
the left-moving fermions and anti-periodic boundary condition on
the right-moving fermions,
\item R-R where we put anti-periodic boundary conditions on
both the left- and the right-moving fermions.
\end{itemize}
Finally, we keep only about (1/4)th of the states in each sector
by keeping only those states in the spectrum which have in them
only even number of left-moving fermions and even number of
right-moving fermions. This is known as the GSO 
projection\cite{GSO}. The
procedure has some ambiguity since in each of the four sectors
we have the
choice of assigning to the ground state either even or odd fermion
number. 
Consistency of string theory rules out most of these
possibilities, but at the end two possibilities remain. These
differ from each other in the following way. In one possibility,
the assignment of the left- and the right-moving fermion number
to the left- and the right-moving Ramond ground states are 
carried out in
an identical manner. This gives type IIB string theory. In the
second possibility the GSO projections in the left- and the
right-moving sector differ from each other. This theory is known
as type IIA string theory. 

Typically states from the Ramond sector are in the spinor
representation of the SO(9,1)
Lorentz algebra, whereas those from the NS
sector are in the tensor representation. Since the product of two
spinor representation gives us back a tensor representation, the
states from the NS-NS and the RR sectors are bosonic, and those
from the NS-R and R-NS sectors are fermionic. It will be
useful to list
the massless bosonic states in these two string theories. Since
the two theories differ only in their R-sector, the NS sector
bosonic states are the same in the two theories. They constitute
a symmetric rank two tensor field, an anti-symmetric rank two tensor
field, and a scalar field
known as the dilaton.\footnote{Although from
string theory we get the spectrum of states, it is useful to organise
the spectrum in terms of fields. In other words the spectrum of
massless fields in string theory is
identical to that of a free field theory with these fields.}
The RR sector massless states of type
IIA string theory consist of a vector, and a rank three
anti-symmetric tensor. On the other hand, the massless states from 
the RR sector of type IIB string theory consist of a 
scalar, a rank two anti-symmetric tensor field, and a
rank four anti-symmetric tensor gauge field satisfying the
constraint that its field strength is self-dual.

The spectrum of both these theories are invariant under
space-time supersymmetry transformations which transform
fermionic states to bosonic states and vice-versa. The
supersymmetry algebra for type IIB theory is known as the chiral
N=2 superalgebra and that of type IIA theory is known as the
non-chiral N=2 superalgebra. Both superalgebras consist of 32
supersymmetry generators.

Often it is convenient to organise the infinite tower of
states in string theory by their
oscillator level defined as follows. 
As has already been pointed out before, the world-sheet degrees
of freedom of the string can be regarded as a collection of
infinite number of harmonic oscillators.
For the creation operator associated with each
oscillator
we define the level as the absolute value of the number of units of
world-sheet momentum that it creates while acting on the vacuum.
The total oscillator level of a state is then the sum of the
levels of all the oscillators that act on the Fock vacuum to
create this state. (The Fock vacuum, in turn, is characterized
by several quantum numbers, which are the momenta conjugate to
the zero modes of various fields $-$ modes carrying zero world-sheet
momentum.)
We can also separately define left- (right-) moving
oscillator level as the contribution to the oscillator level from the
left- (right-) moving bosonic and fermionic fields. Finally, if
$E$ and $P$ denote respectively the
world-sheet energy and momentum\footnote{We should distinguish
between world-sheet momentum, and the
momenta of the (9+1) dimensional theory. The latter are the
the momenta conjugate to the zero modes of various bosonic fields
in the world-sheet theory.} then we
define $L_0=(E+P)/2$ and $\bar L_0=(E-P)/2$. $L_0$ and $\bar L_0$
include contribution from the oscillators as well as from the
Fock vacuum. Thus for example the total contribution to $L_0$
will be given by the sum of the right-moving oscillator level and
the contribution to $L_0$ from the Fock vacuum.

\item Heterotic string theories: The world-sheet theory of the
heterotic string theories consists of eight scalar fields, eight
right-moving Majorana-Weyl fermions and thirty two left-moving
Majorana-Weyl fermions. We have as before NS and R boundary
conditions as well as GSO projection involving the right-moving
fermions. Also as in the case of type II string theories, the NS
sector states transform in the tensor representation and the R
sector states transform in the spinor representation of the
SO(9,1) Lorentz algebra. However, unlike in the case of type II
string theories, in this case the boundary condition on the
left-moving fermions do not affect the Lorentz transformation
properties of the state. Thus bosonic states come from states
with NS boundary condition on the right-moving fermions and
fermionic states come from states with R boundary condition on
the right-moving fermions. 

There are two possible boundary conditions on the left-moving
fermions which give rise to fully consistent string theories.
They are:
\begin{itemize} 
\item SO(32) heterotic string theory: In this case we
have two possible boundary conditions on the left-moving
fermions: either all of them
have periodic boundary condition, or all of them
have anti-periodic boundary condition. In each sector we
also have a GSO projection that keeps only those states in the
spectrum which contain even number of left-moving fermions. 
The massless bosonic states in this theory consist of a symmetric
rank two
field, an anti-symmetric rank two field, a scalar field
known as the
dilaton and a set of 496 gauge fields filling up the adjoint
representation of the gauge group SO(32).

\item
$E_8\times E_8$ heterotic string theory: In this case
we divide the thirty two
left-moving fermions into two groups of sixteen each and use four
possible boundary conditions, 1) all the left-moving
fermions have periodic boundary condition 2) all the left-moving
fermions have anti-periodic boundary condition, 3) all the
left-moving fermions in group 1 have periodic boundary conditions
and all the left-moving fermions in group 2 have anti-periodic
boundary conditions, 4) 
all the
left-moving fermions in group 1 have anti-periodic boundary conditions
and all the left-moving fermions from group 2 have periodic
boundary conditions.  In each sector we
also have a GSO projection that keeps only those states in the
spectrum which contain even number of left-moving fermions from the
first group, and also even number of left-moving fermions from the
second group. 
The massless bosonic states in this theory consist of a symmetric
rank two field, an anti-symmetric rank two field, a scalar
field known as the
dilaton and a set of 496 gauge fields filling up the adjoint
representation of the gauge group $E_8\times E_8$.
\end{itemize}

The spectrum of states in
both the heterotic string theories are invariant under a set of
space-time supersymmetry transformations. The relevant
superalgebra is known as the chiral N=1 supersymmetry algebra,
and has sixteen real generators.

Using the bose-fermi equivalence in (1+1) dimensions, we can
reformulate both the heterotic string theories by replacing the
thirty two left-moving fermions by sixteen left-moving
bosons. In order to get a consistent string theory the momenta
conjugate to these bosons must take discrete values. It turns out
that there are only two consistent ways of quantizing the
momenta, giving us back the two heterotic string theories.

\item Type I string theory: The world-sheet theory of type I
theory is identical to that of type IIB string theory, with the
following two crucial difference.
\begin{itemize}
\item
Type IIB string theory has a symmetry that exchanges the left-
and the right-moving sectors in the world-sheet theory. This
transformation is known as the world-sheet parity transformation.
(This
symmetry is not present in type IIA theory since the GSO
projection in the two sectors are different). In constructing
type I string theory we keep only those states in the spectrum
which are invariant under this world-sheet parity transformation.
\item
In type I string theory we also include open string states in the
spectrum. The world-sheet degrees of freedom are identical to
those in
the closed string sector. Specifying the theory requires us to
specify the boundary conditions on the various fields. We put
Neumann boundary condition on the eight scalars, and appropriate
boundary conditions on the fermions. 
\end{itemize}

The spectrum of massless
bosonic states in this theory consists of a symmetric rank two
tensor and a scalar dilaton
from the closed string NS sector, an anti-symmetric rank two tensor 
from the closed string RR sector, and 496 gauge fields in the
adjoint representation of SO(32) from the open string sector.
This spectrum is also invariant under the chiral N=1
supersymmetry algebra with sixteen real supersymmetry generators.

\end{enumerate}
\begin{figure}[!ht] 
\begin{center}
\leavevmode
\epsfbox{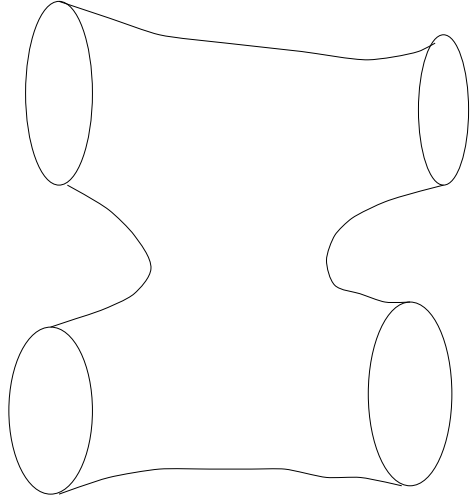}
\end{center}
\caption[]{\small A string world-sheet bounded by four external
strings.}
\label{fp2}
\end{figure}

\subsection{Interactions}

\begin{figure}[!ht] 
\begin{center}
\leavevmode
\epsfbox{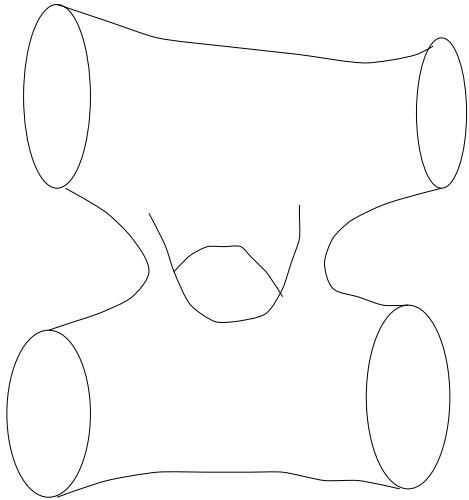}
\end{center}
\caption[]{\small A more complicated string world-sheet.} 
\label{fp3}
\end{figure}

So far we have discussed the spectrum of string theory, but in
order to fully describe the theory we must also describe the
interaction between various particles in the spectrum. In
particular, we would like to know how to compute a scattering
amplitude involving various string states. It turns out that
there is a unique way of introducing interaction in string
theory. Consider for example a scattering involving four external
strings, situated along some specific curves in space-time. 
The prescription for computing the scattering amplitude is
to compute the weighted sum over all possible string world-sheet
bounded by the four strings with weight factor $e^{-S}$, $S$
being the
string tension multiplied by the
generalized area of this surface (taking into account the fermionic 
degrees of freedom of the world-sheet). One such surface is shown
in Fig.\ref{fp2}. If we imagine the time axis running from left
to right, then this diagram represents two strings joining into
one string and then splitting into two strings, $-$ the analog of
a tree diagram in field theory. A more complicated surface is
shown in Fig.\ref{fp3}. This represents two strings joining into
one string, which then splits into two and joins again, and
finally splits into two strings. This is the analog of a one loop
diagram in field theory. The relative normalization between the
contributions from these two diagrams is not determined by any
consistency requirement. This introduces an arbitrary parameter
in string theory, known as the string coupling constant. However,
once the relative normalization between these two diagrams is
fixed, the relative normalization between all other diagrams is
fixed due to various consistency requirement. Thus besides the
dimensionful parameter $\alpha'$,  string theory
has a single dimensionless coupling constant. As we shall see
later, both these parameters can be absorbed into definitions of
various fields in the theory.

What we have described so far is the computation of the
scattering amplitude with fixed locations of the external strings
in space-time. The more relevant quantity is the
scattering amplitude where the external strings are in the 
eigenstates of the energy and momenta operators
conjugate to the coordinates
of the (9+1) dimensional space-time. 
This is done by simply taking the convolution of the
above scattering amplitude with the wave-functions of the strings
corresponding to the external states. In practice there is an
extremely efficient method of doing this computation using the so
called vertex operators. It turns out that unlike in quantum
field theory, all of these scattering amplitudes in string theory
are ultraviolet finite. This is one of the major achievements of
string theory.

Out main interest will be in the scattering involving the
external massless states. The most convenient way to summarize
the result of this computation in any string theory is to
specify the effective action. By definition this
effective action is such that if we compute the {\it tree level}
scattering amplitude using this action, we should reproduce the
S-matrix elements involving the massless states of string theory.
In general such an action will have to contain infinite number of
terms, but we can organise these terms by examining the number of
space-time derivatives that appear in a given term in the action.
Terms with the lowest number of
derivatives constitute the {\it low energy effective action}, $-$
so called because this gives the dominant contribution if we want
to evaluate the scattering amplitude when all the external
particles have small energy and momenta.

The low energy effective action for all five string theories have
been found. The actions for the type IIA and type IIB string
theories correspond to those of two well known supergravity
theories in ten space-time dimensions, called type IIA and type
IIB supergravity theories respectively. On the other hand the
actions for the three heterotic string theories correspond to
another set of
well-known supersymmetric theories in ten dimensions, $-$ $N=1$
supergravity coupled to N=1 super Yang-Mills theory. For type I
and the SO(32) heterotic string theories
the Yang-Mills gauge group is SO(32)
whereas for the $E_8\times E_8$ heterotic string theory the gauge
group is $E_8\times E_8$. The emergence of gravity in all the
five string theories is the most striking result in string theory.
Its origin can
be traced to the existence of the symmetric rank two tensor state
(the graviton) in all these theories.
This, combined with the result on finiteness of scattering
amplitudes, shows that string theory gives us a finite quantum
theory of gravity.
We shall explicitly write down the low energy effective action of
some of the string theories in section \ref{s2}.

The effective action of all five string theories are invariant
under the transformation 
\be \label{epri1}
\Phi\to\Phi -2C, \qquad g_S\to e^Cg_S,
\ee
together
with possible rescaling of other fields. Here $\Phi$ denotes the
dilaton field, $g_S$ denotes the string coupling, and $C$ is an
arbitrary constant. Using this scaling property, $g_S$ can be
absorbed in $\Phi$. Put another way, the dimensionless coupling
constant in string theory is related to the vacuum expectation
value $\langle\Phi\rangle$ of $\Phi$. The perturbative effective
action does not have any potential for $\Phi$, and hence
$\langle\Phi\rangle$ can take arbitrary value. One
expects that in a realistic string theory where supersymmetry is
spontaneouly broken, there will be a potential for $\Phi$, and
hence $\langle\Phi\rangle$ will be determined uniquely.

In a similar vain one can argue that
in string theory even the string
tension, or equivalently the parameter $\alpha'$, has no physical
significance. Since $\alpha'$ has the dimension of (length)$^2$
and is the only dimensionful parameter in the theory, the
effective action will have an invariance under the simultaneous
rescaling of $\alpha'$ and the metric $g_{\mu\nu}$:
\be \label{epri2}
\alpha'\to\lambda\alpha', \qquad g_{\mu\nu}\to \lambda
g_{\mu\nu}\, ,
\ee
together with possible rescaling of other fields. Using this
scaling symmetry $\alpha'$ can be absorbed into the definition of
$g_{\mu\nu}$. We shall discuss these two rescalings in detail in
section \ref{s22}. 

\subsection{Compactification}

So far we have described five different string theories, but they
all live in ten space-time dimensions. Since our world is (3+1)
dimensional, these are not realistic string theories. However one
can construct string theories in lower dimensions using the idea
of compactification. The idea is to take the (9+1) dimensional
space-time
as the product of a $(9-d)$ dimensional compact manifold $\MM$
with euclidean signature
and a $(d+1)$ dimensional Minkowski space
$R^{d,1}$. Then, in the limit
when the size of the compact manifold is sufficiently small so
that the present day experiments cannot resolve this distance,
the world will effectively appear to be $(d+1)$ dimensional.
Choosing $d=3$ will give us a (3+1) dimensional theory.
Of course we cannot choose any arbitrary manifold $\MM$ for
this purpose; it must
satisfy the equations of motion of the effective field
theory that comes out of string theory.
One also normally considers only those manifolds which preserve
part of the space-time supersymmetry of the original ten
dimensional theory, since this guarantees vanishing of the
cosmological constant, and hence consistency of the corresponding
string theory order by order in perturbation theory.
There are many known examples of manifolds satisfying these
restrictions {\it e.g.} tori of different 
dimensions, K3, Calabi-Yau manifolds etc.
Instead of going via the effective action, one can also directly
describe these compactified theories as string theories. For this
one needs to
modify the string world-sheet action in such a way that it
describes string propagation in the new manifold
$\MM\times R^{d,1}$, instead of in
flat ten dimensional space-time. This modifies the world-sheet
theory to an interacting non-linear $\sigma$-model instead of a
free field theory. Consistency of string theory puts
restriction on the kind of manifold on which the string can propagate.
At the end both approaches yield identical results.

The simplest class of compact manifolds, on which we shall focus
much of our attention in the rest of this article, 
are tori $-$ product
of circles. The effect of this compactification
is to periodically identify some
of the bosonic fields in the string world-sheet field
theory $-$ the fields which represent
coordinates tangential to the
compact circles. One effect of this is that the momentum carried
by any string state along any of these circles is quantized in
units of $1/R$ where $R$ is the radius of the circle. But that is
another novel effect: we now have new states that correspond to
strings wrapped around a compact circle. For such a states, as we
go once around the string, we also go once around the compact
circle. These states are known as winding states and play a
crucial role in the analysis of duality symmetries.

\sectiono{Notion of Duality Symmetries in String Theory}
\label{s1}

In this section I shall elaborate the notion of 
duality symmetries, the difficulties in testing them, and the way
of avoiding these difficulties. 
We begin by introducing the notion of duality in string theory.

\subsection{Duality symmetries: Definition and examples}
\label{ssss1}

\begin{figure}[!ht] 
\begin{center}
\leavevmode
\epsfbox{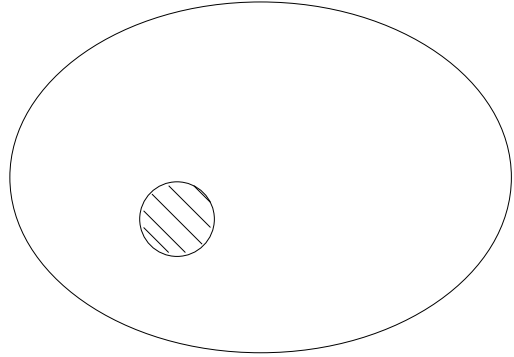}
\end{center}
\caption[]{\small A schematic representation of the moduli space
of a string theory. The shaded region denotes the weak coupling
region, whereas the white region denotes the strong coupling
region.  } \label{f1}
\end{figure}

As was described in the last section, 
there are five consistent string theories
in ten space-time dimensions. 
We also saw that
we can get many different string theories in lower dimensions by
compactifying these five theories on appropriate manifold $\MM$.
Each of these theories is parametrized by a set of parameters
known as moduli\footnote{In string theory these moduli are
related to vacuum expectation values of various dynamical fields 
and are expected to take definite values when supersymmetry is
broken.}
 {\it e.g.}
\begin{itemize}
\item String coupling constant (related to the vacuum expectation value
of the dilaton field),
\item Shape and size of $\MM$ (information contained in the
metric),
\item various other background fields.
\end{itemize}
Inside the moduli space of the theory there is a certain region where the
string coupling is weak and perturbation theory is valid.
Elsewhere the theory is strongly coupled. This situation has been
illustrated in fig.\ref{f1}.
\begin{figure}[!ht] 
\begin{center}
\leavevmode
\epsfbox{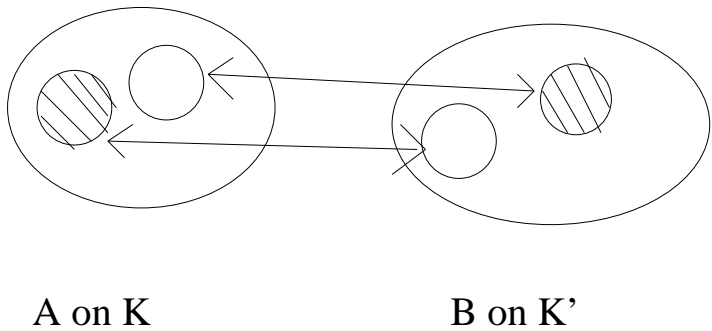}
\end{center}
\caption[]{\small A schematic representation of the duality map 
between the moduli spaces of
two different string theories, $A$ on $K$ and $B$ on
$K'$, where $A$ and $B$ are two of the five string theories in
ten dimensions, and $K$, $K'$ are two compact manifolds. Under
this duality the weak coupling region of the first theory
(denoted by the shaded region) gets mapped to the strong coupling
region of the second theory and vice versa.} \label{f2}
\end{figure}

String duality provides us with an equivalence map between two 
different string theories. In 
general this equivalence relation
maps the weak coupling region of one theory to the
strong coupling region of the second theory and vice versa. This
situation is illustrated in fig.\ref{f2}.
\begin{figure}[!ht] 
\begin{center}
\leavevmode
\epsfbox{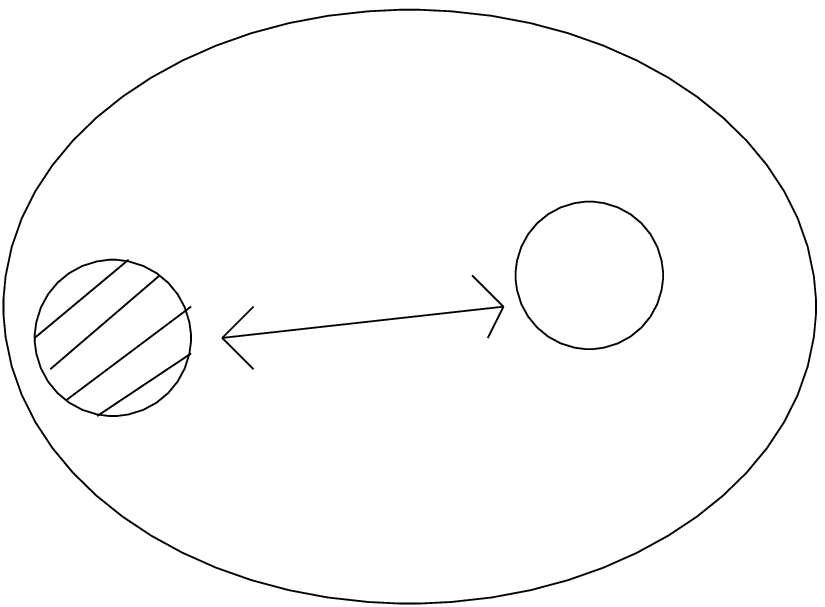}
\end{center}
\caption[]{\small 
Schematic representation of the 
moduli space of a self-dual theory. Duality relates 
weak and strong coupling regions of the same theory.} \label{f3}
\end{figure}
\begin{figure}[!ht] 
\begin{center}
\leavevmode
\epsfbox{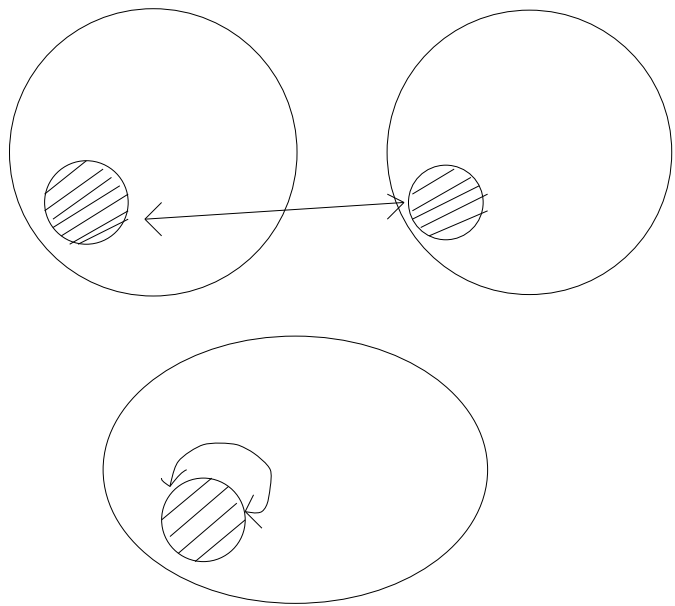}
\end{center}
\caption[]{\small 
Examples of T-duality relating a weakly coupled theory to a
different or the same weakly coupled theory.}
\label{f4}
\end{figure}
\begin{figure}[!ht] 
\begin{center}
\leavevmode
\epsfbox{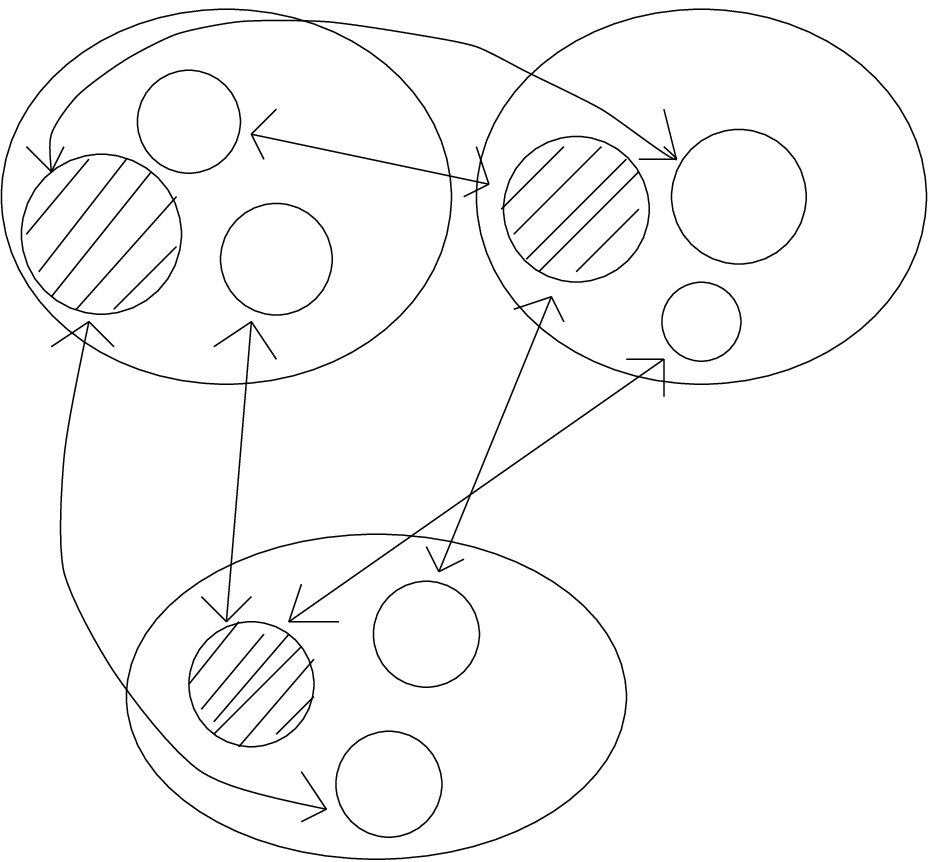}
\end{center}
\caption[]{\small A schematic representation of the moduli spaces
of a chain of theories related by duality. In each case the shaded
region denotes weak coupling region as usual.}
\label{f5}
\end{figure}

Before we proceed, let us give a few examples of dual pairs:
\begin{itemize}
\item Type I and SO(32) heterotic string theories in D=10 are
conjectured to be
dual to each other\cite{WITTEND,DABH,HULLOPEN,POLCWIT}.
\item
Type IIA string theory compactified on K3 and heterotic string theory
compactified on a four dimensional torus $T^4$ are conjectured to
be dual to each
other\cite{HULLTOWN,DUFFSS,WITTEND,SSSD,HARSTRSSD}.\footnote{Throughout
this article a string theory on $\MM$ will mean string 
theory in the background
$\MM\times R^{9-n,1}$ where $n$ is the real dimension of $\MM$,
and $R^{9-n,1}$ denotes $(10-n)$
dimensional Minkowski space.}
\end{itemize}
Under duality, typically
perturbation expansions get mixed up. Thus for example,
tree level results in one theory might include perturbative 
and non-perturbative corrections in the dual theory.
Also under duality,
many of the elementary string states in one theory get mapped
to solitons and their bound states in the dual theory.

Although duality in general relates the weak coupling limit of
one theory to the strong coupling limit of another theory, there
are special cases where the situation is a bit different. For
example,  we can have:
\begin{itemize}
\item Self-duality:
Here duality gives an equivalence relation between different
regions of the moduli space of the same theory, as illustrated in
fig.\ref{f3}.
In this case, duality transformations form a symmetry group
that acts on the moduli space of the theory. For example,
type IIB string theory in D=10 is conjectured to
have an SL(2,Z) self-duality 
group\cite{HULLTOWN}.
\item T-duality: 
In this case duality transformation maps the weak coupling region
of one theory to the weak coupling region of another theory or the
same theory as illustrated in fig.\ref{f4}.
For example, type IIA string theory compactified
on a circle of radius $R$ is dual to IIB string theory compactified
on a circle of radius $R^{-1}$ at the same value of the string
coupling.  Also, 
either of the two heterotic string theories compactified on a circle of
radius $R$ is dual to the same theory compactified on a circle of
radius $R^{-1}$ at the 
same value of the coupling constant. As a result the duality map
does not mix up the perturbation expansions in the two theories.
(For a review of this subject, see \cite{GIVETC}.)
\end{itemize}
In a generic situation duality can relate not just two theories, but a
whole chain of theories, as illustrated in fig.\ref{f5}.
Thus for example, type
IIA string theory compactified on K3 is related to heterotic string
theory compactified on $T^4$. On the other hand, due to the 
equivalence of the SO(32) heterotic and type I string theory in
ten dimensions, SO(32)
heterotic string theory compactified on $T^4$ is
related to type I string theory compactified on $T^4$. Thus these
three theories are related by a chain of duality transformations.

{}From this discussion we see that the presence of duality in
string theory has two important consequences. First of all, 
it reduces the degree of non-uniqueness of string theory, by
relating various apparently unrelated (compactified) string theories.
Furthermore, it
allows us to study a strongly coupled string theory by mapping
it to a weakly coupled dual theory whenever such a dual theory
exists. 

\subsection{Testing duality conjectures} \label{ssss2}

Let us now turn to the question of testing duality.
As we have already emphasized,
duality typically relates a weakly coupled string theory to a
strongly coupled string theory. 
Thus in order to prove / test duality we
must be able to analyze at least one of the theories at strong coupling.
But in string theory we only know how to define the theory
perturbatively at weak coupling. 
Thus it would seem impossible to prove or test any duality
conjecture in string theory.\footnote{
Note that
this problem is absent for T-duality transformations which relates
two weakly coupled string theories, and hence can be
tested using string perturbation theory.
All T-duality symmetries in string theory can be `proved' this
way, at least to all orders in perturbation theory.}
This is where supersymmetry comes to our rescue.
Supersymmetry gives rise to
certain non-renormalization theorems in string theory, due to
which
some of the weak coupling calculations can be trusted even at
strong coupling.
Thus we can focus our attention on such `non-renormalized' quantities
and ask if they are invariant under the proposed duality transformations.
Testing duality invariance of these quantities
provides us with various tests of various duality conjectures, and 
is in fact the basis of all duality conjectures.

The precise content of these non-renormalization theorems depends
on the number of supersymmetries present in the theory.
The maximum number of supersymmetry generators that can be
present in a string theory is 32.
This gives N=2 supersymmetry in ten dimensions, and N=8 
supersymmetry in four dimensions.
Examples of such theories are
type IIA or type IIB string theories compactified on 
$n$ dimensional tori $T^n$.
The next interesting class of theories are those
with 16 supersymmetry generators.  This
corresponds to N=1 supersymmetry in ten dimensions and N=4 
supersymmetry in four dimensions.
Examples of such theories are
type IIA or type IIB string theories compactified on $K3\times T^n$, 
heterotic string theory compactified on $T^n$, etc.
Another class of theories that we shall discuss are those with
eight supersymmetry generators, {\it e.g.} heterotic string
theory on $K3\times T^n$, type IIA or IIB string theory on six
dimensional Calabi-Yau manifolds, etc.
For theories with 16 or more SUSY generators the non-renormalization 
theorems are particularly powerful.
In particular,
\begin{itemize}
\item
Form of the low energy effective action involving the massless states
of the theory is completely fixed by the requirement of supersymmetry
(and the spectrum)\cite{FREREV}.
Thus this effective action cannot get renormalized by string loop 
corrections.
As a result,
any valid symmetry of the theory must be a symmetry of this effective
field theory.

\item
These theories contain special class of states which are invariant
under part of the supersymmetry transformations.
They are known as BPS states, named after Bogomol'nyi, Prasad and 
Sommerfeld.
The mass of a BPS state is completely determined in terms of its charge
as a consequence of the supersymmetry algebra.
Since this relation is derived purely from an analysis of the
supersymmetry algebra, it is not modified by quantum corrections.
Furthermore it can be argued that
the degeneracy of BPS states of a given charge does not
change as we move in the moduli space even from weak to strong coupling
region\cite{WITOLI}.
Thus the spectrum of BPS states can be calculated from weak coupling
analysis and the result can be continued to the strong coupling
region. 
Since any valid symmetry of the theory must be a symmetry of the
spectrum of BPS states, we can use this to design non-trivial
tests of duality\cite{REVB}.
\end{itemize}

For theories with eight supersymmetries the non-renormalization
theorems are less powerful. However, even in this case one can
design non-trivial tests of various duality conjectures. We shall
discuss these in section \ref{s5}.

\sectiono{Analysis of Low Energy Effective Field Theory}
\label{s2}

In this section I shall discuss tests of various dualities in
string theories with $\ge 16$ supersymmetries based on the
analysis of their low energy effective action. As has been
emphasized in the previous section, the form of this low energy
effective action is determined completely by the requirement of
supersymmetry and the spectrum of massless states in the theory.
Thus it does not receive any quantum corrections, and if a given
duality transformation is to be a symmetry of a string theory, 
it must be a symmetry of the corresponding low energy effective
action. Actually, since the low energy {\it effective action} is to
be used only for deriving the equations of motion from this
action, and/or computing the tree level S-matrix elements using
this action, but not to perform a full-fledged path integral,
it is enough that only the equations of motion
derived from this action are invariant under duality
transformations. 
(This also guarantees that the tree level S-matrix elements
computed from this effective action are invariant under the
duality transformations.)
It is not necessary for the action itself to be
invariant. 

Throughout this article we shall denote by $G_{\mu\nu}$ the string
metric $-$ the metric that is used in computing the area of the string
world-sheet embedded in space time for calculating string scattering
amplitudes. 
For a string theory compactified on a $(9-d)$ dimensional
manifold $\MM$, we shall denote by $\Phi$ the shifted dilaton,
related to the dilaton $\Phi^{(10)}$ of the ten dimensional string
theory as
\be \label{edil}
\Phi = \Phi^{(10)} - \ln V\, ,
\ee
where $(2\pi)^{9-d}V$ is the volume of $\MM$ measured in the 
ten dimensional string metric. 
The dilaton is normalized in such a way that 
$e^{\langle\Phi^{(10)}\rangle }$ corresponds to the square of the
closed string coupling constant in ten
dimensions.\footnote{$\Phi$ is related to the more commonly
normalized dilaton $\phi$ by a factor of two: $\Phi=2\phi$.}
$g_{\mu\nu}$ will denote
the canonical Einstein metric which is related to the string metric
by an appropriate conformal rescaling involving the dilaton field, 
\be \label{egmunu}
g_{\mu\nu} = e^{-{2\over d-1}\Phi} G_{\mu\nu}\, .
\ee
We shall always use this metric to raise 
and lower indices. The signature of space-time will be taken as
$(-,+,\cdots +)$. Finally, all fields will be made dimensionless
by absorbing appropriate powers of $\alpha'$ in them.

We shall now consider several examples. The discussion will
closely follow refs.\cite{REVB,HULLTOWN,WITTEND}. For a detailed
review 
of the material covered in this section, see ref.\cite{DELO}.

\subsection{Type I - SO(32) heterotic duality in D=10}
\label{s22}
In SO(32) heterotic string theory, the 
massless bosonic states come from the NS sector of the closed
heterotic string, and contains the metric
$g^{(H)}_{\mu\nu}$, the dilaton $\Phi^{(H)}$, 
the rank two anti-symmetric tensor field $B^{(H)}_{\mu\nu}$, and 
gauge fields $A_\mu^{(H)a}$ ($1\le a\le 496$) in the adjoint
representation of SO(32). 
The low energy dynamics involving these massless
bosonic fields is
described by the N=1 supergravity coupled to SO(32) super Yang-Mills
theory in ten dimensions\cite{HETEROTIC}. 
The action is given by\cite{OPENREF}:
\ben \label{e8a}
S^{(H)}&=& {1\over (2\pi)^7 (\alpha'_H)^4 g_H^2} \int d^{10} x
\sqrt{-g^{(H)}} \Big[ R^{(H)} - 
{1\over 8} g^{(H)\mu\nu} \p_\mu\Phi^{(H)} 
\p_\nu\Phi^{(H)}
\nonumber \\
&& - {1\over 4}
g^{(H)\mu\mu'} g^{(H)\nu\nu'} e^{-\Phi^{(H)}/4} Tr(F^{(H)}_{\mu\nu}
F^{(H)}_{\mu'\nu'}) \nonumber \\
&& - {1\over 12}
g^{(H)\mu\mu'}g^{(H)\nu\nu'}g^{(H)\rho\rho'} e^{-\Phi^{(H)}/2}
H^{(H)}_{\mu\nu\rho}
H^{(H)}_{\mu'\nu'\rho'}\Big]  \, , \nonumber \\
\een
where $R^{(H)}$ is the Ricci scalar,
$F^{(H)}_{\mu\nu}$ denotes the non-abelian gauge field strength,
\be \label{efhet}
F^{(H)}_{\mu\nu}=\p_\mu A^{(H)}_\nu - \p_\nu A^{(H)}_\mu +
\sqrt{2\over \alpha'_H} \, 
[A^{(H)}_\mu, A^{(H)}_\nu]\, ,
\ee
$Tr$ denotes trace in the vector representation of SO(32),
and $H^{(H)}_{\mu\nu\rho}$ is the field strength associated with the
$B^{(H)}_{\mu\nu}$ field:
\ben \label{extwoa}
H^{(H)}_{\mu\nu\rho} &=& \p_\mu B^{(H)}_{\nu\rho} -
{1\over 2}
Tr(A^{(H)}_\mu F^{(H)}_{\nu\rho} - {1\over 3} \sqrt{2\over \alpha'_H}
\, A^{(H)}_\mu [A^{(H)}_\nu, A^{(H)}_\rho]) \nonumber \\
&& +\hbox{cyclic permutations of
$\mu, \nu, \rho$}\, .
\een
$2\pi\alpha'_H$ and $g_H$ are respectively
the inverse string tension and the
coupling constant of the heterotic string theory. 
The rescalings \refb{epri1}, \refb{epri2} take the following form
acting on the complete set of fields:
\ben \label{eppri1}
&& g_H\to e^C g_H, \qquad \Phi^{(H)}\to \Phi^{(H)}-2C, \qquad
g^{(H)}_{\mu\nu}\to e^{C/2} g^{(H)}_{\mu\nu} \nonumber \\
&& B^{(H)}_{\mu\nu}\to B^{(H)}_{\mu\nu}, \qquad A^{(H)a}_\mu\to
A^{(H)a}_\mu\, ,
\een
\ben \label{eppri2}
&& \alpha'_H\to \lambda \alpha'_H, \qquad \Phi^{(H)}\to \Phi^{(H)}, 
\qquad g^{(H)}_{\mu\nu}\to \lambda g^{(H)}_{\mu\nu} \nonumber \\
&& B^{(H)}_{\mu\nu}\to \lambda B^{(H)}_{\mu\nu}, \qquad 
A^{(H)a}_\mu\to \lambda^{1/2} A^{(H)a}_\mu\, ,
\een
Since $g_H$ and $\alpha'_H$ can be changed by this rescaling,
these parameters cannot have a universal significance.
In particular, we can absorb $g_H$ and $\alpha'_H$ into the
various fields by setting $e^{-C}=g_H$ and $\lambda=(\alpha'_H)^{-1}$
in \refb{eppri1}, \refb{eppri2}. This is equivalent to setting
$g_H=1$ and $\alpha'_H=1$. In this notation
the physical coupling constant is given by the vacuum expectation
value of $e^{\Phi^{(H)}/2}$, and 
the ADM mass per unit length of an infinitely long
straight string, measured in the
metric $ e^{\langle\Phi^{(H)}\rangle / 4} g^{(H)}_{\mu\nu}$ that
approaches the string metric $G^{(H)}_{\mu\nu}$
far away from the string,
is equal to $1/2\pi$.  By changing $\langle\Phi^{(H)}\rangle$
we can get all
possible values of string coupling, and using a metric that
differs from the one used here by a constant multiplicative
factor, we can get all possible values of the string tension.

For $\alpha'_H=1$ and $g_H=1$
eqs.\refb{e8a}-\refb{extwoa} take the form:
\ben \label{e8}
S^{(H)}&=& {1\over (2\pi)^7 } \int d^{10} x
\sqrt{-g^{(H)}} \Big[ R^{(H)} - {1\over 8} g^{(H)\mu\nu} \p_\mu\Phi^{(H)} 
\p_\nu\Phi^{(H)}
\nonumber \\
&& - {1\over 4}
g^{(H)\mu\mu'} g^{(H)\nu\nu'} e^{-\Phi^{(H)}/4} Tr(F^{(H)}_{\mu\nu}
F^{(H)}_{\mu'\nu'}) \nonumber \\
&& - {1\over 12}
g^{(H)\mu\mu'}g^{(H)\nu\nu'}g^{(H)\rho\rho'} e^{-\Phi^{(H)}/2}
H^{(H)}_{\mu\nu\rho}
H^{(H)}_{\mu'\nu'\rho'}\Big]  \, , \nonumber \\
\een
\be \label{efheta}
F^{(H)}_{\mu\nu}=\p_\mu A^{(H)}_\nu - \p_\nu A^{(H)}_\mu +
\sqrt 2
[A^{(H)}_\mu, A^{(H)}_\nu]\, ,
\ee
\ben \label{extwo}
H^{(H)}_{\mu\nu\rho} &=& \p_\mu B^{(H)}_{\nu\rho} -
{1\over 2}
Tr\Big(A^{(H)}_\mu F^{(H)}_{\nu\rho} - {\sqrt 2\over 3} 
A^{(H)}_\mu [A^{(H)}_\nu, A^{(H)}_\rho]\Big) \nonumber \\
&& +\hbox{cyclic permutations of
$\mu, \nu, \rho$}\, .
\een

Let us now turn to the type I string theory.
The massless bosonic states in type I theory
come from three different sectors. The
closed string Neveu-Schwarz $-$ Neveu-Schwarz (NS) sector gives the
metric $g^{(I)}_{\mu\nu}$ and the dilaton $\Phi^{(I)}$. The 
closed string Ramond-Ramond (RR) sector gives 
an anti-symmetric tensor field
$B^{(I)}_{\mu\nu}$. Besides these,
there are bosonic fields coming from the NS sector of the open
string. This sector gives rise to gauge fields
$A_\mu^{(I)a}$ ($a=1, \ldots 496$) in the adjoint representation
of the group SO(32). 
(The superscript $(I)$ refers to the fact that these are the
fields in the type I string theory.)
The low energy dynamics is
again described by the N=1 supergravity theory coupled to SO(32)
super Yang-Mills theory\cite{TYPEI}. 
But it is instructive to rewrite the
effective action in terms of the type I variables. 
For suitable choice of the string tension and the coupling
constant,
this is given by\cite{OPENREF}
\ben \label{e7}
S^{(I)}&=& {1\over (2\pi)^7 }
\int d^{10} x
\sqrt{-g^{(I)}} \Big[ R^{(I)} - {1\over 8}
g^{(I)\mu\nu} \p_\mu\Phi^{(I)} 
\p_\nu\Phi^{(I)}
\nonumber \\
&& - {1\over 4}
g^{(I)\mu\mu'} g^{(I)\nu\nu'} e^{\Phi^{(I)}/4} Tr(F^{(I)}_{\mu\nu}
F^{(I)}_{\mu'\nu'}) \nonumber \\ &&
- {1\over 12}
g^{(I)\mu\mu'}g^{(I)\nu\nu'}g^{(I)\rho\rho'} e^{\Phi^{(I)}/2}
H^{(I)}_{\mu\nu\rho}
H^{(I)}_{\mu'\nu'\rho'}\Big]  \, , \nonumber \\
\een
where $R^{(I)}$ is the Ricci scalar,
$F^{(I)}_{\mu\nu}$ denotes the 
non-abelian gauge field strength,
\be \label{efope}
F^{(I)}_{\mu\nu}=\p_\mu A^{(I)}_\nu - \p_\nu A^{(I)}_\mu +
\sqrt 2
[A^{(I)}_\mu, A^{(I)}_\nu]\, ,
\ee
and $H^{(I)}_{\mu\nu\rho}$ is the field strength associated with the
$B^{(I)}_{\mu\nu}$ field:
\ben \label{exone}
H^{(I)}_{\mu\nu\rho} &=& \p_\mu B^{(I)}_{\nu\rho} - {1\over 2}
Tr\Big(A^{(I)}_\mu F^{(I)}_{\nu\rho} - 
{\sqrt 2\over 3} A^{(I)}_\mu [A^{(I)}_\nu, A^{(I)}_\rho]\Big) 
\nonumber \\
&& +\hbox{cyclic permutations of
$\mu, \nu, \rho$}\, .
\een

For both, the type I and the SO(32) heterotic string theory, the
low energy effective action is derived from the string tree level
analysis. However, to this order in the derivatives, the form of
the effective action is determined completely by the requirement
of supersymmetry for a given gauge group. Thus neither action can
receive any quantum corrections.

It is straightforward to see that the
actions \refb{e8} and
\refb{e7} are identical provided we make the identification:
\ben \label{e9}
&& \Phi^{(H)}=-\Phi^{(I)}, \qquad
g^{(H)}_{\mu\nu}=g^{(I)}_{\mu\nu}
\nonumber \\
&& B^{(H)}_{\mu\nu}=B^{(I)}_{\mu\nu}, \qquad A_\mu^{(H)a}=A_\mu^{(I)a}
\, .
\een
This led to the hypothesis that
the type I and the SO(32) heterotic
string theories in ten dimensions are equivalent\cite{WITTEND}.
One can find stronger evidence for this hypothesis by analysing
the
spectrum of supersymmetris states, but the equivalence of the two
effective actions was the reason for proposing this duality in
the first place.

Note the $-$ sign in the relation between $\Phi^{(H)}$ and
$\Phi^{(I)}$ in eq.\refb{e9}.
Recalling that $e^{\langle\Phi\rangle /2}$ 
is the string coupling, we see that  
the strong coupling limit of one theory is related to
the weak coupling limit of the other theory and vice versa.

{}From now on I shall use the unit $\alpha'=1$ for writing
down the effective action of all
string theories. Physically this would mean that
the ADM mass per unit length of a test string, measured in
the metric $e^{2\langle \Phi\rangle /(d-1)}g_{\mu\nu}$ that
agrees with the string metric $G_{\mu\nu}$ defined in
\refb{egmunu} far away from the test string,
is given by $1/2\pi$. In future we shall refer to the ADM mass of a
particle measured in this metric as the mass measured in the
string metric.

\subsection{Self-duality of heterotic string theory on $T^6$}
\label{sh1}

In the previous subsection we have described the massless bosonic
field content of the ten dimensional SO(32) heterotic string
theory. When we compactify it on a six dimensional torus, we can
get many other massless scalar fields from the internal
components of the metric, the anti-symmetric tensor field and the
gauge fields in the Cartan subalgebra of the gauge
group.\footnote{Only the sixteen gauge fields in the Cartan subalgebra
of the gauge group can develop vacuum expectation value since
such vacuum expectation values do not generate any field strength, and
hence do not generate energy density.}  
This gives a total of (21+15+96=132) scalar fields.
It turns out that these scalars can be represented
by a $28\times 28$ matrix valued field $M$
satisfying\footnote{For a review of this construction, see
\cite{REVB}.}
\be \label{eh1}
MLM^T = L, \qquad M^T=M\, ,
\ee
where
\be \label{eh2}
L = \pmatrix{ & I_6 & \cr I_6 && \cr & & -I_{16}}\, .
\ee
$I_n$ denotes an $n\times n$ identity matrix. 
We shall choose a convention in which $M=I_{28}$ corresponds to a
compactification on $(S^1)^6$ with each $S^1$ having 
radius $\sqrt{\alpha'}=1$ measured in the string metric, 
and without any background gauge or
antisymmetric tensor fields.
We can get another
scalar field $a$ by dualizing the gauge invariant
field strength $H$ of the antisymmetrix
tensor field through the relation:
\be \label{eh3}
H^{\mu\nu\rho}= -(\sqrt{-g})^{-1} e^{2\Phi}
\epsilon^{\mu\nu\rho\sigma} \p_\sigma a\, ,
\ee
where $\Phi$ denotes the four dimensional dilaton and $g_{\mu\nu}$
denotes the (3+1) dimensional canonical metric defined in
eqs.\refb{edil}, \refb{egmunu} respectively. It is
convenient to combine the dilaton $\Phi$ and the axion field $a$
into a single complex scalar $\lambda$:
\be \label{eh4}
\lambda = a + i e^{-\Phi} \equiv \lambda_1 + i\lambda_2\, .
\ee
At a generic point in the moduli space, where the scalars $M$ 
take arbitrary vacuum expectation values, the non-abelian gauge
symmetry of the ten dimensional theory is broken to its abelian
subgroup $U(1)^{16}$. Besides these sixteen U(1)
gauge fields we get
twelve other U(1)
gauge fields from components $G_{m\mu}$, $B_{m\mu}$
($4\le m\le 9$, $0\le \mu\le 3$) of the metric and the
anti-symmetric tensor field respectively. Let us denote these
$28$ U(1) gauge fields (after suitable normalization)
by $A_\mu^a$ ($1\le a\le 28$). 
In terms of these
fields, the low energy effective action of the theory is given
by\cite{DEROO,FERR,TEREN,MAHSCH,REVB},\footnote{The normalization
of the gauge fields used here differ from that in ref.\cite{REVB}
by a factor of two. Also there we used $\alpha'=16$ whereas here
we are using $\alpha'=1$.}
\ben \label{eh5}
S &=& {1\over 2\pi} \int d^4 x \sqrt{-g} \Big[ R - g^{\mu\nu}
{\p_\mu\lambda\p_\nu \bar \lambda\over 2(\lambda_2)^2} + {1\over
8} g^{\mu\nu} Tr(\p_\mu M L \p_\nu M L) \nonumber \\
&& -{1\over 4}
\lambda_2 g^{\mu\mu'} g^{\nu\nu'} F^a_{\mu\nu} (LML)_{ab}
F^b_{\mu'\nu'} +{1\over 4}
\lambda_1 g^{\mu\rho} g^{\nu\sigma}
F^a_{\mu\nu} L_{ab}
\wt F^b_{\rho\sigma}\Big]\, , \nonumber \\
\een
where $F^a_{\mu\nu}$ is the field strength associated with 
$A_\mu^a$, $R$ is the Ricci scalar. and
\be \label{eh9}
\wt F^{a\mu\nu} = {1\over 2} (\sqrt{-g})^{-1}
\epsilon^{\mu\nu\rho\sigma} F^a_{\rho\sigma}\, .
\ee

This action is invariant under an O(6,22) 
transformation:\footnote{$O(p,q)$ denotes the group of Lorentz
transformations in $p$ space-like and $q$ time-like dimensions.
(These have nothing to do with physical space-time, which always
has only one time-like direction.) $O(p,q;Z)$ denotes a discrete
subgroup of $O(p,q)$.}
\be \label{eh6}
M \to \Omega M \Omega^T, \qquad A_\mu^a\to \Omega_{ab} A_\mu^b, 
\qquad g_{\mu\nu}\to g_{\mu\nu}, \qquad \lambda\to\lambda \, ,
\ee
where $\Omega$ satisfies:
\be \label{eh7}
\Omega L \Omega^T = L\, .
\ee
An O(6,22;Z) subgroup of this can be shown to be a T-duality
symmetry of the full string 
theory\cite{GIVETC}. This O(6,22;Z) subgroup can be described as
follows. Let $\Lambda_{28}$ denote a twenty eight dimensional
lattice obtained by taking the direct sum of
the twelve dimensional lattice of
integers, and the sixteen dimensional root lattice of
SO(32).\footnote{More precisely we have to take the root lattice
of $Spin(32)/Z_2$ which is obtained by adding to the SO(32) root
lattice the weight vectors of the spinor representations of SO(32)
with a definite chirality.} O(6,22;Z) is defined to be the subset
of O(6,22) transformations which leave $\Lambda_{28}$ invariant, 
{\it i.e.} acting on any vector in $\Lambda_{28}$, produces
another vector in $\Lambda_{28}$.
It will be useful for our future reference to undertstand why
only an O(6,22;Z) subgroup of the full O(6,22) group
is a symmetry of the full
string theory. Since O(6,22;Z) is a T-duality symmetry, this
question can be answered within the context of perturbative
string theory. The point is that although at a generic point in
the moduli space the massless string states do not carry any
charge, there are massive charged states in the spectrum of full
string theory. Since there are 28 charges associated with the 28 U(1)
gauge fields, a state can be characterized by a 28
dimensional charge vector. With appropriate normalization, this
charge vector can be shown to lie in the lattice $\Lambda_{28}$,
{\it i.e.} the charge vector of any state in the spectrum can be 
shown to be an element of the lattice $\Lambda_{28}$. Since the
O(6,22) transformation acts linearly on the U(1) gauge fields, it
also acts linearly on the charge vectors. As a result only those
O(6,22) elements can be genuine symmetries of string theory which
preserve the lattice $\Lambda_{28}$. Any other O(6,22) element,
acting on a physical state in the spectrum, will take it to a
state with charge vector outside the lattice $\Lambda_{28}$.
Since such a state does not exist in the spectrum, such an
O(6,22) transformation cannot be a symmetry of the full string
theory.

In order to see a specific example of a T-duality transformation,
let us consider heterotic string theory compactified on $(S^1)^6$
with one of the circles having radius $R$ measured in the
string metric, and the rest having
unit radius. Let us also assume that there is no background gauge
or anti-symmetric tensor fields. Using the convention of
ref.\cite{REVB} one can show that for this background
\be \label{esps1}
M^{(H)}= \pmatrix{R^{-2} &&&& \cr & I_5 &&&\cr && R^2 && \cr &&&
I_5 &\cr &&&& I_{16}}\, .
\ee
Consider now the O(6,22;Z) transformation with the matrix:
\be \label{esps2}
\Omega = \pmatrix{ 0 & & 1 & \cr & I_5 &&\cr 1 && 0 &\cr &&&
I_{21}}\, .
\ee
Using eq.\refb{eh6} we see that this transforms $M^{(H)}$ to
\be \label{esps3}
M^{(H)}= \pmatrix{R^2 &&&& \cr & I_5 &&&\cr && R^{-2} && \cr &&&
I_5 &\cr &&&& I_{16}}\, .
\ee
Thus the net effect of this transformation is $R\to R^{-1}$. It
says that the heterotic string theory compactified on a circle of
radius $R$ is equivalent to the same theory compactified on a
circle of radius $R^{-1}$. For this reason $R=1$
({\it i.e.} $R=\sqrt{\alpha'}$) is known as the self-dual radius. Other
O(6,22;Z) transformations acting on \refb{esps1} will give rise
to more complicated $M^{(H)}$ corresponding to a 
configuration with background gauge
and / or anti-symmetric tensor fields.

Besides this symmetry, 
the equations of motion derived from this action
can be shown to be invariant under an $SL(2,R)$ transformation of
the form\cite{DEROO,TSW,SONE}
\ben \label{eh8}
&& F^a_{\mu\nu} \to (r\lambda_1 + s) F^a_{\mu\nu}
+ r\lambda_2 (ML)_{ab} \wt F^b_{\mu\nu},
\qquad \lambda\to {p\lambda + q\over
r\lambda + s}, \nonumber \\
&& g_{\mu\nu}\to g_{\mu\nu}, \qquad M\to M\, ,
\een
where $p,q,r,s$ are real numbers satisfying $ps-qr=1$.
The existence of such symmetries (known as hidden non-compact
symmetries) in this and in
other supergravity theories were discovered in early days of
supergravity theories and in fact played a crucial role in the
construction of these theories in the first place\cite{HIDDEN,DEROO}.
Since this SL(2,R) transformation mixes the gauge field strength
with its Poincare dual, it is an electric-magnetic duality
transformation. 
This leads to the conjecture that a subgroup of this continuous
symmetry group is an exact symmetry of string 
theory\cite{FILQ,REY,SONE,SCHONE,STWO,STHREE,SSCH,REVB}.
One might wonder why the conjecture refers
to only a discrete subgroup of SL(2,R) instead of the full
SL(2,R) group as the genuine symmetry group.
This follows from the same logic that was responsible for
breaking O(6,22) to O(6,22;Z); however since the SL(2,R)
transformation mixes electric field with magnetic field, we now
need to take into account the quantization of magnetic charges.
We have already described the
quantization condition on the electric charges. Using the usual
Dirac-Schwinger-Zwanziger rules one can show that in appropriate
normalization, the 28 dimensional magnetic charge vectors also
lie in the same lattice $\Lambda_{28}$. Also with this
normalization convention the
electric and magnetic charge vectors
transform as  doublet under the SL(2,R)
transformation; thus it is clear that the subgroup of SL(2,R) that
respects the charge quantization condition is SL(2,Z). An
arbitrary SL(2,R)
transformation acting on the quantized electric and magnetic
charges will not give rise to electric and magnetic
charges consistent with
the quantization law.
This is the reason behind the
conjectured SL(2,Z) symmetry of heterotic string theory on
$T^6$. Note that since this duality acts non-trivially on the
dilaton and hence the string coupling, this is a non-perturbative
symmetry, and cannot be verified order by order in perturbation
theory. Historically, this is the first example of a concrete
duality conjecture in string theory. Later we shall
review other tests of this duality conjecture.

\subsection{Duality between heterotic on $T^4$ and 
type IIA on K3}
\label{siiahet}

The massless bosonic field content of
heterotic string theory compactified on $T^4$ can be found in a
manner identical to that in heterotic string theory on $T^6$.
Besides the dilaton $\Phi^{(H)}$, we
get many other massless scalar fields from the internal
components of the metric, the anti-symmetric tensor field and the
gauge fields. In this case these scalars can be represented
by a $24\times 24$ matrix valued field $M^{(H)}$ satisfying
\be \label{ess1}
M^{(H)}LM^{(H)T} = L, \qquad M^{(H)T}=M^{(H)}\, ,
\ee
where
\be \label{ess2}
L = \pmatrix{& I_4 & \cr I_4 && \cr && -I_{16}}\, .
\ee
We again use the convention that $M^{(H)}=I_{24}$ corresponds to
compactification on $(S^1)^4$ with each $S^1$ having self-dual
radius ($\sqrt{\alpha'}=1$), without any background 
gauge field or anti-symmetric
tensor field. 
At a generic point in the moduli space, where the scalars $M^{(H)}$ 
take arbitrary vacuum expectation values, we get a $U(1)^{24}$
gauge group, with 16 gauge fields coming from the Cartan
subalgebra of the original gauge group in ten dimensions, and
eight other gauge fields from components $G_{m\mu}$, $B_{m\mu}$
($6\le m\le 9$, $0\le \mu\le 5$) of the metric and the
anti-symmetric tensor field respectively. Here $x^m$ denote the
compact directions, and $x^\mu$ denote the non-compact
directions. Let us denote these
$24$ U(1) gauge fields by $A_\mu^{(H)a}$ ($1\le a\le 24$). 
Finally, let $g^{(H)}_{\mu\nu}$ and $B^{(H)}_{\mu\nu}$ 
denote the canonical metric and the anti-symmetric tensor field
respectively.  In terms of these
fields, the low energy effective action of the theory is given
by, 
\ben \label{ess5}
S_H &=& {1\over (2\pi)^3}
\int d^6 x \sqrt{-g^{(H)}} \Big[ R^{(H)} - {1\over 2}
g^{(H)\mu\nu} \p_\mu\Phi^{(H)} \p_\nu\Phi^{(H)} \nonumber \\ &&
+ {1\over
8} g^{\mu\nu} Tr(\p_\mu M^{(H)} L \p_\nu M^{(H)} L) \nonumber \\
&& - {1\over 4} e^{-\Phi^{(H)}/2} g^{(H)\mu\mu'} g^{(H)\nu\nu'} 
F^{(H)a}_{\mu\nu} (LM^{(H)}L)_{ab}
F^{(H)b}_{\mu'\nu'} \nonumber \\ &&
- {1\over 12} e^{-\Phi^{(H)}} g^{(H)\mu\mu'} g^{(H)\nu\nu'} 
g^{(H)\rho\rho'} H^{(H)}_{\mu\nu\rho} 
H^{(H)}_{\mu'\nu'\rho'} 
\Big]\, ,
\een
where $F^{(H)a}_{\mu\nu}$ is the field strength associated with 
$A_\mu^{(H)a}$, $R^{(H)}$ is the Ricci scalar, and 
$H^{(H)}_{\mu\nu\rho}$ is the 
field strength associated with $B^{(H)}_{\mu\nu}$:
\be \label{eheq}
H^{(H)}_{\mu\nu\rho} = (\p_\mu B^{(H)}_{\nu\rho} + {1\over 2}
A^{(H)a}_\mu L_{ab}
F^{(H)b}_{\nu\rho}) + \hbox{(cyclic permutations of $\mu,\nu,\rho)$}.
\ee
This action is invariant under an O(4,20) transformation:
\ben \label{esh6}
&& M^{(H)} \to \Omega M^{(H)} \Omega^T, \qquad 
A_\mu^{(H)a}\to \Omega_{ab} A_\mu^{(H)b}, 
\qquad g^{(H)}_{\mu\nu}\to g^{(H)}_{\mu\nu}, \nonumber \\
&& B^{(H)}_{\mu\nu}\to B^{(H)}_{\mu\nu},
\qquad \Phi^{(H)}\to \Phi^{(H)} \, ,
\een
where $\Omega$ satisfies:
\be \label{esh7}
\Omega L \Omega^T = L\, .
\ee
Again as in the case of $T^6$ compactification, only an O(4,20;Z)
subgroup of this which preserves the charge lattice $\Lambda_{24}$
is an exact T-duality symmetry of this theory. The lattice
$\Lambda_{24}$ is obtained by taking the direct sum of the 8
dimensional lattice of integers and the root lattice of
$Spin(32)/Z_2$.

Let us now turn to the spectrum of massless bosonic fields in
type IIA string theory on K3.
In ten dimensions the massless bosonic fields in type IIA string
theory are the metric $g_{MN}$, the rank two anti-symmetric
tensor $B_{MN}$ and the scalar dilation $\Phi$ coming from
the NS sector, and a gauge field $A_M$ and a rank three
antisymmetric tensor field $C_{MNP}$ coming from the RR sector.
The low energy effective action of this theory involving the
massless bosonic fields is given by\cite{IIAM}
\ben \label{eiia}
S_{IIA}&=& {1\over (2\pi)^7}
\int d^{10}x \sqrt{-g} \Big[ R - {1\over 8} g^{\mu\nu}
\p_\mu\Phi \p_\nu\Phi \nonumber \\
&& -
{1\over 12} e^{-\Phi/2} g^{\mu\mu'} g^{\nu\nu'}
g^{\rho\rho'} H_{\mu\nu\rho} H_{\mu'\nu'\rho'} 
- {1\over 4} e^{3\Phi/4} g^{\mu\mu'} g^{\nu\nu'} 
F_{\mu\nu} F_{\mu'\nu'} \nonumber \\
&& - {1\over 48}
e^{\Phi/4} g^{\mu\mu'} g^{\nu\nu'} g^{\rho\rho'}
g^{\sigma\sigma'} G_{\mu\nu\rho\sigma} G_{\mu'\nu'\rho'\sigma'}
\nonumber \\
&& - {1\over (48)^2} (\sqrt{-g})^{-1}
\varepsilon^{\mu_0\cdots \mu_9} B_{\mu_0\mu_1}G_{\mu_2\cdots
\mu_5} G_{\mu_6\cdots \mu_9}\Big] \, ,
\een
where $R$ is the Ricci scalar, and
\ben \label{efstre}
F_{\mu\nu} &=& \p_\mu A_\nu - \p_\nu A_\mu, \nonumber \\
H_{\mu\nu\rho} &=& \p_\mu B_{\nu\rho} + \hbox{cyclic
permutations of $\mu$, $\nu$, $\rho$}\, , \nonumber \\
G_{\mu\nu\rho} &=& \p_\mu C_{\nu\rho\sigma}
+ A_\mu H_{\nu\rho\sigma} + (-1)^P\cdot \hbox{cyclic
permutations}\, , 
\een
are the field
strengths associated with $A_\mu$, $B_{\mu\nu}$ and 
$C_{\mu\nu\rho}$ respectively.
Upon compactification on $K3$ we get a new set of scalar fields
from the Kahler and complex structure moduli of K3. 
These can be regarded as deformations of the metric and give a total
of 58 real scalar fields.
We get 22 more scalar fields $\phi^{(p)}$
by decomposing the antisymmetric tensor
field $B_{MN}$ along the twenty two harmonic two forms
$\omega^{(p)}_{mn}$ in K3:
\be \label{ehar}
B_{mn}(x,y) \sim \sum_{p=1}^{22} \phi_p(x) 
\omega^{(p)}_{mn}(y) + \cdots\, .
\ee
Here $\{x^\mu\}$ and $\{y^m\}$ 
denote coordinates along the non-compact and K3
directions respectively.
These eighty scalar fields together parametrize a coset
$O(4,20)/O(4)\times O(20)$ and can be described by a matrix 
$M^{(A)}$ satisfying properties identical to those of $M^{(H)}$
described in \refb{ess1}. This theory also has twenty four U(1)
gauge fields. 
22 of the gauge fields arise from the components of the
three form field $C_{MNP}$:
\be \label{e29}
C_{mn\mu}(x,y) = \sum_{p=1}^{22}
\omega^{(p)}_{mn}(y) \AAA_\mu ^{(p)}(x)
+\ldots \, .
\ee
$\AAA_\mu^{(p)}$ defined in \refb{e29}
behaves as gauge fields in six dimensions. 
One more gauge field comes from the original RR gauge field $A_\mu$.
The last one $\AAA_\mu$ comes from dualizing $C_{\mu\nu\rho}$:
\be \label{e30}
G \sim ~^*(d\AAA)\, ,
\ee
where $~^*$ denotes Poincare dual in six dimensions. Together we
shall denote these gauge fields by $A^{(A)a}_\mu$ for $1\le a\le
24$. Besides these fields, the theory contains the canonical
metric and  the anti-symmetric
tensor field which we shall denote by
$g^{(A)}_{\mu\nu}$ and $B^{(A)}_{\mu\nu}$ respectively. The action
involving these fields is given by,
\ben \label{eiiact}
S_{A} &=& {1\over (2\pi)^3}
\int d^6 x \sqrt{-g^{(A)}} \Big[ R^{(A)} - {1\over
2}
g^{(A)\mu\nu} \p_\mu\Phi^{(A)} \p_\nu\Phi^{(A)} \nonumber \\
&& + {1\over
8} g^{\mu\nu} Tr(\p_\mu M^{(A)} L \p_\nu M^{(A)} L) \nonumber \\
&& - {1\over 4} e^{\Phi^{(A)}/2} g^{(A)\mu\mu'} g^{(A)\nu\nu'} 
F^{(A)a}_{\mu\nu} (LM^{(A)}L)_{ab}
F^{(A)b}_{\mu'\nu'} \nonumber \\ &&
- {1\over 12} e^{-\Phi^{(A)}} g^{(A)\mu\mu'} g^{(A)\nu\nu'} 
g^{(A)\rho\rho'} H^{(A)}_{\mu\nu\rho} 
H^{(A)}_{\mu'\nu'\rho'} \nonumber \\
&& -{1\over 16}\varepsilon^{\mu\nu\rho\delta\epsilon\eta}
(\sqrt{-g^{(A)}})^{-1} B^{(A)}_{\mu\nu} F^{(A)a}_{\rho\delta}
L_{ab} F^{(A)b}_{\epsilon\eta}
\Big]\, ,
\een
where $F^{(A)a}_{\mu\nu}$ is the field strength associated with 
$A_\mu^{(A)a}$, $R^{(A)}$ is the Ricci scalar, and 
$H^{(A)}_{\mu\nu\rho}$ is the field strength associated with
$B^{(A)}_{\mu\nu}$:
\be \label{ehfstr}
H^{(A)}_{\mu\nu\rho} = \p_\mu B^{(A)}_{\nu\rho} + 
\hbox{(cyclic permutations of $\mu,\nu,\rho$)}\, .
\ee
In writing down the above action
we have used the convention that $M^{(A)}=I_{24}$ corresponds
to compactification on a specific reference K3, possibly with
specific background $B_{mn}$ fields. This action
has an O(4,20) symmetry of the form:
\ben \label{eiish6}
&& M^{(A)} \to \Omega M^{(A)} \Omega^T, \qquad 
A_\mu^{(A)a}\to \Omega_{ab} A_\mu^{(A)b}, 
\qquad g^{(A)}_{\mu\nu}\to g^{(A)}_{\mu\nu}, \nonumber \\
&& B^{(A)}_{\mu\nu}\to B^{(A)}_{\mu\nu},
\qquad \Phi^{(A)}\to \Phi^{(A)} \, ,
\een
where $\Omega$ satisfies:
\be \label{eiish7}
\Omega L \Omega^T = L\, .
\ee
An O(4,20;Z) subgroup of this can be shown to be an exact
T-duality symmetry of string theory\cite{KTDUAL}. The lattice
$\Lambda_{24}'$ which is preserved by
this O(4,20;Z) subgroup of O(4,20) is
not the lattice $\Lambda_{24}$ defined earlier, but is in general
an O(4,20) rotation of that lattice:
\be \label{elattice}
\Lambda'_{24} = \Omega_0 \Lambda_{24}\, .
\ee
$\Omega_0$ depends on the choice of the special reference
K3 mentioned earlier.

It is now a straightforward exercise to show that the equations
of motion and the Bianchi identities
derived from \refb{ess5} and \refb{eiiact} are
identical if we use the following map between the heterotic and
the type II variables\cite{SEIOLD,HULLTOWN}:
\ben \label{esiimap}
&& g^{(H)}_{\mu\nu}=g^{(A)}_{\mu\nu}, \qquad M^{(H)} = \wt\Omega
M^{(A)}\wt\Omega^T, \nonumber \\
&& \Phi^{(H)}=-\Phi^{(A)}, 
\qquad A^{(H)a}_\mu=\wt\Omega_{ab} A^{(A)a}_\mu,
\nonumber \\
&& \sqrt{-g^{(H)}} \exp(-\Phi^{(H)}) H^{(H)\mu\nu\rho}
={1\over 6} \varepsilon^{\mu\nu\rho\delta\epsilon\eta}
H^{(A)}_{\delta\epsilon\eta}\, .
\een
where $\wt\Omega$ is an arbitrary O(4,20) matrix. 
This leads to the conjectured equivalence between heterotic
string theory compactified on $T^4$ and type IIA string theory
compactified on $K3$\cite{HULLTOWN}. But clearly the two theories
cannot be equivalent for all $\wt\Omega$ since in the individual
theories the O(4,20) symmetry is broken down to O(4,20;Z).
$\wt\Omega$ can be found (up to an O(4,20;Z) transformation) by
comparing the T-duality symmetry transformations in the two
theories. To do this let us note that according to eq.\refb{esiimap}
a transformation $M^{(H)}\to
\Omega M^{(H)}\Omega^T$ will induce a transformation 
\be \label{eaux1}
M^{(A)}\to (\wt\Omega^{-1}\Omega\wt\Omega) M^{(A)}
(\wt\Omega^{-1}\Omega\wt\Omega)^T\, .
\ee
Thus if $\Omega$ preserves the
lattice $\Lambda_{24}$, $\wt\Omega^{-1}\Omega\wt\Omega$ should
preserve the lattice $\Lambda'_{24}=\Omega_0\Lambda_{24}$.
This happens if we choose:
\be \label{eomfix}
\wt\Omega = \Omega_0^{-1}\, .
\ee

Note again that there is a relative minus
sign that relates $\Phi^{(H)}$ and $\Phi^{(A)}$, showing that
the strong coupling limit of one theory corresponds to the weak
coupling limit of the other theory.

\subsection{SL(2,Z) self-duality of Type IIB in D=10} \label{s21}
As described in section \ref{sspec},
the massless bosonic fields in type IIB string theory come from
two sectors, $-$ Neveu-Schwarz$-$Neveu-Schwarz (NS) and
Ramond-Ramond (RR). The
NS sector gives the graviton described by the
metric $g_{\mu\nu}$, an anti-symmetric tensor field
$B_{\mu\nu}$, and a scalar field $\Phi$ known as the dilaton.
The
RR sector  contributes a scalar field $a$ sometimes called the axion, 
another rank two anti-symmetric tensor field$B'_{\mu\nu}$, 
and a rank four anti-symmetric tensor field$D_{\mu\nu\rho\sigma}$
whose field strength is self-dual. 

It is often convenient to combine the axion and the dilaton into
a complex scalar field $\lambda$ as follows:\footnote{Note that this
field $\lambda$ has no relation to the field $\lambda$ defined in
section \ref{sh1} for heterotic string theory on $T^6$, although
both transform as modulus under the respective SL(2,Z) duality
transformations in the two theories.}
\be \label{e1}
\lambda = a + i e^{-\Phi/2} \equiv \lambda_1 + i\lambda_2 \, .
\ee
The low energy effective action in this theory
can be
determined either from the requirement of supersymmetry, or by explicit
computation in string theory. Actually it turns out that there is
no simple covariant action for this low energy theory, but there are
covariant field equations\cite{IIBSUGRA}, which are
in fact just the equations of motion of type IIB supergravity.
Although in string theory this low energy theory
is derived from the tree
level analysis, non-renormalization theorems tell us that this
is exact to this order in the space-time derivatives. Basically
supersymmetry determines the form of the equations of motion
to this order in the derivatives
completely, and so there is no scope for the quantum corrections
to change the form of the action.

For the sake of brevity, we shall not explicitly write down the
equations of motion. The main point is that 
these equations of motion 
are covariant (in the sense that they transform into each
other) under an SL(2,R) transformation\cite{IIBSUGRA}:
\ben \label{e3}
&&
\lambda\to {p\lambda+q\over r\lambda + s}, \qquad
\pmatrix{B_{\mu\nu}\cr B'_{\mu\nu}} \to
\pmatrix{p & q\cr r & s} \pmatrix{B_{\mu\nu}\cr B'_{\mu\nu}},
\nonumber \\
&& g_{\mu\nu}\to g_{\mu\nu}, \quad D_{\mu\nu\rho\sigma} \to
D_{\mu\nu\rho\sigma}\, ,
\een
where $p,q,r,s$ are real numbers satisfying,
\be \label{e4}
ps -qr =1\, .
\ee

The existence of this SL(2,R) symmetry in the type IIB supergravity
theory led to the conjecture that an SL(2,Z) subgroup of this
SL(2,R), obtained by restricting $p,q,r,s$ to be integers instead
of arbitrary real numbers, is a symmetry of the full string
theory\cite{HULLTOWN}. 
The breaking of SL(2,R) to SL(2,Z) can be seen as follows.
An elementary string is known to carry $B_{\mu\nu}$ charge. In
suitable normalization convention, it carries exactly one unit of
$B_{\mu\nu}$ charge. This means that the
$B_{\mu\nu}$ charge must be quantized in integer units, as the
spectrum of string theory does not contain fractional strings
carrying a fraction of the charge carried by the
elementary string. From \refb{e3} we see that acting on an
elementary string state carrying one unit of $B_{\mu\nu}$ charge,
the SL(2,R) transformation gives a state with $p$ units of
$B_{\mu\nu}$ charge and $r$ units of $B'_{\mu\nu}$ charge. Thus
$p$ must be an integer. It is easy to see that the maximal
subgroup of SL(2,R) for which $p$ is always an integer consists
of matrices of the form
\be \label{egenslt}
\pmatrix{p & \alpha q\cr \alpha^{-1} r & s}\, ,
\ee
with $p,q,r,s$ integers satisfying $(ps-qr)=1$, and $\alpha$ a
fixed constant. Absorbing $\alpha$ into a redefinition of
$B'_{\mu\nu}$ we see
that the subgroup of SL(2,R) matrices consistent with charge
quantization are the SL(2,Z) matrices $\pmatrix{p & q\cr r & s}$
with $p,q,r,s$ integers satisfying $ps-qr=1$.

Note that this argument only shows that SL(2,Z) is the maximal
possible subgroup of SL(2,R) that {\it can be a symmetry of the full
string theory}, but does not prove that SL(2,Z) is a symmetry of
string theory. In particular, since SL(2,Z) acts non-trivially on
the dilaton, whose vacuum expectation value represents the string
coupling constant, it cannot be verified order by order in string
perturbation theory. We shall see later how one can find
non-trivial evidence for this symmetry.

Besides this non-perturbative SL(2,Z) transformation, type IIB
theory has two perturbatively verifiable discrete $Z_2$
symmetries. They are as follows:
\begin{itemize}
\item
$(-1)^{F_L}$: It changes the sign of all the Ramond sector states 
on the left moving sector of the world-sheet.
In particular, acting on the massless bosonic sector fields,
it changes the sign of $a$, $B'_{\mu\nu}$ and
$D_{\mu\nu\rho\sigma}$, but leaves $g_{\mu\nu}$, $B_{\mu\nu}$ and
$\Phi$ invariant. 
\item
$\Omega$: This is the
world-sheet parity transformation mentioned in section \ref{sspec}
that exchanges the left- and
the right-moving sectors of the world-sheet. Acting on the
massless bosonic sector fields,
it changes the sign of $B_{\mu\nu}$, $a$ and
$D_{\mu\nu\rho\sigma}$, leaving the other fields invariant.
\end{itemize}
{}From this description, we see that the
effect of $(-1)^{F_L}\cdot\Omega$ is to change of sign of
$B_{\mu\nu}$ and $B'_{\mu\nu}$, leaving the other massless
bosonic fields invariant. Comparing this with the action of the
SL(2,Z) transformation laws of the massless bosonic sector
fields, we see that
$(-1)^{F_L}\cdot\Omega$
can be identified with the SL(2,Z) transformation:
\be \label{e6}
\pmatrix{-1 & \cr & -1}\, .
\ee
This information will be useful to us later.

Theories obtained by modding out (compactified) type IIB string
theory by a discrete symmetry group, where some of the elements
of the group involve $\Omega$, are
known as orientifolds\cite{ORIENT,GIMPOL}. The simplest example
of an orientifold is type IIB string theory modded out by
$\Omega$. This corresponds to type I string theory. The closed
string sector of type I theory consists of the $\Omega$ invariant
states of type IIB string theory. The open string states of type
I string theory are the analogs of twisted sector states in an
orbifold, which must be added to the theory in order to maintain
finiteness.

\subsection{Other examples} \label{s23}

Following the same procedure, namely, 
studying symmetries of the effective action together with charge
quantization rules, we are led to many other duality conjectures in 
theories with 16 or more supersymmetry generators.
Here we shall list the main series of such duality conjectures.
We begin with the self duality groups of
type II string theories compactified on tori of
different dimensions. As mentioned earlier,
there is a T-duality that relates type IIA on a circle to type
IIB on a circle of inverse radius. Thus for $n\ge 1$, the
self-duality groups of type IIA and type IIB theories compactified
on an $n$-dimensional torus $T^n$ will be
identical. We now list the conjectured
self-duality groups of type IIA/IIB
string theory compactified on $T^n$ for
different values of $n$\cite{HULLTOWN}:
\vbox{
\ben
D=(10-n) & \hbox{Full Duality Group} & \hbox{T-duality Group} 
\nonumber \\
\quad 9  \quad & SL(2,Z) \quad & - \nonumber \\
\quad 8 \quad  & SL(2,Z)\times SL(3,Z) \quad  & SL(2,Z)\times SL(2,Z) \nonumber \\
\quad 7 \quad  & SL(5,Z) \quad  & SO(3,3;Z) \nonumber \\
\quad 6 \quad  & SO(5,5;Z) \quad  & SO(4,4;Z) \nonumber \\
\quad 5 \quad  & E_{6(6)}(Z) \quad  & SO(5,5;Z) \nonumber \\
\quad 4 \quad  & E_{7(7)}(Z) \quad  & SO(6,6;Z) \nonumber \\
\quad 3 \quad  & E_{8(8)}(Z) \quad  & SO(7,7;Z) \nonumber \\
\quad 2 \quad  & \wh{E_{8(8)}}(Z) \quad  & SO(8,8;Z) 
\nonumber \een
}
Note that besides the full duality group, we have also displayed
the T-duality group of each theory which can be verified order by
order in string perturbation theory. $E_{n(n)}$ denotes a
non-compact version of the exceptional group $E_n$ for
$n=6,7,8$, and $E_{n(n)}(Z)$ denotes a discrete subgroup of
$E_{n(n)}$. 
$\wh G$ for any group $G$ denotes
the loop group of $G$ based on the corresponding affine algebra
and $\wh G(Z)$ denotes a discrete subgroup of this loop group.
Note that we have stopped at $D=2$. We could in principle
continue this all the way to $D=1$ where all space-like
directions are compactified. In this case one expects a very
large duality symmetry group based on hyperbolic Lie 
algebra\cite{HYPER},
which is not well understood to this date.

In each of the cases mentioned, the low energy effective field
theory is invariant under the full continuous group\cite{JULIA}, 
but charge
quantization breaks this symmetry to its discrete subgroup. As
noted before, these
symmetries were discovered in the early days of supergravity
theories, and were known as hidden non-compact symmetries.

\medskip

Next we turn to the self-duality conjectures involving
compactified heterotic string theories. Although there are two
distinct heterotic string theories in ten dimensions, upon
compactification on a circle, the two heterotic string theories
can be shown to be related by a T-duality transformation. As a
result, upon compactification on $T^n$,
both of them will have the same self-duality group. We now
display this self-duality group in various dimensions:
\ben
D=(10-n) & \hbox{Full Duality Group} & \hbox{T-duality Group} 
\nonumber \\
\quad 9  \quad & O(1,17,Z) \quad & O(1,17;Z) \nonumber \\
\quad 8  \quad & O(2,18,Z) \quad & O(2,18;Z) \nonumber \\
\quad 7  \quad & O(3,19,Z) \quad & O(3,19;Z) \nonumber \\
\quad 6  \quad & O(4,20,Z) \quad & O(4,20;Z) \nonumber \\
\quad 5  \quad & O(5,21,Z) \quad & O(5,21;Z) \nonumber \\
\quad 4  \quad & O(6,22,Z)\times SL(2,Z) \quad & O(6,22;Z) 
\nonumber \\
\quad 3  \quad & O(8,24,Z) \quad & O(7,23;Z) \nonumber \\
\quad 2  \quad & \wh{O(8,24,Z)} \quad & O(8,24;Z) 
\nonumber \een

Since type I and SO(32) heterotic string theories are conjectured
to be dual to each other in ten dimensions, the second
column of the  above table also
represents the duality symmetry group of type I string theory on
$T^n$. However, in the case of type I string theory, there is no
perturbatively realised self-duality group (except trivial 
transformations which are part of the SO(32) gauge group and the
group of global diffeomorphisms of $T^n$). 

The effective action of type IIB
string theory compactified on $K3$ has an $SO(5,21)$ 
symmetry\cite{SEIOLD},
which leads to the conjecture that an SO(5,21;Z) subgroup of this
is an exact self-duality symmetry of the type IIB string theory on K3.
The conjectured duality between type IIA string
theory compactified on $K3$ and heterotic string theory
compactified on $T^4$ has already been discussed before. 
Due to the equivalence of type IIB on $S^1$ and type IIA on
$S^1$, type IIA on $K3\times T^n$ is equivalent to type IIB on
$K3\times T^n$.
Finally, due to the conjectured duality between type IIA on K3
and heterotic on $T^4$,
type IIA/IIB on $K3\times T^n$ are dual to 
heterotic string theory on $T^{n+4}$ for $n\ge 1$. 
Thus the self-duality symmetry groups in these theories can be
read out from the second column of the previous table displaying
the self-duality groups of heterotic string theory on $T^n$.

Besides the theories discussed here, there are other theories
with 16 or more supercharges obtained from non-geometric
compactification of heterotic/type II string 
theories\cite{FERRKOU,CHL,CP}. The
duality symmetry groups of these theories can again be guessed
from an analysis of the low energy effective field theory and the
charge quantization conditions. Later we shall also describe a
more systematic way of `deriving' various duality conjectures
from some basic set of dualities.

Although in this section I have focussed on duality symmetries of
the low energy effective action which satisfy a
non-renormalization theorem as a consequence of space-time
supersymmetry, this is not the only part of the full effective
action which satisfy such a non-renormalization theorem. Quite
often the effective action contains another set of terms
satisfying non-renormalization theorems. They are 
required for anomaly cancellation, and are known
as Green-Schwarz terms.
Adler-Bardeen theorem guarantees that they are not
renormalized beyond one loop. These terms
have also been used effectively for testing various duality
conjectures\cite{ANOM}, but I shall not discuss it in this article.

\sectiono{Precision Test of Duality: Spectrum of BPS States}
\label{s3}

Analysis of the low energy effective action, as discussed in the
last section, provides us with only a crude test of duality. 
Its value lies in its simplicity. Indeed, most of the duality
conjectures 
in string theory were arrived at by analysing the symmetries of
the low energy effective action.

But once we have arrived at a duality conjecture based on the
analysis of the low energy effective action, we can perform a
much more precise test by analysing the spectrum of BPS states in
the theories.
BPS states are states which are invariant under part of the
supersymmetry
transformation, and are characterized by two important
properties: 
\begin{itemize}
\item
They belong to a supermultiplet which has typically less dimension
than a non-BPS state. This has an analog in the theory of
representations of the Lorentz group, where
massless states form a shorter
representation of the algebra than massive states.
Thus for example
a photon has only two polarizations but a massive vector particle
has three polarizations.
\item
The mass of a BPS state is completely determined by its charge as a
consequence of the supersymmetry algebra.
This relation between the mass and the charge is known as the BPS
mass formula. This statement also has an analog in the theory of
representations of the Lorentz algebra, {\it e.g.}
a spin 1 representation of the
Lorentz algebra containing only two states must be necessarily
massless.
\end{itemize}

We shall now explain the origin of these two 
properties\cite{WITOLI}. 
Suppose the theory has $N$ real supersymmetry generators 
$Q_\alpha$ ($1\le\alpha\le N$).
Acting on a single particle state {\it at rest}, the supersymmetry
algebra takes the form:
\be \label{e10}
\{Q_\alpha, Q_\beta\} = f_{\alpha\beta}(m, \vec Q, \{y\})\, ,
\ee
where $f_{\alpha\beta}$ is a real symmetric matrix which is 
a function of its arguments $m$, $\vec Q$ and $\{y\}$. Here
$m$ denotes the rest mass of the particle, $\vec Q$ denotes
various gauge charges carried by the particle, and $\{y\}$
denotes the coordinates labelling the moduli space of 
the theory.\footnote{Only specific combinations of $\vec Q$ and
$\{y\}$, known as central charges, appear in the algebra.}
We shall now consider the following distinct cases:
\begin{enumerate}
\item $f_{\alpha\beta}$ has no zero eigenvalue. In this case by
taking appropriate linear combinations of $Q_\alpha$ we can
diagonalize $f$. By a further appropriate rescaling of 
$Q_\alpha$, we can bring $f$ into the identity matrix. Thus in
this basis the supersymmetry algebra has the form:
\be \label{e11}
\{Q_\alpha, Q_\beta\} = \delta_{\alpha\beta}.
\ee
This is the $N$ dimensional Clifford algebra. Thus the single
particle states under consideration form a representation of this
Clifford algebra, which is $2^{N/2}$ dimensional. (We are
considering the case where $N$ is even.) Such states would
correspond to non-BPS states.
\item
$f$ has $(N-M)$ zero eigenvalues for some $M<N$. In this case, by
taking linear combinations of the $Q_\alpha$ we can bring the
algebra into the form:
\ben \label{e12}
\{Q_\alpha, Q_\beta\} &=& \delta_{\alpha\beta}, \quad \hbox{for}
\quad 1\le \alpha, \beta \le M\, , \nonumber \\
&=& 0 \quad \hbox{for} \quad \alpha \,\, \hbox{or} \,\, \beta > M\, .
\een
We can form an irreducible representation of this algebra by
taking all states to be annihilated by $Q_\alpha$ for $\alpha>M$.
In that case the states will form a representation of an $M$
dimensional Clifford algebra generated by $Q_\alpha$ for
$1\le\alpha\le M$. This representation is $2^{M/2}$ dimensional
for $M$ even. Since $M<N$, we see that these are lower
dimensional representations compared to that of a generic non-BPS
state. Furthermore, these states are invariant under part of the
supersymmetry algebra generated by $Q_\alpha$ for $\alpha>M$.
These are known as  BPS states. We can get different
kinds of BPS states depending on the value of $M$, {\it i.e.}
depending on the number of supersymmetry generators that
leave the state invariant. 
\end{enumerate}
{}From this discussion it is clear that in order to get a BPS
state, the matrix $f$ must have some zero eigenvalues. This in
turn, gives a constraint involving mass $m$, charges $\vec Q$ 
and the moduli $\{y\}$, and is the origin of the BPS formula
relating the mass and the charge of the particle.

Before we proceed, let us illustrate the preceeding discussion
in the context of a string theory. Consider 
Type IIB string theory compactified on a circle $S^1$.
The total number of supersymmetry generators in this theory is 32.
Thus a
generic non-BPS supermultiplet is $2^{16} = (256)^2$ dimensional.
These are known as long multiplets.
This theory also has 
BPS states breaking half the space-time supersymmetry. For these
states $M=16$ and hence we have
$2^8=256$ dimensional representation of the supersymmetry algebra.
These states are known as ultra-short multiplets. We can also
have BPS states breaking 3/4 of  the space-time supersymmetry
($M=24$). These will form a
$2^{12}=256\times 16$ dimensional representation, and are
known as short multiplets. 
In each case there is a specific relation between the mass and
the various charges carried by the state. We shall discuss this
relation as well as the origin of these BPS states
in more detail later.

As another example, consider heterotic string theory compactified
on an $n$-dimensional torus $T^n$. The original theory has 16
supercharges. Thus a generic non-BPS state will belong to a
$2^8=256$ dimensional representation of the supersymmetry
algebra. But if we consider states that are invariant under half
of the supercharges, then they belong to a $2^4=16$ dimensional
representation of the supersymmetry algebra. This is known as
the short representation of this superalgebra. 
We can also have states that break 3/4 of the
supersymmetries.\footnote{It turns out that these states can
exist only for $n\ge 5$. This constraint arises due to the fact
that the unbroken supersymmetry generators must form a
representation of the little group SO($9-n$) of a massive
particle in $(10-n)$ dimensional space-time.}
These belong to a 64 dimensional representation of
the supersymmetry algebra known as intermediate states.

BPS states are further characterized by the property that the
degeneracy of BPS states with a given set of charge quantum
numbers is independent of the value of the moduli fields $\{y\}$.
Since string coupling is also one of the moduli of the theory,
this implies that the degeneracy at any value of the string
coupling is the same as that at weak coupling. This is the key
property of the BPS states that makes them so useful in testing
duality, so let us review the argument leading to this 
property\cite{WITOLI}. 
We shall discuss this in the context of the specific example of
type IIB string theory compactified
on $S^1$, but it can be applied to any
other theory.
Suppose the theory has an ultra-short multiplet at some point in the
moduli space. 
Now let us change the moduli. The question that we shall be
asking is:
can the ultra-short multiplet become a long (or any other)
multiplet as we change the moduli?
If we assume that the total number of states does not change
discontinuously, then this is clearly 
not possible since other multiplets have different number of
states. Thus as long as the spectrum varies smoothly with the
moduli (which we shall assume), 
an ultra-short multiplet stays ultra-short as we move in the
moduli space\cite{SEIWIT}.
Furthermore, as long as it stays ultra-short, its mass is determined by
the BPS formula. Thus we see that the degeneracy of ultra-short
multiplets cannot change as we change the moduli of the theory.
A similar argument can be given for other multiplets as well.
Note that for this argument to be strictly valid, we require 
that the mass of
the BPS state should stay away from the continuum, since 
otherwise the counting of states is not a well defined procedure.
This requires that the mass of a BPS state should be strictly
less than the total mass of any set of two or more particles
carrying the same total charge as the BPS state.

Given this result, we can now adapt the following strategy to
carry out tests of various duality conjectures using the
spectrum of BPS states in the theory:
\begin{enumerate}
\item Identify BPS states in the spectrum of elementary string states.
The spectrum of these BPS states can be trusted at all values of the
coupling even though it is calculated at weak coupling.
\item
Make a conjectured duality transformation.
This typically takes a BPS state in the spectrum of elementary 
string states to another BPS state, but with quantum 
numbers that are not present in the spectrum of elementary string
states. Thus these states must arise as solitons /composite states.
\item
Try to explicitly verify the existence of these solitonic states with
degeneracy as predicted by duality.
This will provide a non-trivial test of the corresponding duality
conjecture.
\end{enumerate}
We shall now illustrate this procedure with the help of
specific examples. We shall mainly follow
\cite{SBPS,POLD,WITTDB}.

\subsection{SL(2,Z) S-duality in heterotic on $T^6$ and
multi-monopole moduli spaces} \label{s32}

As discussed in section \ref{sh1}, heterotic string theory
compactified on $T^6$ is conjectured to have an SL(2,Z) duality
symmetry. In this subsection we shall see how one can test this
conjecture by examining the spectrum of BPS states.

Since the BPS spectrum does not change as we change the moduli,
we can analyse the spectrum near some
particular point in the moduli space.  As discussed in section
\ref{sh1}, at a generic point in the moduli space the unbroken
gauge group is $U(1)^{28}$. But there are special points in this
moduli space where we get enhanced non-abelian gauge
group\cite{NARAIN}. Thus for example, if we set the internal
components of the original ten dimensional gauge fields to zero,
we get unbroken $E_8\times E_8$ or SO(32) gauge symmetry.
Let us consider a special point in
the moduli space where an SU(2)
gauge symmetry is restored. This can be done for example by
taking a particular $S^1$ in $T^6$ to be orthogional to all other
circles, taking the components of the gauge fields along this 
$S^1$ to be zero, and taking the radius of this $S^1$ to be the
self-dual radius. In that case the effective field
theory at energies much below the string scale will be described
by an N=4 supersymmetric SU(2) gauge theory, together with a set
of decoupled N=4 supersymmetric U(1) gauge theories and N=4
supergravity. 
The conjectured SL(2,Z) duality of the heterotic
string theory will require the N=4 supersymmetric SU(2) gauge
theory to have this SL(2,Z) symmetry.\footnote{Independently of
string theory, the existence of a strong-weak coupling duality in
this theory was conjectured earlier\cite{MONOLI,OSBORN}.}
Thus by testing the
duality invariance of the spectrum of this N=4 supersymmetric
SU(2) gauge theory we can test the conjectured SL(2,Z) symmetry
of heterotic string theory.

The N=4 supersymmetric SU(2) gauge theory has a vector, six
massless
scalars and four massless Majorana fermions in the adjoint
representation of SU(2)\cite{OSBORN}.
The form of the lagrangian is fixed completely by the requirement
of $N=4$ supersymmetry up to two independent parameters $-$ the
coupling constant $g$ that determines the strength of all
interactions (gauge, Yukawa, scalar self-interaction etc.), and
the vacuum angle $\theta$ that multiplies the topological term
$Tr(F\wt F)$ involving the gauge field. With the choice of
suitable normalization convention, $g$ and $\theta$ are related
to the vacuum expectation value of the field $\lambda$ defined in
\refb{eh4} through the relation:
\be \label{eyy1}
\langle\lambda\rangle = {\theta\over 2\pi} + i 
{4\pi \over g^2}\, .
\ee
The potential involving the six adjoint representation
scalar fields $\phi_m^\alpha$
($1\le\alpha\le 3, 1\le m\le 6$) is proportional to
\be \label{eyy2}
\sum_{m<n}\sum_\alpha
(\epsilon^{\alpha\beta\gamma}\phi^\beta_m\phi^\gamma_n)^2\, .
\ee
This vanishes for
\be \label{eyy3}
\phi_m^\alpha = a_m \delta_{\alpha 3}\, .
\ee
Vacuum expectation values of $\phi_m^\alpha$ of the form
\refb{eyy3} does not break supersymmetry, but breaks the gauge
group $SU(2)$ to $U(1)$. 
The parameters $\{a_m\}$
correspond to the vacuum expectation values of a subset of the
scalar 
moduli fields $M$ in the full string theory. We shall work in a
region in the moduli space where $a_m\ne 0$ for some $m$,
but the scale of breaking of
SU(2) is small compared to the string scale ($|a_m|<< 
(\sqrt{\alpha'})^{-1})$ for
all $m$), so that gravity is
still decoupled from this gauge theory. The BPS states in the
spectrum of elementary particles in this theory are the heavy
charged bosons $W^\pm$ and their superpartners. These break half
of the 16 space-time supersymmetry generators and hence form a
$2^{8/2}=16$ dimensional representation of the supersymmetry
algebra. These states can be found 
explicitly in the spectrum of
elementary string states from the sector containing strings with
one unit of winding and one unit of momentum along the special
$S^1$ that is responsible for the enhanced SU(2) gauge symmetry. 
As we approach the point in the moduli
space where this special $S^1$ has self-dual radius,
these states become
massless and form part of the SU(2) gauge multiplet.

When SU(2) is broken to U(1) by the vacuum expectation value of
$\phi_m$, the
spectrum of solitons in this theory is characterized by two
quantum numbers, the electric charge quantum number $n_e$ and the
magnetic charge quantum number $n_m$, normalized so that $n_e$
and $n_m$ are both integers. We shall denote such a
state by $\pmatrix{n_e\cr n_m}$. In this notation
the elementary $W^+$ boson
corresponds to a $\pmatrix{1\cr 0}$ state. By studying the action
of the SL(2,Z) transformation \refb{eh8} on the gauge fields, we
can easily work out its action on the charge quantum numbers
$\pmatrix{n_e\cr n_m}$\cite{REVB}. The answer is
\be \label{ehh1}
\pmatrix{n_e\cr n_m}\to \pmatrix{p & q\cr r & s} \pmatrix{n_e\cr
n_m}\, ,
\ee
for appropriate choice of sign convention for $n_e$ and $n_m$.
Thus acting on an $\pmatrix{1\cr 0}$ state it produces a 
$\pmatrix{p\cr
r}$ state. From the relation $ps-qr=1$ satisfied by an SL(2,Z)
matrix, we can easily see that $p$ and $r$ are relatively prime.
Furthermore for every $p$ and $r$ relatively prime, we can find
integers $q$ and $s$ satisfying $ps-qr=1$. Thus SL(2,Z) duality
predicts that {\it for every $p$ and $r$ relatively prime, the
theory must contain a unique short multiplet with charge quantum
numbers $\pmatrix{p\cr r}$\cite{SBPS}.}

We can now directly examine the solitonic sector of the theory to
check this prediction. The theory contains classical monopole 
solutions which break half of the supersymmetries of the
original theory. These solutions are non-singular everywhere, and 
in fact, for a given $r$, there is a $4r$ parameter non-singular
solution with $r$ units of total magnetic 
charge\cite{MONMOD,AH}. These $4r$ 
parameters correspond to the bosonic collective excitations of this
system\cite{MANTON}. In order to
study the spectrum of BPS solitons, we need to quantize these
collective excitations and look for
supersymmetric ground states of the corresponding quantum
mechanical system. Each solution also has infinite number of
vibrational modes with non-zero frequency, but excitations
of these modes are not relevant for finding supersymmetric ground
states.

States with $r=1$ come from one monopole solution. This has four
bosonic collective coordinates, three of which correspond to the 
physical position of the monopole in the three dimensional space,
and the fourth one is an angular variable describing the U(1)
phase of the monopole. The momenta conjugate to 
the first three coordinates
correspond to the components of the physical momentum of the
particle. These can be set to zero by working in the rest frame
of the monopole. The fourth coordinate is periodically identified and
hence its conjugate momentum is quantized in integer units. This
integer $p$ corresponds to the electric charge quantum number $n_e$.
Thus the states obtained by quantizing the bosonic sector of the
theory has charge quantum numbers $\pmatrix{p\cr 1}$ for all 
integer $p$.

The degeneracy comes from quantizing the fermionic sector. There
are eight fermionic zero modes, which describe the result of
applying the eight broken supersymmetry generators on the
monopole solution. These form an eight dimensional Clifford
algebra. Thus the ground state has $2^4=16$-fold degeneracy,
exactly as predicted by SL(2,Z)\cite{OSBORN}.

Let us now turn to the analysis of states 
with $r>1$\cite{SBPS}. As has
already been said, this system has $4r$ bosonic collective
coordinates, which, when the monopoles are far away 
from each other, correspond to the
spatial location and the U(1) phase of each of the $r$ monopoles.
The total number of fermionic collective coordinates can be
computed from an index theorem and is equal to $8r$\cite{INDEX}.
We can divide this set into the `center of mass' coordinates
containing
four bosonic and eight fermionic coordinates, and the `relative
coordinates' containing $4(r-1)$ bosonic and $8(r-1)$
fermionic coordinates. The quantization of the center of mass
system gives states carrying charge quantum numbers
$\pmatrix{p\cr r}$ with 16-fold degeneracy, $p$ being the
momentum conjugate to the overall U(1) phase.
This shows that the degeneracy is always a multiple of 16,
consistent with the fact that a short multiplet is 16-fold
degenerate. At this stage $p$
can be any integer, not necessarily
prime relative to $r$. However, since the total wave-function is
a product of the wave-function of the center of mass system and
the relative system, 
in order to determine the number of short multiplets
for a given value of $p$, we need to turn to the quantum
mechanics of the relative coordinates. 

It turns out that the bosonic coordinates in the relative
coordinate system describe a non-trivial $4(r-1)$ dimensional
manifold, known as the relative moduli space of $r$ 
monopoles\cite{MANTON,AH,GIBMAN}.
The quantum
mechanics of the bosonic and fermionic relative coordinates 
can be regarded as that of a supersymmetric particle moving in
this moduli space. 
There are several subtleties with this system. They are listed 
below:
\begin{itemize}
\item
First of all, the center of mass and the relative coordinates do
not completely decouple, although they decouple locally. 
The full moduli space has the structure\cite{AH}:
\be \label{ecmre}
(R^3\times S^1\times \MM_r)/ Z_r\, ,
\ee
where $R^3$ is parametrized by the center of mass location, $S^1$ 
by the
overall U(1) phase, and $\MM_r$ by the relative
coordinates. There is an identification of points in the product
space $R^3\times S^1\times \MM_r$ by a $Z_r$ transformation that
acts as a shift by $2\pi/r$ on $S^1$ and as a diffeomorphism on
$\MM_r$ without any fixed point\cite{AH,GIBMAN}. Due to this
identification, the total wave-function
must be invariant under this $Z_r$ transformation. Since the
part of the wave-function involving the coordinate of $S^1$ picks
up a phase $\exp(2\pi i p/r)$ under this $Z_r$, we see that the
wave-function involving the relative coordinates must pick up a
phase of $\exp(-2\pi i p/r)$ under this $Z_r$ transformation.
\item Normally the part of the wave-function involving the
relative coordinates will be a function on $\MM_r$. But it turns
out that the effect of the $8(r-1)$ fermionic degrees of freedom
in the quantum mechanical system makes the wave-function a
differential form of arbitrary rank on $\MM_r$\cite{GAUNT,BLUM}.
\item Finally, among all the possible states, the ones saturating
Bogomol'nyi bound correspond to harmonic differential forms on
$\MM_r$. This can be understood as follows. It can be shown that
the Hamiltonian of the relative coordinates correspond to the
Laplacian on $\MM$. Also it turns out that the BPS mass formula
is saturated by contribution from the center of mass
coordinates. Hence in order to get a BPS state, the part of the
wave-function involving the relative coordinates must be an
eigenstate of the corresponding Hamiltonian with zero eigenvalue
{\it i.e.} it must be a harmonic form on $\MM_r$.
Thus for every harmonic differential form we get a
short multiplet, since the fermionic degrees of freedom
associated with the center of mass coordinates supply the
necessary 16-fold degeneracy.
\end{itemize}
Thus the existence of a short multiplet of charge quantum numbers
$\pmatrix{p\cr r}$ would require the existence of a harmonic form
on $\MM_r$ that picks up a phase of $\exp(2\pi ip/r)$ under the
action of $Z_r$. According to the prediction of SL(2,Z) 
{\it such a
harmonic form should exist only for $p$ and $r$ relatively prime,
and not for other values of $p$\cite{SBPS}.}

For $r=2$ the relevant harmonic form can be constructed 
explicitly\cite{MANSCH,GIBRUB,SBPS},
therby verifying the existence of the states predicted by SL(2,Z)
duality.
For $r>2$ the analysis is more complicated since the metric in
the multimonopole moduli space is not known. However general
arguments showing the existence of the necessary harmonic forms
has been given\cite{SEG,PORR}.

Besides the BPS states discussed here, the spectrum of elementary
string states in the heterotic string theory on $T^6$ contains
many other BPS states. In the world-sheet theory, a generic state
is created by applying oscillators from the left- and the
right-moving sector on the Fock vacuum. The Fock vacuum, in turn,
is characterized by a pair of vectors $(\vec k_L, \vec k_R)$
specifying the charges (momenta) associated with the six
right-handed and twenty two left-handed currents on the
world-sheet. From the viewpoint of the space-time theory, these
28 components of $(\vec k_L, \vec k_R)$ are just appropriate
linear combinations of the charges
carried by the state under the 28 U(1) gauge fields. The tree
level mass formula for an elementary string state in the NS
sector is given by,\footnote{In this and all subsequent mass
formula $\lambda_2$ should really be interpreted as the vacuum
expectation value of $\lambda_2$.}
\be \label{eheto}
m^2 = {4\over \lambda_2} \Big[ {\vec k_R^2\over 2} + N_R -{1\over
2}\Big] = {4\over \lambda_2} \Big[ {\vec k_L^2\over 2} + N_L
-1\Big]\, ,
\ee
where $N_R$ and $N_L$ denote respectively the oscillator levels
of the state in the right- and the left-moving sectors of the
world-sheet. In the above equation the terms in the square
bracket denote the total contribution to $L_0$ and $\bar L_0$
from the oscillators, the internal momenta, and the vacua
in the right- and
the left-moving sectors respectively. 
Normally we do not have the factor of
$\lambda_2^{-1}$ in the mass formula since the formula refers to
the ADM mass measured in the string metric 
$G_{\mu\nu}=\lambda_2^{-1}g_{\mu\nu}$.
But here (and in the rest of the article) we
quote the ADM mass measured in the canonical metric
$g_{\mu\nu}$. This is
more convenient for discussing duality invariance of the
spectrum, since it is $g_{\mu\nu}$ and not $G_{\mu\nu}$ that
remains invariant under a duality transformation. 
The additive factor of $-1/2$ and
$-1$ can be interpreted as
the contributions to $L_0$ and $\bar L_0$ from the
vacuum. (In the covariant formulation these can be traced to the
contributions from the world-sheet ghost fields).

It turns out that of the full set of elementary string states,
only those states which satisfy the constraint\cite{DABHAR} 
\be \label{ehett}
N_R={1\over 2}\, ,
\ee
correspond to BPS states (short multiplets). From
eqs.\refb{eheto} we see that for these states
\be \label{ehetth}
N_L = {1\over 2} (\vec k_R^2 - \vec k_L^2) + 1\, .
\ee
The degeneracy $d(N_L)$ of short multiplets
for a given set of $\vec k_L$, $\vec k_R$ is 
determined by the number
of ways a level $N_L$ state can be created out of the Fock vacuum
by the 24 left-moving bosonic oscillators (in the light-cone
gauge) $-$ 8 from the transverse bosonic coordinates of
the string and
16 from the bosonization of the 32 left-moving fermions on the
world-sheet $-$ 
and is given by the formula:
\be \label{ehetf}
\sum_{N_L=0}^\infty d(N_L) q^{N_L} = \prod_{n=1}^\infty {1\over
(1-q^n)^{24}}\, .
\ee
The BPS states discussed earlier $-$ the ones which can be
regarded as the massive gauge bosons of a spontaneously broken
non-abelian gauge theory $-$ correspond to the $N_L=0$ states in
this classification. From eq.\refb{ehetf} we see that we have
only one short multiplet for states with this quantum number;
this is consistent with their description as heavy gauge bosons
in an N=4 supersymmetric gauge theory. The next interesting class
of states are the ones with $N_L=1$. From \refb{ehetf} we see that
they have degeneracy 24.\footnote{In counting degeneracy we are
only 
counting the number of short multiplets, and ignoring the trivial
factor of 16 that represents the degeneracy within each short
multiplet.} An SL(2,Z) transformation relates these states to
appropriate magnetically charged states with $r$ units of
magnetic charge and $p$ units of electric charge for $p$ and $r$
relatively prime. Thus the SL(2,Z) self-duality symmetry of the
heterotic string theory predicts the existence of 24-fold
degenerate solitonic states with these charge quantum numbers.

Verifying the existence of these solitonic states turns out to be
quite difficult\cite{GAUHAR}. The main problem is that unlike the
$N_L=0$ states, the solitonic states (known as H-monopoles) which
are related to the $N_L=1$ states by SL(2,Z) duality turn out to
be singular objects, and hence we cannot unambiguously determine
the dynamics of collective coordinates of these solitons just
from the low energy effective field theory. Nevertheless, the
problem has now been solved for
$r=1$\cite{WITTSM,PORRH,SETHSTTWO,BLU}, and one finds that these
solitons have exactly the correct degeneracy 24.

Similar analysis based on soliton solutions of low energy
supergravity theory has been used to test many other duality
conjectures\cite{HULLTOWN,SSSD,HARSTRSSD,DABH,HULLOPEN,SCHSLT}. 
One of the main problems with this approach has been
that unlike the example discussed in this section, most of these
other solutions are either singular, or has strong curvature at
the core where the low energy approximation breaks down. As a
result, analysis based on these solutions has been of
limited use. The
situation changed after the advent of D-branes, to which we now
turn.

\subsection{SL(2,Z) duality in type IIB on $S^1$ and
D-branes} \label{s31}

As discussed earlier, type IIB string theory in ten dimensions
has a conjectured SL(2,Z) duality symmetry group. In this section
I shall discuss the consequence of this conjectured symmetry for
the spectrum of BPS states in type IIB string theory compactified
on a circle $S^1$. For details, see \cite{WITTDB,DASMATFR}.

The spectrum of elementary string states in this theory
are characterized by two charges $k_L$ and $k_R$
defined as:
\be \label{e13}
k_L=(k\lambda_2^{1/4}/R-wR/\lambda_2^{1/4})/\sqrt{2}, 
\qquad k_R=(k\lambda_2^{1/4}/R+wR
/\lambda_2^{1/4})/\sqrt{2}\, ,
\ee
where $R$ denotes the radius of $S^1$ measured in the ten
dimensional canonical metric,
$k/R$ denotes the momentum along $S^1$ with $k$ being an integer, and 
$w$, also an integer, denotes the number of times the elementary
string 
is wound along $S^1$. 
As usual we have set $\alpha'=1$.
In the world-sheet theory describing first quantized string
theory, $k_L$ and $k_R$ denote the left and the right-moving
momenta respectively.
There are infinite tower of states with this
quantum number, obtained by applied appropriate oscillators, both
from the left- and the right-moving sector of the world-sheet, on
the Fock vacuum of the world-sheet theory carrying these quantum
numbers. 
The mass formula for any state in this tower, measured in the
ten dimensional canonical metric, is given by:
\be \label{e14}
m^2 = {2\over \sqrt{\lambda_2}} (k_L^2 + 2 N_L) = {2\over 
\sqrt{\lambda_2}}
(k_R^2+ 2 N_R)\, ,
\ee
where $N_L,N_R$ denote oscillator levels on the
left- and the right- moving
sector of the world-sheet respectively.\footnote{We have stated
the formula in the RR sector, but due to space-time supersymmetry
we get identical spectrum from the NS and the R sectors.} In normal
convention, one does not
have the factors of $(\lambda_2)$ in the mass
formula, but here it comes due to the fact that we are using the 
{\it ten dimensional}
canonical metric instead of the string metric to define the mass
of a state. (Note that if we had used the nine dimensional
canonical metric as defined in eqs.\refb{edil}, 
\refb{egmunu}, there will
be an additional multiplicative factor of $R^{-2/9}$ in the
expression for $m^2$.)

Most of these states are not BPS states as they are
not invariant under any part of the supersymmetry transformation.
It turns out that in order to be invariant under half of the
space-time supersymmetry coming from the left- (right-) moving
sector of the world-sheet, $N_L$ ($N_R$) must vanish\cite{DABHAR}. 
Thus a state with $N_L=N_R=0$ will preserve half of the total
number of supersymmetries and will 
correspond to ultra-short multiplets. 
{}From eq.\refb{e14} we see
that mass formula for these states takes the form:
\be \label{e15}
m^2 = {2k_L^2\over \sqrt{\lambda_2}} = {2k_R^2\over 
\sqrt{\lambda_2}}\, .
\ee
This is the BPS mass formula for these ultra-short multiplets. 
This requires
$k_L=\pm k_R$ or, equivalently, $k=0$ or $w=0$. On the other
hand, a state with 
either $N_L=0$ or $N_R=0$ will break (3/4)th of the total number
of supersymmetries in the theory, and will correspond to short
multiplets. If, for definiteness, we consider states with
$N_R=0$, then the BPS mass formula takes the form:
\be \label{e16}
m^2 = {2k_R^2\over \sqrt{\lambda_2}}\, .
\ee
$N_L$ is determined in 
terms of $k_L$ and $k_R$ through the relation:
\be \label{e16a}
N_L = {1\over 2} (k_R^2-k_L^2) = wk\, .
\ee
There is no further constraint on $w$ and $k$. Although we
have derived these mass formulae by directly analysing the
spectrum of elementary string states, they can also be derived by
analyzing the supersymmetry algebra, as indicated earlier.

One can easily calculate the degeneracy of these states by
analyzing the spectrum of elementary string states in detail. For
example, for the states with $N_L=N_R=0$, there is a 16-fold
degeneracy of states in each (left- and right-) sector of the
world-sheet, $-$ 8 from the NS sector and 8 from the R sector.
Thus the net degeneracy of such a state is $16\times 16=256$,
showing that there is a unique ultra-short multiplet carrying given
charges $(k_L,k_R)$. The degeneracy of short multiplets can be
found in a similar manner. Consider for example states with
$N_R=0$, $N_L=1$. 
In this case there is a $16$-fold degeneracy coming from
the right-moving sector of the world-sheet. There is an
8-fold degeneracy from the Ramond sector
Fock vacuum of the left-moving
sector. There is also an extra degeneracy factor in the
left-moving Ramond sector due to the
fact that there are many oscillators that can act
on the Fock vacuum of the world-sheet theory
to give a state at oscillator level $N_L=1$. 
For example we get eight states by acting with
the transverse bosonic oscillators $\alpha^i_{-1}$ ($1\le i\le
8)$, and eight states by acting with the transverse fermionic
oscillators $\psi^i_{-1}$.\footnote{Since $\psi^i_{-1}$ has fermion
number one, it has to act on the Fock
vacua
with odd fermion number in order that the states
obtained after acting with $\psi^i_{-1}$
on the vacua satisfy GSO projection.} This gives total
degeneracy factor of 8$\times$16 in the left-moving Ramond sector. 
Due to supersymmetry, we get an identical factor from the left-moving
NS sector as well.
Thus we get a state with total degeneracy $16\times 16\times
16$, $-$ 16 from the right moving sector, and $16\times 16$ from
the left-moving sector $-$ 
which is the correct degeneracy of a single
short multiplet. Similar counting can be done for higher values
of $N_L$ as well.
It turns out that the total number of short multiplets $d(N_L)$ with
$N_R=0$ for some given value of $N_L\ge 1$ is given by the formula:
\be \label{e17}
\sum_{N_L} d(N_L) q^{N_L} = {1\over 16} \, \, 
\prod_{n=1}^\infty \Big({1+q^n\over 1-q^n}\Big)^8\, .
\ee
The $(1+q^n)^8$ and $(1-q^n)^8$ factors in the numerator and the
denominator are related respectively to the fact that in the
light-cone gauge there are 8 left-moving fermionic fields
and 8 left-moving bosonic fields on the world-sheet. 
The overall factor of
(1/16) is due to the fact that the lowest level state
is only 256-fold
degenerate but a single short multiplet requires $16\times 256$
states.

Let us first consider the ultra-short multiplet with $k=0$, $w=1$.
These states have mass
\be \label{e18}
m^2 = {R^2\over \lambda_2}\, .
\ee
It is well known that an elementary string acts as a source of
the $B_{\mu\nu}$ field (see {\it e.g.} ref.\cite{DABHAR}). Thus
in the (8+1) 
dimensional theory obtained by compactifying type IIB on $S^1$,
the $w=1$ state will carry one unit of $B_{9\mu}$ gauge field
charge. Now, under SL(2,Z)
\be \label{e19}
\pmatrix{B_{9\mu}\cr B'_{9\mu}}\to \pmatrix{p & q\cr r & s}
\pmatrix{B_{9\mu}\cr B'_{9\mu}} \, .
\ee
This converts the $w=1$ state, which we shall denote by
$\pmatrix{1 \cr 0}$ reflecting the $\pmatrix{B_{9\mu}\cr
B'_{9\mu}}$ charge carried by the state, 
to a $\pmatrix{p\cr r}$ state, {\it i.e.} a state
carrying $p$ units of
$B_{9\mu}$ charge and $r$ units of $B'_{9\mu}$ charge.
The condition
$ps - qr=1$ implies that the pair of integers $(p,r)$ are relatively
prime. 
Thus SL(2,Z) duality of type IIB string theory
predicts that $\forall (p,r)$ relatively prime,
the theory must have a unique ultra-short multiplet with $p$ units of
$B_{9\mu}$ charge and $r$ units of $B'_{9\mu}$ charge\cite{SCHSLT}.
The BPS mass formula for these states
can be derived by analysing the
supersymmetry algebra, as indicated earlier, and is given by,
\be \label{e20}
m^2 = {R^2\over \lambda_2} |r\lambda-p|^2\, .
\ee
Note that this formula is invariant under the
SL(2,Z) transformation:
\be \label{e21}
\lambda\to {a\lambda +b\over c\lambda +d}, \qquad
\pmatrix{p\cr r}\to \pmatrix{a & b\cr c & d} \pmatrix{p\cr r}\, ,
\ee
where $\pmatrix{a & b\cr c & d}$ is an SL(2,Z) matrix.

A similar prediction for the spectrum of BPS states
can be made for short multiplets as well. In this case the state
is characterized by three integers $p$, $r$ and $k$ 
reflecting the $B_{9\mu}$, $B'_{9\mu}$ and $G_{9\mu}$ charge
(momentum along $S^1$) respectively.
Let us 
denote by $d(k,p,r)$ the degeneracy of such short multiplets.
For $(p,r)$ relatively prime, an SL(2,Z) transformation relates
these to elementary string states with one unit of winding and
$k$ units of momentum along $S^1$. Such states have degeneracy
$d(k)$ given in eq.\refb{e17}.
Then by following the same logic as before, we see that the
SL(2,Z) duality  predicts that for $(p,r)$ relatively prime,
$d(k,p,r)$ is independent of $p$ and $r$ and depends
on $k$ according to the relation:
\be \label{e22}
\sum_k d(k,p,r) q^k = {1\over 16} \prod_{n=1}^\infty
\Big({1+q^n\over 1 - q^n}\Big)^8 \, .
\ee
In other words, there should be a
Hagedorn spectrum of short multiplets with charge 
$\pmatrix{p\cr r}$.

A test of SL(2,Z) symmetry involves 
explicitly verifying the existence of these states. To see what
such a test involves, recall that $B'_{\mu\nu}$ arises in the RR
sector of string theory.
In type II theory, all elementary string states are neutral under RR
gauge fields as can be seen by computing a three point function
involving any two elementary string states and an RR sector
gauge field. Thus
a state carrying $B'_{9\mu}$ charge must arise as a soliton. The
naive approach will involve constructing such a soliton solution
as a solution to the low energy supergravity equations of motion,
quantizing its zero modes, and seeing if we recover the correct
spectrum of BPS states. However, in actual practice, when one
constructs the solution carrying $B'_{\mu\nu}$ charge, it turns
out to be singular. 
Due to this fact it is difficult to proceed further along
this line, as
identifying the zero modes of a singular solution is not a well
defined procedure. In particular we need to determine what
boundary condition the modes must satisfy at the singularity.
Fortunately, in this theory, there is a novel way of constructing
a soliton solution that avoids this problem. This construction
uses Dirichlet (D-) branes\cite{POLD,DOLD}.
In order to
compute the degeneracy of these solitonic states, we
must understand the definition and some of the the properties of
these D-branes. This is the subject to which we now turn.
\begin{figure}[!ht] 
\begin{center}
\leavevmode
\epsfbox{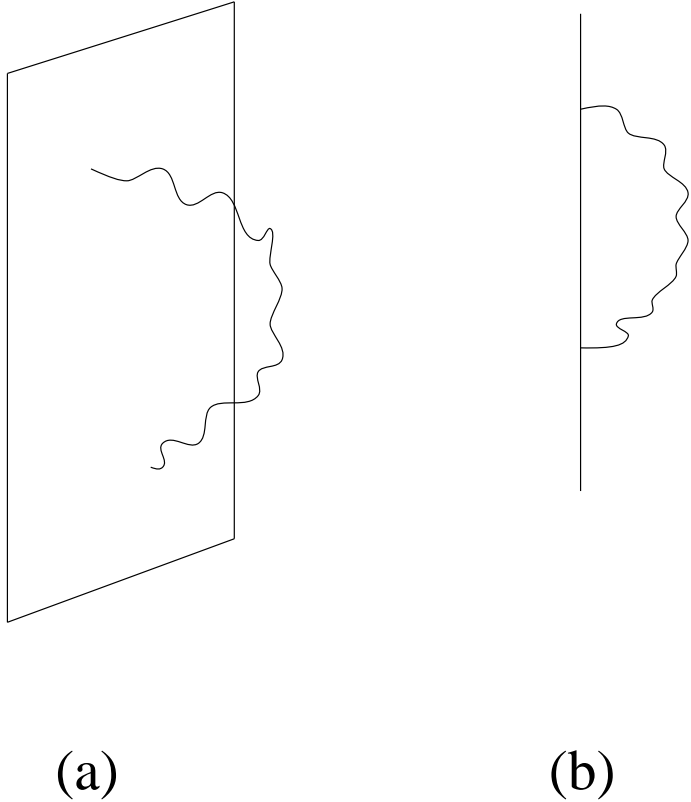}
\end{center}
\caption[]{\small Open string states with ends attached to a
(a) Dirichlet membrane, (b) Dirichlet string.}
\label{f6}
\end{figure}

Normally type IIA/IIB string
theory contains closed string states only.
But we can postulate existence of solitonic extended objects in these
theories such that in the presence of these solitons, there can be
open string states whose ends lie on these extended objects (see
Fig.\ref{f6}).
This can in fact be taken to be the defining relation for these
solitons, with the open string states with ends lying on the soliton
corresponding to the (infinite number of) vibrational modes 
of the soliton.
Of course, one needs to ensure that the soliton defined this way
satisfy all the properties expected of a soliton solution in this
theory {\it e.g.} partially unbroken supersymmetry, existence of
static multi-soliton solutions etc. 
Since open strings satisfy Dirichlet boundary condition in directions
transverse to these solitons, these solitons are called D-branes.
In particular, we shall call a D-brane with Neumann boundary
condition in
$(p+1)$ directions (including time) and Dirichlet 
boundary condition in
$(9-p)$ directions a Dirichlet $p$-brane, since it can be
regarded as a soliton extending along $p$ space-like directions
in which we have put Neumann boundary condition. (Thus a 0-brane
represents a particle like object, a 1-brane a string like
object, and a 2-brane a membrane like object.)
To be more explicit, let us consider the following boundary
condition on the open string:
\ben \label{e23}
X^m(\sigma=0,\pi) &=& x_0^m \quad \hbox{for} \quad (p+1) \le m \le
9\, , \nonumber \\
\p_\sigma X^\mu(\sigma=0, \pi) &=& 0 \quad \hbox{for} \quad 0\le
\mu \le p\, ,
\een
where $\sigma$ denotes the spatial direction on the string
world-sheet.
The boundary conditions on the world-sheet fermion fields are
determined from \refb{e23} using various consistency requirements
including world-sheet supersymmetry that relates the world-sheet 
bosons and fermions.
Note that these boundary conditions break translational
invariance along $x^m$. Since we want the full theory to be
translationally invariant, the only possible interpretation of
such a boundary condition is that there is a $p$ dimensional
extended object situated at $x^m =x_0^m$ that is responsible for
breaking this translational invariance. We call this
a Dirichlet $p$-brane located at $x^m=x_0^m$ ($p+1\le m
\le 9$), and extended along $x^1, \ldots x^p$. 

Let us now summarize some of the important properties of D-branes
that will be relevant for understanding the
test of SL(2,Z) duality in type IIB string theory:
\begin{itemize}
\item
The Dirichlet $p$-brane in IIB is invariant under half of the space-time
supersymmetry transformations for odd $p$.
To see how this property arises, let us denote by
$\epsilon_L$  and $\epsilon_R$ the space-time supersymmetry
transformation parameters in
type IIB string theory, originating in the left- and the
right-moving sector of the world-sheet theory respectively.
$\epsilon_L$ and $\epsilon_R$ satisfy the chirality constraint:
\be \label{e24}
\Gamma^0\cdots \Gamma^9\epsilon_L=\epsilon_L, \qquad
\Gamma^0\cdots \Gamma^9\epsilon_R=\epsilon_R\, ,
\ee
where $\Gamma^\mu$ are the ten dimensional gamma matrices.
The open string boundary conditions \refb{e23} together with the
corresponding boundary conditions on the world-sheet fermions
give further restriction on $\epsilon_L$ and $\epsilon_R$ of the
form\cite{POLD}:
\be \label{e25}
\epsilon_L=\Gamma^{p+1}\ldots \Gamma^9\epsilon_R\, .
\ee
It is easy to see that
the two equations \refb{e24} and \refb{e25}
are compatible only for odd $p$. Thus in type IIB string theory
Dirichlet $p$-branes are invariant under half of the space-time
supersymmetry transformations for odd $p$. An
identical argument shows that in type IIA string theory we have
supersymmetric Dirichlet $p$-branes only for even $p$ since in
this theory eq.\refb{e24} is replaced by,
\be \label{e26}
\Gamma^0\cdots \Gamma^9\epsilon_L=\epsilon_L, \qquad
\Gamma^0\cdots \Gamma^9\epsilon_R=-\epsilon_R\, .
\ee

\item
Type IIB (IIA) string theory contains a $p$-form gauge field for
even (odd) $p$. For example, in type IIB string theory these
$p$-form gauge fields correspond to the scalar $a$, the rank two
anti-symmetric tensor field $B'_{\mu\nu}$
and the rank four anti-symmetric tensor field
$D_{\mu\nu\rho\sigma}$. It can be shown that a
Dirichlet $p$-brane carries one unit of charge under the RR 
$(p+1)$-form gauge field\cite{POLD}. More precisely, if
we denote by $C_{\mu_1\cdots \mu_q}$ the $q$-form gauge potential,
then a Dirichlet $p$-brane extending along $1\cdots p$ direction
acts as a source of $C_{01\cdots p}$. (For $p=5$ and 7 these
correspond to magnetic dual potentials of $B'_{\mu\nu}$ and $a$
respectively.)
This result can be obtained by computing the one point function
of the vertex operator for the field $C$ in the presence of a
D-brane. The relevant string world-sheet diagram has been
indicated in Fig.\ref{f7}. We shall not discuss the details of
this computation here.
\begin{figure}[!ht] 
\begin{center}
\leavevmode
\epsfbox{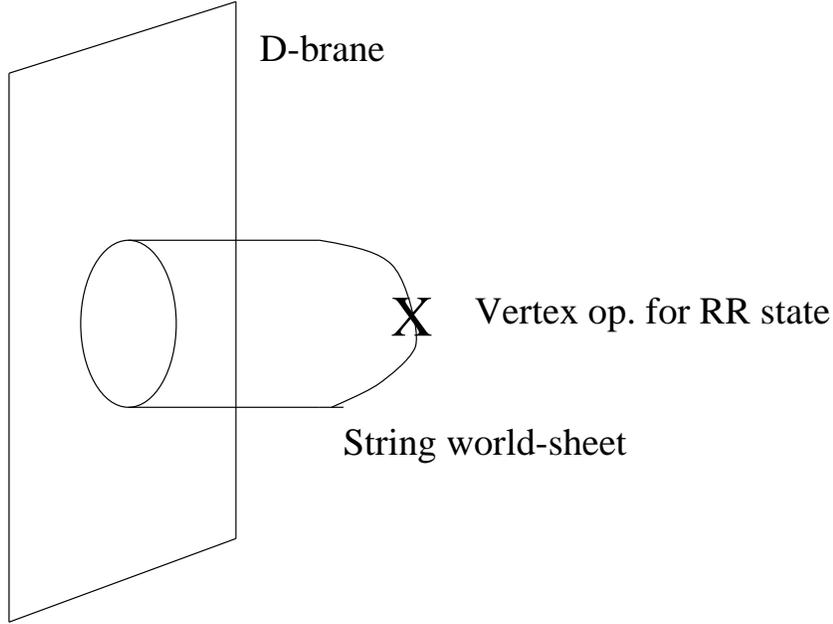}
\end{center}
\caption[]{\small  The string world-sheet diagram relevant for
computing the coupling of the RR gauge field to the D-brane.
It corresponds to a surface of the topology of a hemisphere with
its boundary glued to the D-brane. The vertex operator of the
RR-field is inserted at a point on the hemisphere.
}
\label{f7}
\end{figure}
\end{itemize}

{}From this discussion it follows that 
a Dirichlet 1-brane (D-string) in type IIB theory carries one 
unit of charge under the RR 2-form field $B'_{\mu\nu}$. This
means that in type IIB on $S^1$ (labelled by the coordinate $x^9$)
a D-string wrapped around the $S^1$ describes a particle charged
under $B'_{9\mu}$. This then is a
candidate soliton carrying charge quantum numbers $\pmatrix{0\cr 1}$
that is related to the $\pmatrix{1\cr 0}$ state via SL(2,Z)
duality. As we had seen earlier, SL(2,Z) duality predicts that
there should be a unique ultra-short multiplet carrying charge
quantum numbers $\pmatrix{0\cr 1}$. 
Thus our task now is as follows:
\begin{itemize}
\item Quantize the collective coordinates of this soliton.
\item Verify if we get an ultra-short multiplet in this quantum theory.
\end{itemize}
Since the D-string is a one dimensional object, the dynamics of
its collective coordinates should be described by a (1+1)
dimensional field theory.
As we had discussed earlier, all the vibrational modes of the
D-string are given by the open string states with ends attached
to the D-string. In particular, the zero frequency modes
(collective modes) of the
D-string that are relevant for analyzing the spectrum of BPS states
correspond to {\it massless} open string states propagating
on the D-string.  By analyzing the spectrum of these open string
states one finds that the collective coordinates in
this case correspond to 
\begin{itemize}
\item
8 bosonic fields $y^m$ denoting the location of this string in eight
transverse directions.
\item
A U(1) gauge field.
\item
8 Majorana fermions.
\end{itemize}
It can be shown that the
dynamics of these collective coordinates is described by a (1+1)
dimensional
supersymmetric quantum field theory which is the dimensional
reduction of the N=1 supersymmetric U(1) gauge theory from (9+1) to
(1+1) dimensions.
Normally in (1+1) dimension gauge fields have no dynamics.
But here since the space direction is compact,
$y\equiv \ointop A_1 dl$ is a physical variable. Furthermore, the
compactness of U(1) makes $y$ to be periodically identified
($(y\equiv y+ a)$ for some $a$). 
Thus the momentum $p_y$ conjugate to $y$ is quantized
($p_y=2\pi k/a$ with $k$ integer.)
It can be shown that\cite{WITTDB}
this momentum, which represents electric flux along the D-string, is
actually a source of $B_{9\mu}$ charge! Thus if we restrict to
the $p_y=0$ sector then these states carry $\pmatrix{0\cr 1}$
charge quantum numbers as discussed earlier, but by taking
$p_y=2\pi k/a$, we can get states carrying charge quantum numbers
$\pmatrix{k\cr 1}$ as well.

Due to the compactness of the space direction, we can actually
regard this as a quantum mechanical system instead of a (1+1)
dimensional quantum field theory.
It turns out
that in looking for ultra-short multiplets, we can ignore all
modes carrying momentum along $S^1$.
This corresponds to dimensionally reducing the theory to (0+1)
dimensions. The degrees of freedom of this quantum mechanical
system are:
\begin{itemize}
\item
8 bosonic coordinates $y^m$,
\item
1 compact bosonic coordinate $y$,
\item
16 fermionic coordinates.
\end{itemize}
A quantum state is labelled by the momenta conjugate to $y^m$ (ordinary
momenta) and an integer labelling momentum
conjugate to $y$ which can be identified with
the quantum number $p$ labelling $B_{9\mu}$ charge.
The fermionic coordinates satisfy the sixteen dimensional
Clifford algebra. Thus 
quantization of the fermionic coordinates gives
$2^8=256$ -fold degeneracy, which is precisely the correct
degeneracy for 
a ultra-short multiplet.
This establishes the existence of all the required states of charge
$\pmatrix{p\cr 1}$ predicted by SL(2,Z) symmetry.
\begin{figure}[!ht]
\begin{center}
\leavevmode
\epsfbox{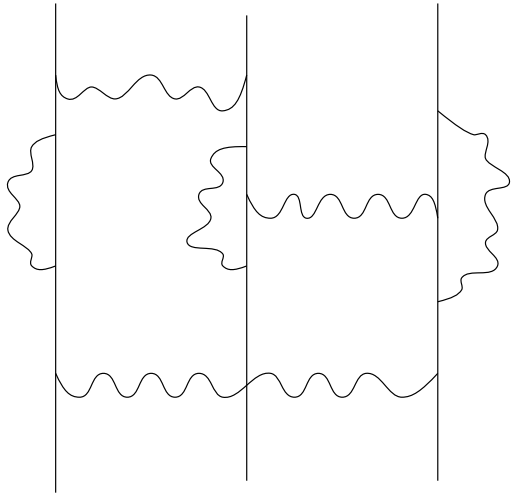}
\end{center}
\caption[]{\small  Possible open string states in the presence of
three parallel D-strings.
}
\label{fa1}
\end{figure}

What about $\pmatrix{p\cr r}$ states with $r>1$?
These carry $r$ units of $B'_{9\mu}$ charge and hence
must arise as a bound state of $r$ D-strings wrapped along $S^1$.
Thus the first question we need to ask is:
what is the (1+1) dimensional quantum field theory
governing the dynamics of this system?
In order to answer this question we need to study the dynamics of
$r$ D-strings. This system can be described as easily as a single
D-string: instead of allowing open strings to end on a single
D-string, we allow it to end on any of the $r$ D-strings situated
at
\be \label{eds1}
x^m = x^m_{(i)}\, , \qquad 2\le m\le 9, \qquad 1\le i\le r\, ,
\ee
where $\vec x_{(i)}$ denotes the location of the $i$-th
D-string. The situation is illustrated in Fig.\ref{fa1}.
Thus the dynamics of this system will now be described not only
by the open strings starting and ending on the same D-string, but
also by open strings whose two ends lie on two different
D-strings.

For studying the spectrum of BPS states we need to focus our
attention on the massless open string states. First of all, for
each of the $r$ D-strings we get a U(1) gauge field, eight scalar
fields and eight Majorana fermions from open strings with both
ends lying on that D-string. But we can get extra massless states
from open strings whose two ends lie on two different D-strings
when these two D-strings coincide. It turns out that for $r$
coincident D-strings the dynamics of massless strings on the
D-string world-sheet is given by
the dimensional reduction to (1+1) dimension
of N=1 supersymmetric $U(r)$ gauge theory
in ten dimensions, or equivalently, 
N=4 supersymmetric $U(r)$ gauge theory in four 
dimensions\cite{WITTDB}.
Following a logic similar to that in the case of a single D-string, 
one can show that the 
problem of computing the
degeneracy of $\pmatrix{p\cr r}$ states reduces to the
computation of certain Witten index in this quantum theory.
We shall not go through the details of this analysis, but just
state the final result. It turns out that {\it
there is a unique ultra-short multiplet for every pair of
integers $(p,r)$ which are relatively prime, precisely as
predicted by $SL(2,Z)$}\cite{WITTDB}!

A similar analysis can be carried out for the short multiplets
that carry momentum $k$ along $S^1$ besides
carrying the $B$ and $B'$
charges $p$ and $r$\cite{WITTDB,DASMATFR}. 
In order to get these states from the
D-brane spectrum, we can no longer dimensionally reduce the
(1+1) dimensional theory to (0+1) dimensions.
Instead we need to take into account the modes 
of the various fields of the (1+1) dimensional field theory
carrying momentum along the internal $S^1$. The BPS states come
from configurations where only the left- (or right-) 
moving modes on $S^1$ are
excited.  The calculation of the degeneracy $d(k,p,r)$ of BPS states
carrying given charge quantum numbers $(p,r,k)$ is done by
determining in how many ways the total momentum $k$ can be
divided among the various left-moving bosonic and fermionic
modes. This counting problem turns out to be identical to the one used
to get the Hagedorn spectrum of BPS states in the elementary
string spectrum, except that the elementary string is replaced
here by the solitonic D-string. Naturally, we get back the
Hegedorn spectrum for $d(k,p,r)$ as well. Thus
the answer agrees exactly with that predicted by SL(2,Z)
duality. This provides us with a 
test of the conjectured SL(2,Z) symmetry of type IIB on $S^1$.

The method of using D-branes to derive the dynamics of 
collective coordinates has been used to verify the
predictions of other duality conjectures involving various string
compactifications. Among them are
self-duality of type II string theory on
$T^4$\cite{BSVO,SUD,VAFUD,SETHSTTWO}, 
the duality between heterotic on $T^4$ and 
type IIA on K3\cite{BSV},
the duality between type I and SO(32) heterotic string
theory\cite{POLCWIT}, etc.

\subsection{Massless solitons and tensionless strings}

An interesting aspect of the conjectured duality between the
heterotic string theory on $T^4$ and type IIA string theory on K3
is that at special points in the moduli space the heterotic
string theory has enhanced non-abelian gauge symmetry {\it e.g.} 
$E_8\times E_8$ or SO(32) in the absence of vacuum expectation
value of the internal components of the gauge fields, SU(2) at
the self-dual radius etc. Perturbative type IIA string theory on
K3 does not have any such gauge symmetry enhancement, since the
spectrum of elementary string states does not contain any state
charged under the U(1) gauge fields arising in the RR sector.
Thus, for example, we do not have the $W^\pm$ bosons that are
required for enhancing a U(1) gauge group to SU(2). At first
sight this seems to lead to a contradiction. However upon closer
examination one realises that this cannot really be a
problem\cite{HTS}. 
To see this let us consider a point in the moduli space of
heterotic string theory on $T^4$ where the
non-abelian gauge symmetry is broken. At this point the would be
massless gauge bosons of the non-abelian gauge theory acquire
mass by Higgs mechanism, and appear as BPS states in the abelian
theory. As we approach the point of enhanced gauge symmetry,
the masses of these states vanish. Since the masses of BPS states
are determined by the BPS formula, the vanishing of the masses
must be a consequence of the BPS formula.
Thus if we are able to find
the images of these BPS states on the type IIA side as appropriate
D-brane states, then the masses of these D-brane states must also
vanish as we approach the point in the moduli space where the
heterotic theory has enhanced gauge symmetry. These massless
D-brane solitons will then provide the states necessary for
enhancing the gauge symmetry.

To see this more explicitly, let us examine the BPS formula. It
can be shown that in the variables defined in section
\ref{siiahet} the BPS formula is given by,
\be \label{eiibps}
m^2 = e^{-\Phi^{(A)}/2} \alpha^T (L M^{(A)}L +L)\alpha\, ,
\ee
where $\alpha$ is a 24 dimensional vector belonging to the
lattice $\Lambda'_{24}$, and represents the U(1) charges carried
by this particular state. For each $\vec\alpha$ we can assign an
occupation number $n(\vec\alpha)$ which gives the number of BPS
multiplets carrying this specific set of charges. Since $M^{(A)}$
is a symmetric O(4,20) matrix, we can express this as
$\Omega^{(A)T}\Omega^{(A)}$ for some O(4,20) matrix
$\Omega^{(A)}$, and rewrite eq.\refb{eiibps} as
\be \label{eiibpsa}
m^2 = e^{-\Phi^{(A)}/2} \alpha^T L\Omega^{(A)T}
(I_{24} +L)\Omega^{(A)} L \alpha\, .
\ee
As can be seen from eq.\refb{ess2}, $(I_{24}+L)$ has 20 zero
eigenvalues. As we vary $M^{(A)}$ and hence $\Omega^{(A)}$, the
vector $\Omega^{(A)}L\alpha$ rotates in the twenty four
dimensional space. If for some $\Omega^{(A)}$ it is aligned along
one of the eigenvectors of $(I_{24}+L)$ with zero eigenvalue, we
shall get massless solitons provided the occupation number
$n(\vec \alpha)$ for this specific $\vec\alpha$ is non-zero.

Although this argument resolves the problem at an abstract level,
one would like to understand this mechanism directly by analysing
the type IIA string theory,
since, after all, we do not encounter massless solitons very
often in physics. This has been possible through the work of
\cite{WITTEND,STROMSM,BSV}. For simplicity let us focus on the case
of enhanced SU(2) gauge symmetry. First of all, one finds that
at a generic point in the moduli space where SU(2) is broken, the
images of the $W^\pm$ bosons in the type IIA theory are given by
a D-2 brane wrapped around a certain 2-cycle (topologically
non-trivial two dimensional surface) inside K3, the + and the $-$
sign of the charge being obtained from two different orientations 
of the D-2 brane. Since the two
tangential directions on the D-2 brane are directed along the two
internal directions of K3 tangential to the 2-cycle, this object
has no extension in any of the five non-compact spatial
directions, and hence behaves like a particle.\footnote{These
states were analyzed in detail in \cite{DOUGENH}}. It turns
out that as we
approach the point in the moduli space where the theory on the
heterotic side develops enhanced SU(2) gauge symmetry, the K3 on
which type IIA theory is compactified becomes singular. At this
singularity the area of the topologicaly non-trivial
2-cycle mentioned above goes to zero.  As a result, the mass of
the wrapped D-2 brane, obtained by multiplying the tension of the D-2
brane by the area of the two cycle, vanishes. This
gives us the massless solitons that are required
for the gauge symmetry
enhancement. A similar mechanism works for getting other gauge
groups as well. In fact it turns out that there is a one to one
correspondence between the enhanced gauge groups, which are
classified by A-D-E dynkin diagram, and the singularity type of
K3, which are also classified by the A-D-E dynkin 
diagram\cite{WITTEND}. This establishes an explicit
physical relationship between A-D-E singularities and A-D-E lie
algebras.

The appearance of enhanced gauge symmetry in type IIA on K3 poses
another puzzle. Let us compactify this theory on one more circle.
Since such a compactification does not destroy gauge symmetry,
this theory also has enhanced gauge symmetry when the K3 becomes
singular. But type IIA on $K3\times S^1$ is T-dual to type IIB on
$K3\times S^1$; thus type IIB on $K3\times S^1$ must also develop
enhanced gauge symmetry when K3 develops singularities. Does this
imply that type IIB on K3 also develops enhanced gauge symmetry
at these special points in the K3 moduli space? This does not
seem possible,
since type IIB string theory does not have any D-2
brane solitons which can be wrapped around the collapsed two
cycles of K3.
It turns out that instead of acquiring enhanced gauge symmetry,
type IIB string theory acquires tensionless strings at these
special points in the K3 moduli space\cite{WITTENTEN}. These
arise from taking a D-3 brane of type IIB string theory, and
wrapping it on a two cycle of K3. Thus two of the tangential
directions of the three brane are directed along the internal
directions of K3, and the third direction of the three brane is
along one of the non-compact spatial directions. Thus from the
point of view of the (5+1) dimensional theory such a
configuration will appear as a string. The tension of this string
is given by the product of the tension (energy per unit three 
volume) of the three brane and the area of the two cycle on which
the three brane is wrapped. Thus as we approach the singular point on
the K3 moduli space where the area of the two cycle vanishes, the
tension of the string goes to zero. In other words, we get
tensionless strings. Upon further compactification on a circle we
get massless particles from configurations where this tensionless
string is wound around the circle. These are precisely the
massless gauge bosons required for the gauge symmetry enhancement
in type IIB on $K3\times S^1$.

\sectiono{Interrelation Between Different Duality Conjectures}
\label{s4}

In the last three sections
we have seen many different duality conjectures and have learned how to
test these conjectures.
We shall now see that many of these conjectures are not independent, but
can be `derived' from each other. There are several different ways
in which dualities can be related to each other. We shall discuss
them one by one. The material covered in this section is taken
mainly from \cite{VAFWITFI,DUORBI,UNITY}.

\subsection{Combining non-perturbative  and T- dualities}
\label{s41}

Suppose a string theory A compactified on a manifold $K_A$ has a
conjectured duality symmetry group $G$.
Now further compactify this theory on some manifold $\MM$.
Then the theory A on $K_A\times \MM$ is expected to have the following
set of duality symmetries:
\begin{itemize}
\item
It inherits the original duality symmetry group $G$ of A on $K_A$.
\item It also has a perturbatively verifiable $T$-duality group.
Let us call it $H$.
\end{itemize}
Quite often $G$ and $H$ do not commute and together
generate a much bigger
group\cite{STHD,HULLTOWN}.
In that case,
the existence of this bigger group of symmetries can be regarded as
a consequence of the duality symmetry of A on $K_A$ and T-dualities.

We shall illustrate this with a specific example\cite{HULLTOWN}. 
We have seen that in ten dimensions type IIB string theory has a
conjectured duality group SL(2,Z) that acts non-trivially on the
coupling constant. From the table given in section \ref{s23} we
see that type IIB on $T^n$ also has a T-duality group
$SO(n,n;Z)$, whose existence can be verified order by order in
string
perturbation theory. It turns out that typically these two duality
groups do not commute, and in fact generate the full duality
symmetry group of type IIB on $T^n$ as given in the table of
section \ref{s23}. Thus we see that the existence of the full
duality symmetry group of type IIB on $T^n$ can be infered from
the SL(2,Z) duality symmetry of the ten dimensional type IIB
string theory, and the perturbatively verifiable T-duality
symmetries of type IIB on $T^n$.

\subsection{Duality of dualities} \label{s42}

Suppose two theories are conjectured to be
dual to each other, and each theory in turn has a conjectured
self-duality group.
Typically part of this self duality group is T-duality, and
the rest involves non-trivial transformation of the coupling
constant.
But quite often the non-perturbative duality transformations in one theory 
correspond to T-duality in the dual theory and vice versa.
As a result, 
the full self duality group in both theories follows from the 
conjectured duality between the two theories.

Again we shall illustrate this with an example\cite{DUFFSS,WITTEND}. 
Let us
start with the conjectured duality between heterotic on
$T^4$ and type IIA on K3. Now let us 
compactify both theories further on a two dimensional torus $T^2$. 
This produces a dual pair of theories: 
type IIA on $K3\times T^2$ and heterotic on $T^6$.
Now,
heterotic on $T^6$ has a T-duality group $O(6,22;Z)$ that can be
verified using heterotic perturbation theory. On the other hand ,
type IIA on $K3\times T^2$ has a T-duality group $O(4,20;Z)\times
SL(2,Z)\times SL(2,Z)'$ that can be verified using type II 
perturbation theory.
The full conjectured duality group in both  theories is
$O(6,22,Z)\times SL(2,Z)$.
\begin{figure}[!ht] 
\begin{center}
\leavevmode
\epsfbox{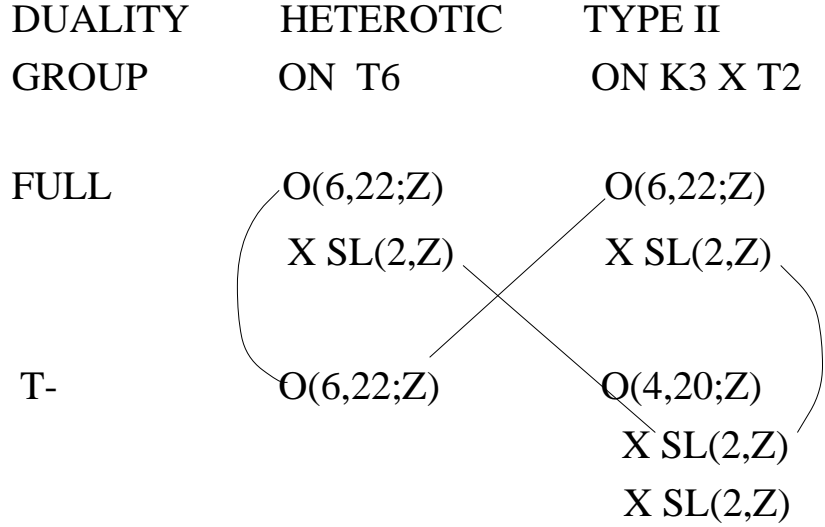}
\end{center}
\caption[]{\small  The embedding of the T-duality groups in the
full duality group in heterotic on $T^6$ and type IIA on
$K3\times T^2$.
}
\label{f8}
\end{figure}

Now the question we would like to address is, how are the T-duality
symmetry groups in the two theories embedded in the full
conjectured
$O(6,22;Z)\times SL(2,Z)$ duality group? This has been
illustrated in Fig.\ref{f8}. In particular we find that the
SL(2,Z) factor of the full duality group is a subgroup of the
T-duality group in type IIA on $K3\times T^2$, and hence can be
verified in this theory order by order in perturbation theory. On
the other hand, the O(6,22;Z) factor of the duality group appears
as a T-duality symmetry of the heterotic string theory, and hence
can be verified order by order in perturbation theory in this
theory. Thus assuming that T-duality in either theory is a valid
symmetry, and the duality between the heterotic on $T^4$ and type
IIA on K3, we can establish the existence of the self-duality
group O(6,22;Z)$\times$SL(2,Z) in heterotic on $T^6$ and type IIA
on K3$\times T^2$.

Using the results of this and the previous subsection, we see
that so far among all the conjectured non-perturbative 
duality symmetries, the independent ones are:
\begin{enumerate}
\item SL(2,Z) of type IIB in D=10,
\item type I $\leftrightarrow$ SO(32) heterotic in D=10, and
\item
IIA on K3 $\leftrightarrow$ heterotic on $T^4$.
\end{enumerate}
We shall now show how to `derive' 3) from 1) and 2).

\subsection{Fiberwise duality transformation} \label{s43}
\begin{figure}[!ht] 
\begin{center}
\leavevmode
\epsfbox{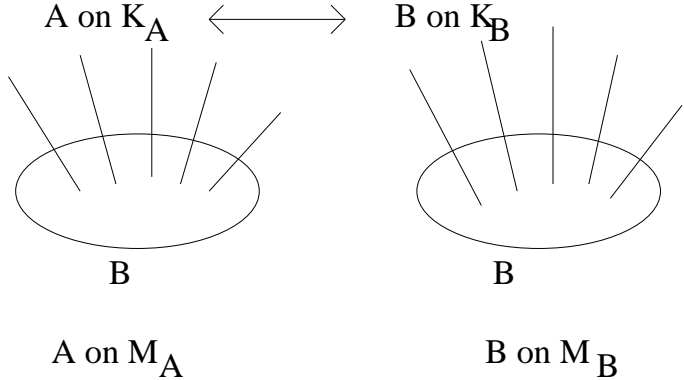}
\end{center}
\caption[]{\small  Application of fiberwise duality
transformation. In each local neighbourhood of the base manifold
$\BB$, the two theories are equivalent due to the equivalence of
the theories living on the fiber ($\times$ any manifold). Thus we
would expect the theories $A$ on $\MM_A$ and $B$ on $\MM_B$ to be
equivalent.}
\label{f9}
\end{figure}

In this subsection we shall describe the idea of constructing
dual pairs of theories using fiberwise duality
transformation\cite{VAFWITFI}.
Suppose (Theory A on $\KK_A$) has been conjectured to be dual to
(Theory B on $\KK_B$). Here
$A$ and $B$ are two of the five different string theories in
D=10, and 
$\KK_A$, $\KK_B$ are two different manifolds (in general).
This duality involves a precise map between the moduli spaces of
the two theories.
Now construct a pair of new manifolds $\MM_A$, $\MM_B$ by starting from
some other manifold $\BB$, and erecting at every point on $\BB$ a
copy of $\KK_A$, $\KK_B$.
The moduli of $\KK_A$, $\KK_B$ vary slowly over $\BB$ and are related
to each other via the duality map that relates (A on $\KK_A$) to (B on
$\KK_B$). 
Then we would expect a duality 
\begin{center}
Theory A on $\MM_A$ $\leftrightarrow$ Theory B on $\MM_B$
\end{center}
by applying the duality transformation fiberwise.
This then gives rise to a new duality conjecture.
This situation has been illustrated in Fig.\ref{f9}. 

Now suppose that
at some isolated points (or subspaces of codimension $\ge$ 1)
on $\BB$ the fibers $\KK_A$ and $\KK_B$ degenerate.
(We shall see
some explicit examples of this later.)
Is the duality between ($A$ on $\MM_A$) and ($B$ on $\MM_B$)
still valid? We might expect that
even in this case the duality between the two theories
holds since the singularities
occur on subspaces of `measure zero'.
Although there is no rigorous argument as to why this should be
so,  this appears to be the case in all known examples.
Conversely, assuming that this is the case, we can derive the
existence of many new duality symmetries from a given duality
symmetry. 
\begin{figure}[!ht] 
\begin{center}
\leavevmode
\epsfbox{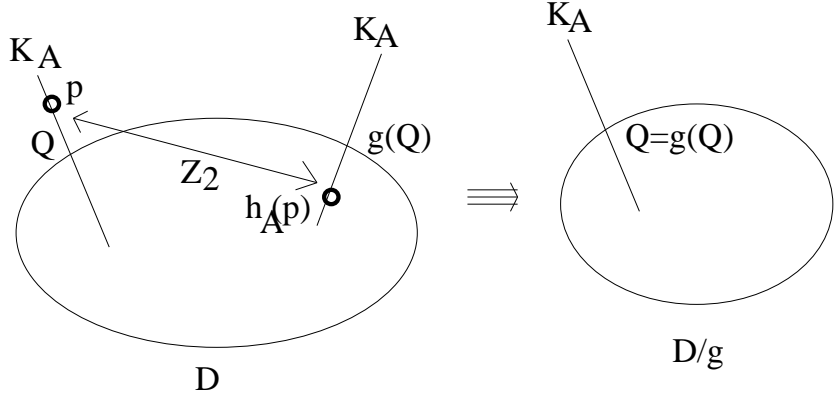}
\end{center}
\caption[]{\small  Representation of a  $Z_2$ orbifold as a
fibered space. The $Z_2$ transformation relates the point $(Q,p)$
on $\DDA\times\KK_A$ to the point $(g(Q),h_A(p))$.}
\label{f10}
\end{figure}

A special case of this construction involves $Z_2$ 
orbifolds.
Suppose we have a dual pair
(A on $\KK_A$) $\leftrightarrow$ (B on $\KK_B$).
Further suppose that (A on $\KK_A$) has a $Z_2$ symmetry generated by
$h_A$.
Then the dual theory must also have a $Z_2$ symmetry generated by
$h_B$.
$h_A$ and $h_B$ are mapped to each other under duality.
Now compactify both theories on another manifold $\DD$ with a $Z_2$
isometry generated by $g$, and compare the two quotient theories
\begin{center}
(A on $K_A \times \DD/h_A\cdot g$) and (B on $K_B\times \DD/h_B\cdot g$) 
\end{center}
$(\KK_A\times \DD/h_A\cdot g)$ is obtained from the product manifold
$\KK_A\times \DDA$ by identifying points that are related by the
$Z_2$ transformation $h_A\cdot g$. This situation is illustrated in
Fig. \ref{f10}. As shown in this figure, $(\KK_A\times \DDA/h_A\cdot g)$
admits a fibration with base $\DD/g$ and
fiber $\KK_A$. In particular,
note that since $h_A\cdot g$ takes a point $(p\in \KK_A, Q\in \DD)$ to
$(h_A(p), g(Q))$,
if we focus our attention on a definite point $Q$ on $\DD$, then there
is no identification of the points in the copy of $\KK_A$ that is
sitting at $Q$.
This shows that the fiber is $\KK_A$ and {\it not} $\KK_A/h_A$.
As we go from $Q$ to $g(Q)$, which is a closed cycle on $\DD/g$, the fiber
gets twisted by the transformation $h_A$.

The second theory, B on $(\KK_B\times \DD)/(h_B\cdot g)$ has an
identical structure.
Thus we can now apply fiberwise duality transformation to derive a new
duality:
\begin{center}
(A on $K_A \times \DD/h_A\cdot g$) 
$\leftrightarrow$ (B on $K_B\times \DD/h_B\cdot g$) 
\end{center}

Note that if $P_0\in \DD$ is a fixed point of $g$ ({i.e.}
if $g(P_0)=P_0$)
then in $\KK_A\times\DDA/h_A\cdot g$ there is an identification of points
$(p,P_0)$ and $(h_A(p),P_0)$. Similarly in 
$\KK_B\times\DDA/h_B\cdot g$ there is an identification of points
$(p',P_0)$ and $(h_B(p'),P_0)$.
Thus
at $P_0$ the fibers degenerate to $\KK_A/h_A$ and $\KK_B/h_B$
respectively.
At these points
the argument in support of duality between the two theories 
breaks down.
However, as we have discuused earlier,
since these are points of `measure zero' on $\DD$, we would expect
that the two quotient theories are still dual to each other\cite{DUORBI}.
We shall now illustrate this construction in the context of a
specific example.

We start with type IIB string theory in ten dimensions. This
has a conjectured SL(2,Z) symmetry.
Let $S$ denote the SL(2,Z) element
\be \label{e27}
\pmatrix{0 & 1\cr -1 & 0}\, .
\ee
Recall that this theory also has two global discrete symmetries 
$(-1)^{F_L}$ and $\Omega$.
The action of $S$, $(-1)^{F_L}$ and $\Omega$
on the massless bosonic fields in this theory
were described in section \ref{s21}.
{}From this one can explicitly compute the
action of $S(-1)^{F_L}S^{-1}$ on these massless fields.
This action turns out to be identical to that of $\Omega$.
A similar result holds for their action on the massless fermionic
fields as well. Finally, since the action of $S$ on the massive
fields is not known, one can define this action in such a way
that the actions of $S (-1)^{F_L} S^{-1}$ and $\Omega$ are
identical on all states.  This gives:
\be \label{e28}
S (-1)^{F_L} S^{-1} = \Omega\, .
\ee
 
We are now ready to apply our formalism. We 
take (A on $\KK_A$) to be type IIB in D=10, 
(B on $\KK_B$) to be type IIB in D=10 transformed by $S$,
$h_A$ to be $(-1)^{F_L}$,  $h_B$ to be $\Omega$, 
$\DDA$ to be $T^4$, and $g$ to be the transformation 
$\II_4$ that changes the sign of all the
coordinates on $T^4$. This gives the duality:
\begin{center}
(IIB on $T^4$/$(-1)^{F_L}\cdot\II_4$) $\leftrightarrow$
(IIB on $T^4$/$\Omega\cdot\II_4$)
\end{center}
Note that in this case the fibers $\KK_A$ and $\KK_B$ are points,
but this does not prevent us from applying our method of
constructing dual pairs. Also there are sixteen fixed points on
$T^4$ under $\II_4$ where the application of fiberwise duality
transformation breaks down, but as has been argued before, we
still expect the duality to hold since these are points of
measure zero on $T^4$.

We shall now bring this duality into a more familiar form via 
T-duality transformation.
Let us make $R\to(1/R)$ duality transformation on one of the circles
of $T^4$ in the theory on the left hand side.
This converts type IIB theory to type IIA.
This also transforms $(-1)^{F_L}\cdot\II_4$ to $\II_4$, which can
be checked by
explicitly studying the action of these transformations
on the various massless fields. 
Thus the theory on the left hand side is T-dual to type IIA 
on $T^4/\II_4$. This of course is just 
a special case of type IIA on K3.

Let us now take the theory on the right hand side and make $R\to (1/R)$
duality transformation on all four circles.
This takes type IIB theory to type IIB theory.
But this transforms $\Omega\cdot \II_4$ to $\Omega$, which can
again be seen by studying the action of these transformations on
the massless fields. 
Thus the theory on the right is T-dual to type IIB on
$T^4/\Omega$. Since type I string theory can be regarded as type
IIB string theory modded out by $\Omega$, we see that the theory
on the right hand side is type I on $T^4$.
But by (heterotic - type I) duality in ten dimensions this 
is dual to heterotic on $T^4$.
Thus we have `derived' the duality
\begin{center}
(Type IIA on K3) $\leftrightarrow$ (Heterotic on $T^4$)
\end{center}
\noindent from other conjectured dualities in D=10. Although this
way the duality has been established only at a particular point
in the moduli space (the orbifold limit of K3), the argument can
be generalized to establish this duality at a generic point in
the moduli space as well\cite{DUORBI}.

There are many other applications of fiberwise duality
transformation. Some of them will be discussed later in this
review.

\subsection{Recovering higher dimensional dualities from lower
dimensional ones} \label{srecov}

So far we have discussed methods of deriving dualities involving
compactified string theories by starting with the duality
symmetries of string theories in higher dimensions. But we
can also proceed in the reverse direction. Suppose a string
theory compactified on a manifold $\MM_1\times \MM_2$ has a
self-duality symmetry group $G$. Now consider the limit when the
size of $\MM_2$ goes to infinity. A generic element of $G$,
acting on this configuration, will convert this configuration to
one where $\MM_2$ has small or finite size. However, there may be
a subgroup $H$ of $G$ that commutes with this limit, {\it i.e.}
any element of this subgroup, acting on a configuration where
$\MM_2$ is big, gives us back a configuration where $\MM_2$ is
big. Thus we would expect that $H$ is the duality symmetry group
of the theory in the decompactification limit, {\it i.e.} of the
original string theory compactified on $\MM_1$. The same argument
can be extended to the case of a pair of dual theories.

{\it A priori} this procedure does not appear to be very useful,
since one normally likes to derive more complicated duality
transformations of lower dimensional theories from the simpler
ones in the higher dimensional theory. But we shall now show how
this procedure can be used to derive the SL(2,Z) duality symmetry
of type IIB string theory from the conjectured duality between
type I and SO(32) heterotic string theories, and T-duality
symmetries of the
heterotic string theory. We shall describe the main
steps in this argument, for details, see \cite{UNITY}. We start
with the duality between type I on $T^2$ and heterotic on $T^2$
that follows from the duality between these theories in ten
dimensions. Now heterotic string theory on $T^2$ has a T-duality
group O(2,18;Z). We shall focus our attention on an $SL(2,Z)
\times SL(2,Z)$ subgroup of this T-duality group. One of these
two SL(2,Z) factors is associated with the global diffeomorphism
of $T^2$, and the other one is associated with the $R\to (1/R)$
duality symmetries on the two circles. By the `duality of
dualities' argument, this must also be a symmetry of type I on
$T^2$. Since type I string theory can be regarded as type IIB
string theory modded out by the world-sheet parity transformation
$\Omega$ discussed in section \ref{s21}, we conclude that
$SL(2,Z)\times SL(2,Z)$ is a subgroup of the self-duality group
of type IIB on $T^2/\Omega$. Let us now make an $R\to (1/R)$
duality transformation on both the circles of this
$T^2$. This
converts type IIB on $T^2$ to
type IIB theory compactified on a dual $T^2$, and
$\Omega$ to $(-1)^{F_L}\cdot\Omega\cdot\II_2$, where
$\II_2$ denotes the reversal of orientation of both the circles
of $T^2$. (This can be seen by studying the action of various
transformations on the massless fields.)
Geometrically, this model describes type IIB string
theory compactified on the surface of a tetrahedron (which is 
geometrically $T^2/\II_2$), with an added twist of
$(-1)^{F_L}\cdot\Omega$ 
as we go around any of the four vertices of the tetrahedron (the
fixed points of $\II_2$). (This theory will
be discussed in more detail in section \ref{s8}).
Thus we conclude that $SL(2,Z)\times SL(2,Z)$ is a
subgroup of the self-duality group of type IIB on
a tetrahedron. Now take the limit where the size of the
tetrahedron goes to infinity. It turns out that both the 
SL(2,Z) factors commute with this limit. One of these SL(2,Z)
groups becomes part of the diffeomorphism group of type IIB string
theory and does not correspond to anything new, but the other
SL(2,Z) factor represents the S-duality
transformation discussed in section \ref{s21}.
Since this limit gives
us back the decompactified type IIB string theory, we conclude
that type IIB string theory in ten dimensions has a self-duality
group SL(2,Z).

Thus we see that all the dualities discussed so far can be
`derived' from a single duality conjecture, $-$ the one between
type I and SO(32) heterotic string theories in ten dimensions.
In the next section
we shall see more examples of dualities which can be
derived from the ones that we have already discussed.

\sectiono{Duality in Theories with Less than Sixteen
Supersymmetry Generators}
\label{s5}

So far our discussion has been focussed on theories with 16 or
more supersymmetry charges. As was pointed out in section
\ref{s2}, for these theories the non-renormalization theorems for
the low energy effective action and the spectrum of BPS states
are particularly powerful. This makes it easy to test duality
conjectures involving these theories. In this section we shall
extend our discussion to theories with eight supercharges.
Examples of such theories are provided by
N=2 supersymmetric theories in four dimensions. We shall see that
these theories have a very rich structure, and although the
non-renormalization theorems are less powerful here, they are
still powerful enough to provide us with some of the most striking
tests of duality conjectures involving these theories.
The material covered in this section is based mainly on
refs.\cite{KACVAF,FHSV,VAFWITFI}.

\subsection{Construction of dual pair of theories with eight
supercharges}

For definiteness we shall focus our attention on N=2
supersymmetric theories in four dimensions.
There are several ways to get theories with $N=2$ supersymmetry
in four dimensions. Two of them are:
\begin{enumerate}
\item
Type IIA/IIB on Calabi-Yau 3-folds: 
In our convention an $n$-fold describes an $n$ complex or $2n$
real dimensional manifold.  In ten dimensions type II theories
have 32 supersymmetry generators.
Compactification on a Calabi-Yau  3-fold breaks 3/4 of the
supersymmetry.
Thus  we are left with 8 supersymmetry
generators in D=4, giving rise to N=2 supersymmetry.
\item
Heterotic string theory on $K3\times T^2$:
In ten dimensions heterotic string theory has sixteen
supersymmetry generators.
Compactification on $K3\times T^2$ breaks half of the
supersymmetry. Thus
we have a theory with eight supersymmetry generators, again
giving N=2 supersymmetry in four dimensions. It is also possible
to construct more general class of four dimensional
heterotic string theories with the same number of
supersymmetries where the background does not have the product
structure $K3\times T^2$\cite{BANDIX,KACVAF}.
\end{enumerate}

The question we would like to ask is:
is it possible to construct pairs of N=2 supersymmetric
type II and heterotic string 
compactifications in four dimensions which will be
non-perturbatively dual to each other? Historically such dual
pairs were first constructed by trial and error\cite{KACVAF} 
and then a more
systematic approach was 
developed\cite{FHSV,VAFWITFI,LERCH,ALDFON,SCHLYN}. 
However we shall begin by
describing the systematic approach, and then describe how one
tests these dualities. 
The
systematic construction of such dual pairs can be carried out by
application of fiberwise duality transformation as described in the 
last section. The steps involved in this construction are as
follows:
\begin{figure}[!ht] 
\begin{center}
\leavevmode
\epsfbox{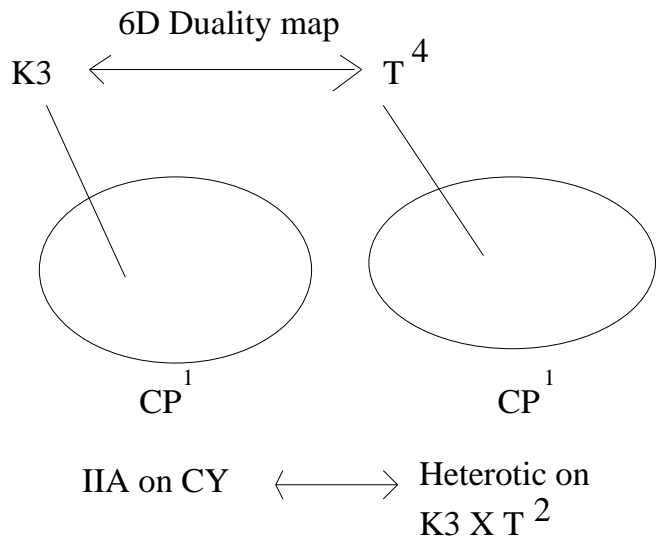}
\end{center}
\caption[]{\small  
Construction of dual pair of N=2 supersymmetric string theories
in four dimensions from the dual pair of theories in six
dimensions.}
\label{f11}
\end{figure}
\begin{itemize}
\item Start from the conjectured duality
(Type IIA on K3) $\leftrightarrow$ (Heterotic on $T^4$).
\item Choose a $CP^1$ base.
\item 
Construct a Calabi-Yau 3-fold by fibering K3 over the base
$CP^1$.
One can construct a whole class of Calabi-Yau manifolds this way by
choosing different ways of varying K3 over $CP^1$.
\item 
For type IIA on each such Calabi-Yau 3-fold we can get a dual heterotic
compactification by replacing the type IIA on $K3$ by heterotic
on $T^4$ on each fiber according to the duality map. This gives
heterotic string theory
on a manifold obtained by varying $T^4$ on $CP^1$
according to the duality map.
Typically this manifold turns out to be $K3\times T^2$ or some variant 
of this.
This model is expected to be dual to the type IIA string theory
on the Calabi-Yau manifold that we started with.
Thus we get a duality map
\begin{center}
(Type IIA on CY) $\leftrightarrow$ (Heterotic on K3$\times T^2$)
\end{center}
\end{itemize}
This construction has been illustrated in Fig.\ref{f11}.
Note that the original duality map 
gives a precise relationship between the moduli
of type IIA on K3 and heterotic on $T^4$.
On the heterotic side the moduli involve background gauge fields on $T^4$
besides the shape and size of $T^4$.
Thus for a specific Calabi-Yau, knowing how K3 varies over $CP^1$,
we can find out how on the heterotic side the background gauge fields
on $T^4$ vary as we move along $CP^1$.
This gives the gauge field configuration on $K3\times T^2$.
Different Calabi-Yau manifolds will give rise to
different gauge fields on K3$\times T^2$.

We shall illustrate this procedure with the example of a pair of
$Z_2$ orbifolds of the form\cite{FHSV}:
\begin{center}
(IIA on $K3\times T^2/h_A\cdot g)$ 
$\leftrightarrow$ (Heterotic on
$T^4\times T^2/h_B\cdot g)$
\end{center}
where 
$g$ acts on $T^2$ by changing the sign of both its coordinates, 
$h_A$ is a specific involution of K3 known as the Enriques
involution, and 
$h_B$ is the image of this transformation on the heterotic side.
By our previous argument relating orbifolds to fibered spaces,
these two theories are expected to be dual to each other via
fiberwise duality transformation. 
$K3\times T^2/(h_A\cdot g)$ can be shown to describe a Calabi-Yau
manifold. Thus the theory on the left-hand side corresponds to
type IIA string theory compactified on this Calabi-Yau manifold. 
In order to determine the theory on the heterotic side, we need
to determine $h_B$. We shall now describe this procedure
in some detail. 

In order to determine $h_B$, we need to study the relationship
between the fields appearing in
type IIA on K3 and heterotic on $T^4$. 
The low energy effective action of both the theories and the origin
of the various massless fields in these theories were discussed in
section \ref{siiahet}. We shall focus our attention on the gauge
fields. As discussed there,
in the type IIA on K3, 22 of the gauge fields come from
decomposing the three form field along the harmonic two forms on
K3.
Now, $h_A$, being a geometric transformation on K3, has known action on
the harmonic forms $\omega^{(p)}$.
For this particular example, $h_A$ corresponds to 
\begin{itemize}
\item exchanging ten of the
$\omega^{(p)}$ with ten others and  
\item
changing the sign of two more
$\omega^{(p)}$.
\end{itemize}
This translates into a similar action on the fields  $\AAA^{(p)}_\mu$ 
defined in eq.\refb{e29}. Furthermore, $h_A$ leaves the
other two gauge fields, coming from the ten dimensional gauge
field $A_\mu$ and the dual of $C_{\mu\nu\rho}$ invariant.
We can now translate this into an action on the gauge fields in
heterotic on $T^4$.
It turns out that the action on the heterotic side is given by:
\begin{itemize}
\item exchanging the gauge fields in the two $E_8$ factors\, ,
\item exchanging ($G_{9\mu}$, $B_{9\mu}$) with 
($G_{8\mu}$, $B_{8\mu}$), and, 
\item  changing the sign of 
($G_{7\mu}$  and $B_{7\mu}$).
\end{itemize}
This translates into the following geometric action in
heterotic string theory on $T^4$:\footnote{Here we are regarding
this theory as the $E_8\times E_8$ heterotic string theory
compactified on $T^4$. By the duality between the two heterotic
string theories upon compactification on a circle, this is
equivalent to SO(32) heterotic string theory compactified on
$T^4$.}
\begin{itemize}
\item exchange of two $E_8$ factors in the gauge group,
\item
$x^8\leftrightarrow x^9$, 
\item $x^7\to -x^7$.
\end{itemize}
This is $h_B$.\footnote{
We need to add to this
a non-geometrical shift involving half of a lattice vector in
$\Lambda_{24}$ in order to get a modular
invariant theory on the heterotic side. This transformation
is not visible in perturbative type IIA theory.}
It turns out that modding out heterotic string
theory on $T^6$ by the transformation
$h_B\cdot g$ produces an N=2 supersymmetric theory. 
Thus this construction gives a type II -
heterotic dual pair with N=2 supersymmetry. 

Using the idea of fiberwise duality transformation we can construct 
many more examples of heterotic - type IIA dual pairs in 
four dimensions with N=2 supersymmetry.
Quite often using mirror symmetry\cite{MIRREV}
we can also relate this to IIB 
string theory on a mirror Calabi-Yau manifold.

\subsection{Test of duality conjectures involving theories with
eight supercharges}

Given such a dual pair of theories constructed by application
of fiberwise duality transformation, the next question will be:
how do we test if these theories are really dual to each other?
After all, as we have seen, there is no rigorous proof that
fiberwise duality transformation always produces a correct dual
pair of theories, particularly when the fiber degenerates at some
points / regions in the base. Unlike in the case of theories with
sixteen supercharges, one cannot directly compare the tree level
low energy effective action in the
two theories, as they undergo quantum corrections in general. 
Furthermore, in this theory the spectrum of BPS saturated states
can change discontinuously as we move in the moduli 
space\cite{SEIWIT}.
Hence the spectrum computed at weak coupling cannot always be
trusted at strong coupling.
Nevertheless there are some non-renormalization theorems which allow
us to test these proposed dualities, as we shall now describe.

Matter multiplets in N=2 supersymmetric theories in four
dimensions are of two types. (For a review, see \cite{SEIWIT}.)
They are
\begin{itemize}
\item vector multiplet containing 
one vector, one complex scalar, and two Majorana fermions, and
\item hypermultiplet containing
two complex scalars and two Majorana fermions.
\end{itemize}
Let us 
consider a theory at a generic point in the moduli space
where the massless matter fields include
only abelian gauge fields and neutral hypermultiplets. 
Let 
$\vec \phi$ denote the complex scalars in the vector multiplet,
and 
$\vec \psi$ denote the complex scalars in the hypermultiplet.
The N=2 supersymmetry requires that 
there is no coupling between the vector and the hypermultiplets
in those terms in
the low energy effective action $S_{eff}$ which contain at most
two space-time derivatives\cite{DECOUP}. 
Thus the scalar kinetic
terms appearing in the lagrangian density associated with
$S_{eff}$ must be of the form:
\be \label{e31}
G^V_{m\wb n} (\vec \phi) \p_\mu\phi^m \p^\mu \wb{\phi^n}
+ G^H_{\alpha\wb\beta}(\vec\psi) \p_\mu\psi^\alpha
\p^\mu\wb{\psi^\beta}\, ,
\ee 
where $G^V$ and $G^H$ are appropriate metrics in the vector and
the hypermultiplet moduli spaces. The kinetic terms of the
vectors and the fermionic fields are related to these scalar
kinetic terms by the requirement of N=2 supersymmetry.

This decoupling between the hyper- and the vector- multiplet
moduli spaces 
by itself is not of much help,
since each term may be independently modified by quantum
corrections.\footnote{There are however strong restrictions on
what kind of metric $G_V$ and $G_H$ should describe. In
particular $G_V$ must describe a special Kahler 
geometry\cite{SPL,DECOUP}, whereas
$G_H$ must describe a quaternionic geometry\cite{HYPERM}. However, these
restrictions do not fix $G_V$ and $G_H$ completely.}
But in string theory we have some extra 
ingredient\cite{KACVAF,VAFWITFI}.
Recall that the coupling constant in string theory
involves the dilaton. 
Thus quantum corrections to a given term must involve a coupling to the
dilaton.
Now consider the following two special cases.
\begin{enumerate}
\item The dilaton belongs to a hypermultiplet.
Then there can be no correction to the vector multiplet kinetic term 
since such corrections will give a coupling between the dilaton and the
vector multiplet.
\item The dilaton belongs to a vector multiplet. In this case
the same argument shows that 
there can be no correction to the hypermultiplet kinetic term.
\end{enumerate}
In type IIA/IIB string theory
on Calabi-Yau manifold the dilaton belongs to a hypermultiplet.
Thus in these theories
the vector multiplet kinetic term, calculated at the tree level, is
exact.
On the other hand in heterotic on $K3\times T^2$, the dilaton is in the
vector multiplet.
Thus the hypermultiplet kinetic term, calculated at the tree level, is
exact.
Using this information we can adopt the following 
strategy for testing duality.\footnote{Here we describe the test
using the
vector multiplet kinetic term, but a similar
analysis should be possible with the hypermultiplet kinetic term
as well.}
\begin{enumerate}
\item Take a type II - heterotic dual pair and calculate the 
vector multiplet
kinetic term exactly from the tree level analysis on the type II side.
\item
Using the map between the fields in the type II and the heterotic theory,
we can rewrite the exact vector multiplet kinetic term in terms of
the heterotic variables.
\item
In particular the heterotic variables include the heterotic dilaton
$\Phi_H$ which is in the vector multiplet.
So we can now expand the exact answer in powers of $e^{\Phi_H}$ and
compare this answer with the explicit calculations in heterotic string
perturbation theory.
Typically the expansion involves tree, one loop, and non-perturbative
terms. (There is no perturbative contribution in the heterotic
theory beyond one loop due to some Adler-Bardeen type 
non-renormalization theorems.)
Thus one can compare the expected tree and one loop terms, calculated
explicitly in the heterotic string theory, with the expansion of the
exact answer.
\end{enumerate}
The results of the above calculation in heterotic and type II
string theories agree in all the cases 
tested\cite{KACVAF,KAPL,NARA}!
This agreement is quite remarkable, since 
the one loop calculation is highly non-trivial on the heterotic side,
and involves integrals over the moduli space of the torus.
Indeed, 
the agreement between the two answers is a consequence of highly
non-trivial mathematical identities.

Given that the tree and one loop results in the heterotic string
theory agree with the expansion of the exact result on the type
II side, one might ask if a similar agreement can be found for
the non-perturbative contribution from the heterotic string
theory as well. 
{}From the exact answer calculated from the type II side we know
what this contribution should be.
But we cannot calculate it directly
on the heterotic side, since there is no
non-perturbative formulation of string theory.
However, one can
take an appropriate limit in which the stringy effects on the
heterotic side disappear and the theory reduces to some
appropriate $N=2$ supersymmetric quantum field
theory.\footnote{This is in the same spirit as in the case of
toroidal compactification of heterotic string theory, where,
by going near a special point in the moduli space, we can
effectively get an N=4 supersymmetric Yang-Mills theory.}
Thus now the calculation of these non-perturbative effects on the 
heterotic side reduces to a calculation in the N=2 supersymmetric
field theory.
This can be carried out using the method developed by
Seiberg and Witten\cite{SEIWIT}.  Again 
there is perfect agreement with the results from the type II 
side\cite{NONPER}.
Besides providing a non-trivial test of string duality, 
this also shows that the
complete Seiberg-Witten\cite{SEIWIT}
results (and more) are contained in the classical
geometry of Calabi-Yau spaces!

\sectiono{M-theory} \label{smtheory}

So far we have discussed dualities that relate known string
theories. However, sometime analysis similar to those that lead
to various duality conjectures can also lead to the discovery
of new theories. One such theory is a conjectured theory
living in eleven dimensions. This theory is now known as
M-theory. In this section we shall give a brief description of
this theory following
refs.\cite{WITTEND,SCHM,ASPINM,DASMUK,WITTOR,HORWIT,MORBI}.

\subsection{M-theory in eleven dimensions}
\begin{figure}[!ht] 
\begin{center}
\leavevmode
\epsfbox{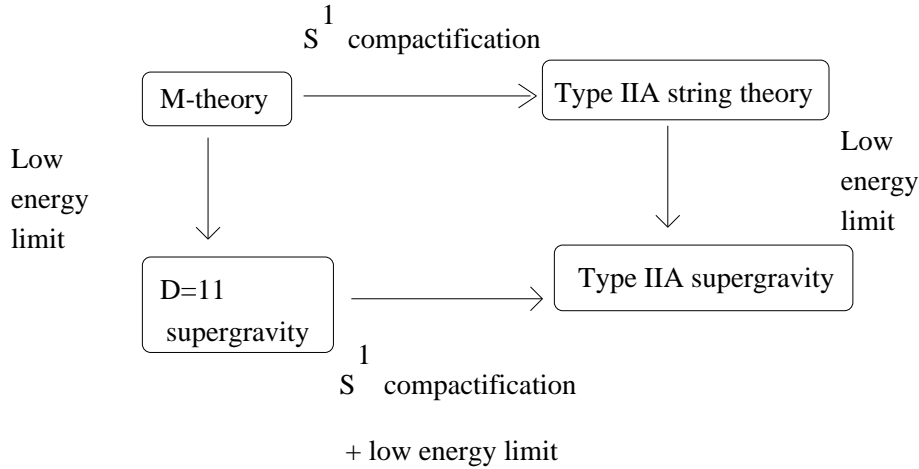}
\end{center}
\caption[]{\small  
The relationship between M-theory and various other supergravity
/ string theories.}
\label{fb2}
\end{figure}
The arguments leading to the existence of M-theory goes as
follows\cite{TOWNM,WITTEND}. 
Take type IIA string theory in ten dimensions.
The low  energy effective action of this theory
is non-chiral N=2 supergravity in ten dimensions.
It is well known that this
can be obtained from the dimensional reduction of $N=1$ supergravity
in eleven dimensions\cite{IIAM}. 
More specifically, the relationship between
the two theories is as follows.
The bosonic fields in $N=1$ supergravity theory in eleven
dimensions consist of the metric $g^{(S)}_{MN}$ and a rank three
anti-symmetric tensor field $C^{(S)}_{MNP}$ ($0\le M,N\le 10$). 
The bosonic part of
the action of this theory is given by\cite{CRJL}
\ben \label{esugra}
S_{SG} &=&  {1\over (2\pi)^8}\int d^{11} x \bigg[\sqrt{-g^{(S)}}
\Big(R^{(S)} - {1\over 48}
G^{(S)2}\Big)\nonumber \\
&& - {1\over (12)^4}\varepsilon^{\mu_0\cdots\mu_{10}} 
C^{(S)}_{\mu_0\mu_1\mu_2} G^{(S)}_{\mu_3\cdots \mu_8}
G^{(S)}_{\mu_7\cdots \mu_{10}}\bigg]
\, ,
\een
where $G^{(S)}\sim dC^{(S)}$ is the four form 
field strength associated with
the three form field $C^{(S)}$. In writing down the above
equation we have set the eleven dimensional Planck mass to unity
(or equivalently we can say that we have absorbed it into a
redefinition of the metric.)
Let us now compactify this supergravity theory
on a circle of radius $R (\sim \sqrt{g^{(S)}_{10,10}})$ 
measured in the supergravity metric $g^{(S)}_{MN}$ 
and ignore (for the time being) the Kaluza-Klein modes
carrying momentum in the internal direction.
Then the effective action in the dimensionally reduced theory agrees
with that of type IIA string theory given in \refb{eiia}
under the identification\cite{IIAM}:
\ben \label{e32}
&& \sqrt{g^{(S)}_{10,10}} = e^{\Phi/3}, \qquad  
g^{(S)}_{\mu\nu} \simeq e^{-\Phi/12} g_{\mu\nu} 
 \qquad 
g^{(S)}_{10\mu} \simeq e^{2\Phi/3} A_\mu, \nonumber \\
&& C^{(S)}_{\mu\nu\rho} \simeq  C_{\mu\nu\rho}, \qquad
C^{(S)}_{10\mu\nu} \simeq B_{\mu\nu}, \qquad (0\le \mu,\nu\le 9)
\, . 
\een
Here $\simeq$ denotes equality up to  additive
terms involving second and higher powers in fields.
We are using the convention that $\Phi=0$ corresponds to
compactification on a circle of unit radius.
Note that as the radius $R\big(\sim\sqrt{g^{(S)}_{10,10}}\big)$
approaches $\infty$, $\Phi\to\infty$. 
This corresponds to 
strong coupling limit of the type IIA string theory. 
This leads one to the conjecture\cite{TOWNM,WITTEND} that
{\it in the strong coupling limit type IIA string theory approaches 
an 11 dimensional
Lorentz invariant theory, whose low energy limit is 11-dimensional
N=1 supergravity.}
This theory has been called M-theory. 
The situation is illustrated in Fig.\ref{fb2}.
Part of the conjecture is
just the definition of
M-theory as the strong coupling limit of
type IIA string theory. The
non-trivial part of the conjecture is that
it describes a Lorentz invariant theory in
eleven dimensions. 

The evidence for the existence of an eleven dimensional theory,
as discussed so far, has been analogous to the evidence for
various duality conjectures
based on the comparison of their low energy
effective action. One might ask if there are more precise tests
involving the spectrum of BPS states.
There are indeed such tests.
M theory on $S^1$ will have Kaluza-Klein modes 
representing states in the eleven dimensional N=1 supergravity multiplet
carrying momentum along the compact $x^{10}$ direction.
These are BPS states, and can be shown to belong to the
256 dimensional ultra-short representation of the supersymmetry
algebra.
The charge quantum number characterizing such a state is the
momentum ($k/R$) along $S^1$.
Thus for every integer $k$ we should find such BPS states in type 
IIA string theory in ten dimensions.
In M-theory these states carry $k$ units of 
$g^{(S)}_{10\mu}$ charge.
Since $g^{(S)}_{10\mu}$ gets mapped to $A_\mu$ under the M-theory -
IIA duality, these states 
must carry $k$ units of $A_\mu$ charge in type IIA string theory.
If we now recall that in type IIA string
theory $A_\mu$ arises in the RR sector, we see that these states
cannot come from elementary string states, as 
elementary string excitations are neutral under RR sector gauge
fields.
However Dirichlet 0-branes in this theory do carry $A_\mu$ charge.
In particular the state with 
$k=1$ corresponds to a single Dirichlet zero brane.
As usual, the collective coordinate dynamics of the 0-branes is
determined from the dynamics of 
massless open string states with ends lying on the D0-brane, and
in this case is described by the
dimensional reduction of N=1 super-Maxwell theory from 
(9+1) to (0+1) dimensions.
This theory has sixteen fermion zero modes whose 
quantization leads to a $2^8=256$ fold degenerate state.
Thus we see that we indeed have 
an ultra-short multiplet with unit $A_\mu$ charge, as predicted
by the M-theory - IIA duality conjecture.

What about states with $k>1$?
In type IIA string
theory these must arise as bound states of $k$ D0-branes.
Dynamics of collective coordinates of $k$ D0 branes
is given by the
dimensional reduction of N=1 supersymmetric U(k) gauge theory from
(9+1) to (0+1) dimensions.
Thus the number of ultra-short multiplets with $k$-units of $A_\mu$
charge is determined in terms of the number of normalizable
supersymmetric ground states of this quantum mechanical system.
Finding these bound states
is much more difficult than the bound state
problems discussed earlier. The 
main obstacle to this analysis is that
a charge $k$ state has the same energy as $k$ charge 1
states at rest. 
Thus the bound states we are looking for sit at the bottom of a
continuum. Such states are difficult to study. For $k=2$ such a
bound state with the correct degeneracy has been found\cite{SETHSTEO}. 
The
analysis for higher $k$ still remains to be done.

The analysis can be simplified by
compactifying $M$-theory on $T^2$ and considering the Kaluza-Klein
modes carrying $(k_1,k_2)$ units of momenta along the two
$S^1$'s. Assuming that the two $S^1$'s are orthogonal, and have
radii $R_1$ and $R_2$ respectively, the mass of such a state, up
to a proportionality factor, is
\be \label{emass}
\sqrt{\Big({k_1\over R_1}\Big)^2 + 
\Big({k_2\over R_2}\Big)^2}\, . 
\ee
For $(k_1,k_2)$ relatively prime, such a state has strictly
less energy than the sum of the masses of any other set of states
with the same total charge\cite{SMARG}.
Thus one should be able to find these states in type
IIA string theory on $S^1$ (which, according to the conjecture,
is equivalent to M-theory on $T^2$) without
encountering the difficulties mentioned earlier.
By following the same kind of 
argument, these states can be shown to be
in one to one correspondence to a class of supersymmetric vacua
in a (1+1) dimensional supersymmetric gauge theory compactified
on a circle.\footnote{In fact, these states are related via an 
$R\to (1/R)$ duality transformation to the ultra-short multiplets
in type IIB on $S^1$ discussed in section \ref{s31}.}
All such states have been found with degeneracy as predicted by
the M-theory - IIA duality.

There are also other consistency checks on the proposed M-theory
- IIA duality.
Consider $M$-theory on $T^2$.
According to M-theory - type IIA duality, it is dual to IIA on $S^1$.
But we know that IIA on $S^1$ is related by T-duality to IIB on
$S^1$. 
Thus we have a duality between M-theory on $T^2$ and IIB on
$S^1$.
Now IIB on $S^1$ has an SL(2,Z) strong-weak
coupling duality inherited from
ten dimensional type IIB string theory.
Thus one might ask, 
what does it correspond to in $M$-theory on $T^2$?
One can find the answer by using the known map between the
massless fields in the two theories, and the action of SL(2,Z) in
type IIB string theory. It turns out that this SL(2,Z) symmetry
in M-theory is 
simply the group of global diffeomorphisms of 
$T^2$\cite{SCHSLT,ASPINM,SCHM}.
Thus we again have an example of `duality of dualities'.
The SL(2,Z) of IIB is a non-perturbative symmetry.
But in M-theory on $T^2$ it is simply a consequence of the 
diffeomorphism invariance of the 11-dimensional theory.

Turning this analysis around we see that 
this also supports the ansatz that M-theory, defined as the strong
coupling limit of IIA, is a fully Lorentz invariant theory in
eleven dimension. The argument goes as follows:
\begin{itemize}
\item First of all, from Lorentz invariance of type IIA string
theory we know that we have
Lorentz invariance in coordinates $x^0, \ldots x^9$ when all the
coordinates $x^0,\ldots x^9$ are non-compact.
\item
Then from the conjectured SL(2,Z) duality symmetry
of type IIB string theory we know that we have an exchange
symmetry between the 9th and the 10th coordinate of M-theory when
these coordinates are compact.
In the limit when the radius of both the compact circles are
taken to be large, this would mean that we should have
Lorentz invariance in coordinates $x^0, \ldots x^{10}$.
\end{itemize}

\subsection{Compactification of M-theory}

Given the existence of M-theory, we can now construct new vacua of the
theory by compactifying M-theory on various manifolds. (For a
review of compactification of eleven dimensional supergravity,
see \cite{DUFFNREV}.
For example, we can consider M-theory compactified on
K3, Calabi-Yau, and various orbifolds.
These can all be regarded as appropriate strong coupling limits of
type IIA compactification on the same manifold.
But in general these cannot be regarded as perturbative string
vacua. 
The essential feature of this strong coupling
limit is the emergence of Lorentz
invariance in one higher dimension. For example,
M-theory on a Calabi-Yau manifold gives
a five dimensional theory with N=1 supersymmetry\cite{MCY}.
Such a theory cannot be constructed by conventional compactification of
type IIA string theory at weak coupling.

Of course in many cases these non-perturbative vacua are related
to perturbative string vacua by conjectured duality relations.
These duality conjectures can be arrived at by using arguments
very similar to those used in arriving at string duality
conjectures.  Some
examples of such conjectured dualities are given 
below\cite{HORWIT,DASMUK,WITTOR,MORBI}:
\ben 
\hbox{M-theory on} && \nonumber \\
S^1/Z_2 \qquad  &\leftrightarrow&  \hbox{($E_8\times E_8$) heterotic in
D=10} \nonumber \\
K3 \qquad  &\leftrightarrow& \hbox{Heterotic/Type I on $T^3$} \nonumber
\\
T^5/Z_2 \qquad
&\leftrightarrow& \hbox{IIB on K3} \nonumber \\
T^8/Z_2 \qquad &\leftrightarrow& \hbox{Type I/Heterotic on $T^7$}
\nonumber \\
T^9/Z_2 \qquad &\leftrightarrow& \hbox{Type IIB on
$T^8/Z_2$}\nonumber 
\een
In each case $Z_2$ acts by reversing the sign of all the
coordinates of $T^n$; for odd $n$ this is also accompanied by a
reversal of sign of $C^{(S)}_{MNP}$.
Each of these duality conjectures satisfy the consistency condition
that the theory on the
right hand side, upon further compactification on a circle,
is dual to type IIA string theory compactified
on the manifold on the left hand side.

The duality between M-theory on $S^1/Z_2$ and the $E_8\times E_8$
heterotic string theory is particularly amusing.
Here the $Z_2$ transformation acts by reversing the orientation
of $S^1$, together with a change of sign of the three form field
$C^{(S)}_{MNP}$.
$S^1/Z_2$ denotes a real line segment bounded by the two fixed 
points on $S^1$. 
It turns out that the two $E_8$ gauge multiplets arise from `twisted
sector' 
of the theory and sit at the two ends of this line segment.
The supergravity sector, on the other hand, sits in the bulk.
Now in the conventional heterotic string compactification on
Calabi-Yau spaces, all the observed gauge bosons and charged particles
come from one $E_8$ and are neutral under the second $E_8$\cite{CAND}.
The second $E_8$, known as the hidden sector or the shadow
world, is expected to be
responsible for supersymmetry breaking.
In the M-theory picture these two sectors are physically separated in
space.  In other words, the
real world and the shadow world live at two ends of 
the line and interact only
via the exchange of supergravity multiplets propagating in the 
bulk\cite{WITTSH}!
It has been suggested that this physical separation 
could be as large as a 
millimeter\cite{KAPEE}! This limit comes from the analysis of
the fifth force experiment, since if this dimension is too large,
we should have inverse cube law for the gravitational force instead
of inverse square law. No such direct limit comes from the
inverse square law of gauge interaction, since gauge fields live
on the boundary of $S^1/Z_2$ and hence do not get affected by the
existence of this extra dimension.

Many of the listed duality conjectures involving M-theory (in fact
all except the first one)
can be derived by fiberwise duality
transformation\cite{MORBI}.\footnote{The duality between
$E_8\times E_8$ heterotic string theory and M-theory on $S^1$ can
be `derived' from other known duality conjectures by taking the
infinite radius limit of a lower dimensional duality
relation\cite{UNITY}.}
Let us for example consider the duality
\begin{center}
(M theory on $T^5/Z_2$) $\leftrightarrow$ (type IIB on K3)
\end{center}
The $Z_2$ generator is $\II_5\cdot\sigma$ where
$\II_5$ changes the sign of all five coordinates $(x^6,\ldots x^{10})$
on $T^5$, and 
$\sigma$ denotes the transformation $C^{(S)}_{MNP}\to -C^{(S)}_{MNP}$. 
Let us express this as $(\II_1\cdot \sigma)\cdot \II_4$ where
$\II_1$ changes the sign of $x^{10}$, and 
$\II_4$ changes the sign of $(x^6,\ldots x^9)$. 
We now use the result of fiberwise duality transformation:
\begin{center}
(A on $K_A\times \DD/(h_A\cdot g)$) $\equiv$
(B on $K_B\times \DD/(h_B\cdot g)$)
\end{center}
by choosing 
A on $K_A$ to be M-theory on $S^1$, 
B on $K_B$ to be  type IIA string theory, 
$h_A$ to be $\II_1\cdot \sigma$, 
$h_B$ to be
$(-1)^{F_L}$ (this can be shown to be the image of $h_A$ in the type
IIA string theory),
$\DD$ to be $T^4$ spanned by $x^6,\ldots x^9$,
and  $g$ to be $\II_4$. 
Thus we get the duality
\begin{center}
M-theory on $(S^1\times T^4/\II_1\cdot\sigma\cdot\II_4)$
$\leftrightarrow$ IIA on $T^4/(-1)^{F_L}\cdot \II_4$
\end{center}
The theory on the left hand side is M-theory on $T^5/Z_2$.
On the other hand, if we 
take the theory on the right hand side
and make an $R\to(1/R)$ duality
transformation on one of the circles, it converts
\begin{itemize}
\item type IIA theory to type IIB theory, and
\item
$(-1)^{F_L}\cdot\II_4$ into $\II_4$.
\end{itemize}
Thus the theory on the right is dual to type IIB on $T^4/\II_4$, which
is a special case of type IIB on $K3$.
Thus we get the duality:
\begin{center}
(M-theory on $T^5/Z_2$) $\leftrightarrow$ (IIB on K3)
\end{center}
This duality was first conjectured in \cite{DASMUK,WITTOR}.

As in the case of type IIA string theory, we can get
non-perturbative enhancement of gauge symmetries in M-theory when
the compact manifold develops singularities\cite{WITTEND,WITTBR,ENH}. 
M-theory contains classical membrane and five-brane soliton solutions 
carrying electric and magnetic charges of $C^{(S)}_{MNP}$
respectively\cite{DUFFREV}.  The
extra massless states required for this symmetry enhancement
come from membranes wrapped around the collapsed two cycles of
the singular manifold.

\sectiono{F-Theory} \label{s8}

Just as M-theory can be used to describe non-perturbative
compactification of type IIA string theory, 
F-theory describes non-perturbative compactification of type IIB 
string theory\cite{VAFAF,VMORF}. 
However, unlike M-theory, it does not correspond
to a Lorentz invariant higher dimensional theory, although, as we
shall see, an
auxiliary manifold with two extra dimensions plays 
a crucial role in the
construction of F-theory compactification. This section will be
based mainly on refs.\cite{VAFAF,VMORF,FTHEORY}.

\subsection{Definition of F-theory}
\begin{figure}[!ht] 
\begin{center}
\leavevmode
\epsfbox{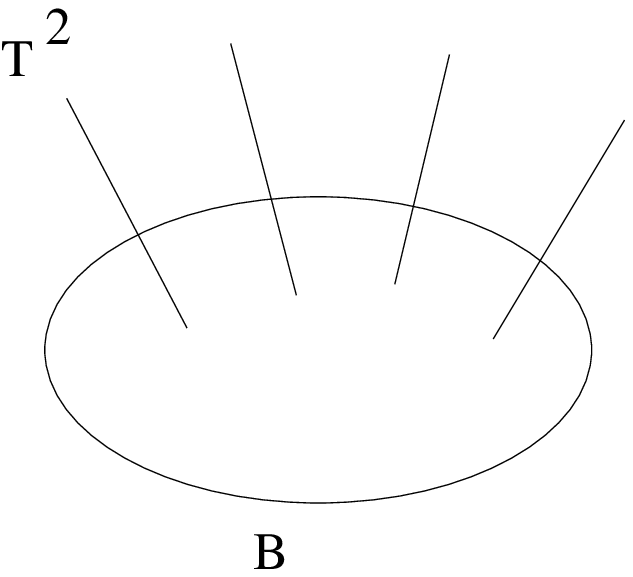}
\end{center}
\caption[]{\small  
An elliptically fibered manifold $\MM$ with base $\BB$.}
\label{f12}
\end{figure}
In 
conventional perturbative type IIB compactification one takes the
dilaton-axion field $\lambda$ (defined in section \ref{s21})
to be constant. 
F-theory is a novel way of compactifying type IIB theory that avoids
this restriction and allows the string coupling to vary over the
compact manifold. 
The starting point in this
construction is an elliptically fibered Calabi-Yau
manifold defined as follows. 
Let $\BB$ be a manifold (which we shall call
base manifold) of real dimension $d$ and $\MM$ be 
another manifold of real dimension $d+2$,
obtained by erecting at every point of $\BB$ a copy of a torus $T^2$,
with the moduli of $T^2$ varying over $\BB$.
This situation is illustrated in Fig.\ref{f12}.
$\MM$ is called an elliptically fibered manifold. We shall choose
the base $\BB$ and the fibration in such a way that $\MM$
describes a Calabi-Yau manifold.
Let $\vec z$ denote the  complex coordinate on $\BB$,
and $\tau(\vec z)$ denote the
complex structure of $T^2$ as a function of $\vec z$.
Then F-theory on $\MM$ is defined to be type IIB string theory 
compactified on $\BB$ with
\be \label{e33}
\lambda(\vec z)=\tau(\vec z)\, .
\ee
Note that the size of the fiber torus does not appear in
eq.\refb{e33}. Thus F-theory on a manifold $\MM$ is insensitive
to a subset of the moduli of $\MM$ which describe how the size of
the fiber torus varies over the base.

In order that $\MM$ is well defined, 
$\lambda=\tau(\vec z)$ must come 
back to its original value only up to an SL(2,Z) transformation as we
move along a closed cycle on $\BB$.
Due to the presence of this non-perturbative duality transformation in
the monodromy group, conventional type IIB perturbation theory cannot be
used to describe this system.
In particular, there are points in $\BB$ where $Im(\lambda)$ is of order 
unity, and hence type IIB theory is strongly coupled. For
example, suppose that
$Im(\lambda)$ is large in one region of the manifold,
and also that as we go around a closed curve 
starting from this region, there is an SL(2,Z) transformation by
the matrix $\pmatrix{p & q\cr r & s}$ so that
$\lambda$ comes back to a value near $p/r$.
Then at some point on the curve $Im(\lambda)$ must be
finite, and hence the string theory is strongly coupled.
\begin{figure}[!ht] 
\begin{center}
\leavevmode
\epsfbox{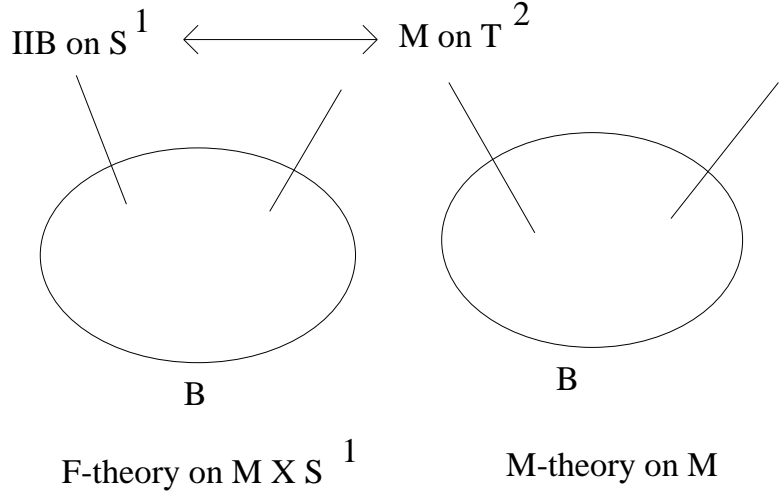}
\end{center}
\caption[]{\small  
Fiberwise application of duality to relate F-theory on $\MM\times
S^1$ to M-theory on $\MM$.}
\label{f13}
\end{figure}

{}From this note it would seem that although F-theory describes a
novel way of compactifying type IIB string theory, we cannot
extract any information about such a theory, since string
perturbation theory cannot be used to analyse this system.
However, it turns out that we can learn quite a lot about these
theories by using various known duality relations. 
For this
consider F-theory on $\MM\times S^1$, {\it i.e.} type IIB theory on
$\BB\times S^1$ with $\lambda(\vec z)=\tau(\vec z)$.
Now, as we have discussed in the last section,
type IIB theory on $S^1$ is dual to M-theory on $T^2$, with $\lambda$
being the modular parameter of $T^2$.
Thus we can now apply fiberwise duality transformation
illustrated in Fig.\ref{f13} to relate the F-theory
compactification on $\MM\times S^1$ to an M-theory compactification. 
Under this duality the modulus of $T^2$ on the right hand side must be
set equal to $\lambda(\vec z)$ of the theory on the left, which,
according to eq.\refb{e33}, is just $\tau(\vec z)$.
This means that the
manifold on the right hand side is the original manifold $\MM$
that we started with. 
This gives the duality 
\begin{center}
F-theory on $\MM\times S^1$ $\leftrightarrow$ M-theory on $\MM$
\end{center}
Thus many of the properties of F-theory on $\MM$ ({\it e.g.} the number
of supersymmetries, spectrum of massless states, etc.) can be studied from
that of M-theory on $\MM$ and then taking appropriate limit in which
the size of $S^1$ on the F-theory side goes to $\infty$.
Using this one can show for example that
if $\MM$ is K3, then  we preserve half of the space-time
supersymmetries of the type IIB theory, whereas if $\MM$ is a Calabi-Yau
manifold then we preserve 1/4 of the supersymmetry.

In order to gain more insight into various F-theory
compactifications, we need to 
develop a convenient formalism for describing elliptically fibered
manifolds.
Our starting point will be the equation describing a torus:
\be \label{e34}
y^2 = x^3 + f x + g\, ,
\ee
where $x$ and $y$ are complex variables, and 
$f$ and $g$ are complex parameters. 
For every pair of constants
$f$ and $g$ the above equation describes a one complex
dimensional surface, which can be shown to be a torus.
The modular parameter $\tau$ of the torus is given by
\be \label{e35}
j(\tau) = { 4 . (24 f)^3 \over 27 g^2 + 4f^3}\, ,
\ee
where $j(\tau)$ is a known function of $\tau$. In fact it is the
unique modular invariant function
with a single pole at $\tau=i\infty$ and zeroes at
$\tau=e^{i\pi/3}$. The overall normalization of $j$ is chosen in
such a way that $j(i)=(24)^3$.

In order to describe elliptically fibered manifold on some base
$\BB$ we simply
need to make $f$ and $g$ depend on the coordinates of $\BB$.
We shall illustrate this with the help of an
elliptically fibered K3. Here we choose the base
$\BB$ to be $CP^1$. 
Thus the elliptically fibered manifold $\MM$ is
described by the equation
\be \label{e36} 
y^2 = x^3 + f(z) x + g(z)\, ,
\ee
where $z$ is the coordinate on $CP^1$. It can be shown that $\MM$
describes a K3 manifold provided
$f(z)$ is a polynomial of degree 8, and 
$g(z)$ is a polynomial of degree 12 in $z$.
The modular parameter of the torus varies over the base $CP^1$ according
to the relation:
\be \label{e37}
j(\tau(z)) = { 4 . (24 f(z))^3 \over 27 g(z)^2 + 4f(z)^3}\, .
\ee

By definition
F-theory on this elliptically fibered
K3 is type IIB string theory compactified
on $CP^1$ with:
\be \label{e38}
\lambda = \tau(z)\, .
\ee
In order to specify the background completely we also need to specify the
metric on the base $CP^1$. This can be
calculated from the low energy effective field
theory when the size of $\BB$ is sufficiently large, 
and the answer is\cite{STRCOS}
\be \label{e39}
ds^2 =  F(\tau(z),\bar\tau(\bar z)) \, dz d\bar z \,
\Big(\prod_{i=1}^{24} (z-z_i)^{-1/12} (\wb z-\wb z_i)^{-1/12}\Big) 
\, ,
\ee
where $z_i$ are the zeroes of
$\Delta\equiv (4f^3 + 27 g^2)$, and 
\be \label{e40}
F(\tau,\bar\tau)
= (\tau_2) \eta(\tau)^2 \bar\eta(\bar\tau)^2\, .
\ee
$\eta(\tau)$ denotes the Dedekind function and $\tau_2$ is the imaginary
part of $\tau$.

Similarly we can describe more complicated F-theory compactifications
by choosing more complicated base $\BB$.
For example, consider the base $CP^1\times CP^1$ labelled by a pair
of complex coordinates $(z,w)$\cite{VMORF}.
We can get an elliptically fibered manifold on this base by the equation
\be \label{e41}
y^2 = x^3 + f(z,w) x + g(z,w) \, .
\ee
In order that this manifold is Calabi-Yau, we need
$f(z,w)$ to be a polynomial of degree (8,8) in ($z,w$), and
$g(z,w)$ to be a polynomial of degree (12,12) in ($z,w$).
F-theory on such a manifold is by definition a configuration 
of $\lambda(z,w)$ described by the equation
\be \label{e42}
j(\lambda(z,w)) = { 4 . (24 f)^3 \over 27 g^2 + 4f^3}\, .
\ee
\subsection{Dualities involving F-theory}
\begin{figure}[!ht] 
\begin{center}
\leavevmode
\epsfbox{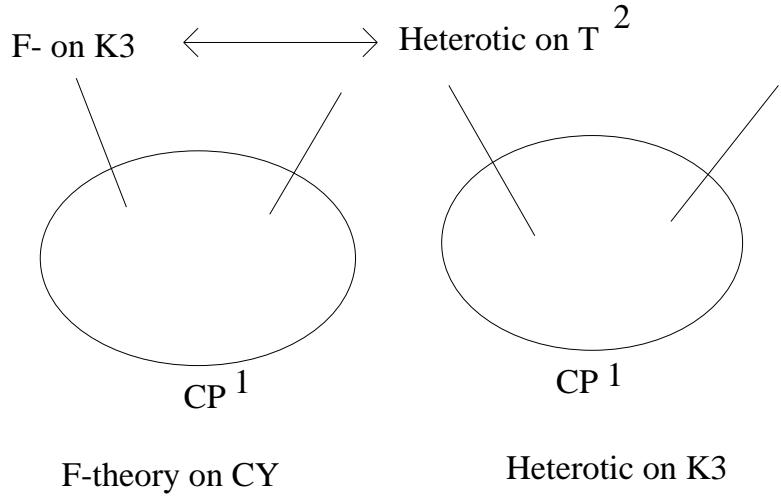}
\end{center}
\caption[]{\small  
Fiberwise application of duality to relate F-theory on Calabi-Yau
to heterotic string theory on K3.}
\label{f14}
\end{figure}
There are many conjectured dualities involving F-theory
compactifications. 
Some examples are given below\cite{VMORF}:
\ben 
\hbox{F-theory on K3} \qquad &\leftrightarrow& \qquad \hbox{Heterotic
on $T^2$} \nonumber \\
\hbox{F-theory on CY 3-fold} \quad &\leftrightarrow& \quad 
\hbox{Heterotic on K3 }\nonumber
\een
All the duality conjectures involving F-theory have the following
property:

\noindent{\it
If F-theory on $\MM$ is dual to some string theory S compactified
on a manifold $\KK$, then M-theory on $\MM$ must be dual to the
same string theory compactified on $\KK\times S^1$, and type IIA
on $\MM$ must be dual to the same string theory compactified on
$\KK\times T^2$. These results follow from the duality between
type IIB on $S^1$ and M-theory on $T^2$, and that between
M-theory on $S^1$ and type IIA string theory.}

All conjectured dualities involving F-theory on Calabi-Yau
3-folds and more
complicated manifolds can be derived from the fiberwise duality 
transformation\cite{VMORF,FMW,BERET}. 
For example, for Calabi-Yau 3-fold, this is done by 
representing the Calabi-Yau
3-fold as K3 fibered over $CP^1$, and replacing
F-theory on $K3$ by heterotic on $T^2$ fiberwise. This has been
illustrated in Fig.\ref{f14}. The theory on the right hand side
of this figure represents heterotic string theory on K3 with
apropriate gauge field background.

We shall now show how to derive the parent duality
\begin{center}
(F-theory on K3) $\leftrightarrow$ (Heterotic on $T^2$)
\end{center}
from other known duality conjectures\cite{FTHEORY}.
Recall that for this background:
\ben \label{e43}
j(\lambda(z)) &=& { 4 . (24 f(z))^3 \over 27 g(z)^2 + 
4f(z)^3}\nonumber \\
ds^2 &=&  F(\lambda,\bar\lambda)\, dz d\bar z \,
\Big(\prod_{i=1}^{24} (z-z_i)^{-1/12} (\wb z-\wb z_i)^{-1/12}
\Big) \, ,
\een
where $f(z)$ and $g(z)$ are polynomials of degree 8 and 12
respectively, and $z_i$ are the zeroes of $\Delta\equiv (4f^3+27
g^2)$.
The strategy is to 
try to go to a special point in the moduli space where $\lambda$,
instead of varying over $CP^1$, becomes a constant.
At this special point the theory reduces to a conventional
compactification of type IIB string theory.
Examining eq.\refb{e43} we see that this
requires $f^3/g^2$ to be a constant.
If we now recall that
$f$ is a polynomial of degree 8 in $z$ and
$g$ is a polynomial of degree 12 in $z$, we see that
for $f^3/g^2$ to be constant, we need
\be \label{e44}
f=\alpha \phi^2, \qquad g=\phi^3\, ,
\ee
where $\phi$ is a polynomial of degree 4 in $z$, and
$\alpha$ is a constant. Using the
freedom of an overall rescaling of $\phi$ which does not change
the value of $\lambda$, we can take 
\be \label{e45}
\phi=\prod_{m=1}^4 (z-z_m)\, .
\ee
This gives 
\be \label{e46}
\Delta \equiv 4f^3+27 g^2 = (4\alpha^3+27) \prod_{m=1}^4(z-z_m)^6\, ,
\ee
\be \label{e47}
j(\lambda) = 4. (24\alpha)^3/(4\alpha^3+27)\, ,
\ee
and
\be \label{e48}
ds^2 = F(\lambda,\bar \lambda)\,  dz d\bar z \,
\prod_{m=1}^4 (z-z_m)^{-1/2} 
(\wb z-\wb z_m)^{-1/2}\, .
\ee
\begin{figure}[!ht] 
\begin{center}
\leavevmode
\epsfbox{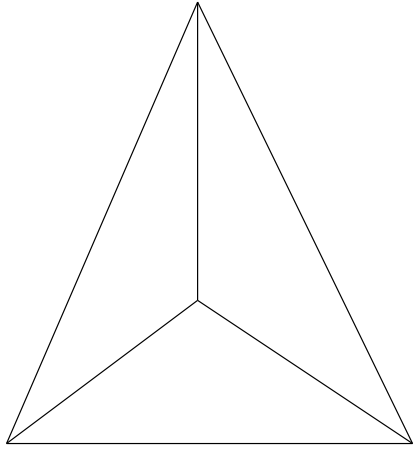}
\end{center}
\caption[]{\small  
$CP^1$ with a flat metric except at four points.}
\label{f15}
\end{figure}

Since $\lambda$ is now a constant, the metric can be simplified by
going to a new coordinate system $w$ defined by:
\be \label{e49}
dw = 
\prod_{m=1}^4 (z-z_m)^{-1/2} dz\, .
\ee 
Then
\be \label{e50}
ds^2 = C \, dw \, d\bar w\, ,
\ee
where $C=F(\lambda,\bar\lambda)$ is a constant.
Thus the metric is flat! But this poses a puzzle,  since we
know that the base is $CP^1$, and that we cannot put a flat
metric on $CP^1$ since it has non-zero Euler number.
The resolution to this puzzle comes from noting
that if $w_m$ are
the images of $z=z_m$ in the $w$ plane, then,
near $z=z_m$, 
\be \label{e51}
(w-w_m) \sim (z-z_m)^{1/2}\, .
\ee
This gives rise to a
deficit angle of $\pi$ at each of these four points.
Thus the base has flat metric everywhere except for conical
singularities at these four
points. This represents a regular tetrahedron as shown in
Fig.\ref{f15}.
This can be also be
identified as the orbifold $T^2/\II_2$.
Here $w$ is the complex
coordinate on $T^2$, and $\II_2$ denotes the transformation
$w\to -w$. 
$z$ on the other hand, is the  coordinate on $T^2/\II_2$. 
The points $z_m$ are the fixed points of $\II_2$. 
\begin{figure}[!ht] 
\begin{center}
\leavevmode
\epsfbox{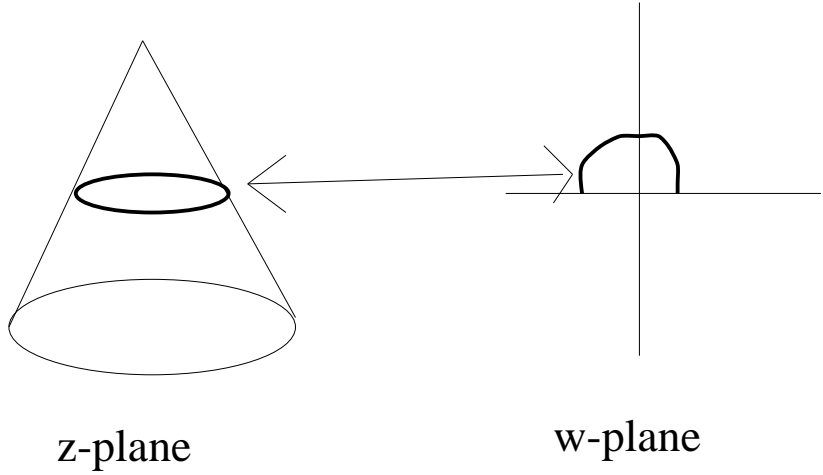}
\end{center}
\caption[]{\small  
A closed curve around $z=z_m$
in the $z$-plane and its image in the $w$-plane.}
\label{f16}
\end{figure}

This analysis would suggest that at this special point in the moduli
space F-theory on K3 has reduced to type IIB on $T^2/\II_2$ with
constant $\lambda$ given in eq.\refb{e47}.
However, there is a further subtlety.
Recall that going once around a fixed point in the
$z$ plane corresponds to 
going from $w$ to $-w$ as illustrated in Fig.\ref{f16}.
The relevant question to ask would be:
is there any twist by some internal symmetry transformation $g$
of type IIB theory as we go around a fixed point of $z$?
If there is such a twist, then the $Z_2$ orbifold group will be 
generated by $w\to -w$, accompanied by the transformation $g$.
In order to 
find $g$ we need to study the effect of going around the point
$z=z_m$ once.
To do this, recall the equation describing this particular K3:
\be \label{e52}
y^2 = x^3 + \alpha x \prod_{m=1}^4 (z - z_m)^2 +
\prod_{m=1}^4 (z - z_m)^3\, .
\ee
Let us take $z$ around $z_1$ once through the parametrization
\be \label{e53}
\prod_{m=1}^4 (z-z_m)= e^{2\pi it}\, \prod_{m=1}^4
(z_{initial}-z_m)
\, ,
\ee
and continuously changing $t$ from 0 to 1.
Also during this change, 
focus on a point on the fiber torus and follow its trajectory.
This can be achieved by choosing:
\be \label{e54}
x= x_{initial} e^{2\pi it}, \qquad y = y_{initial} 
e^{3\pi i t}\, .
\ee
This point lies on the surface \refb{e52} for all values of $t$
if the initial point $(x_{initial},y_{initial},z_{initial})$ lies
on this surface.
At the end of this process when $t=1$,
we do not return to the original point but
to $(x_{initial}, -y_{initial}, z_{initial})$.
To see what
this transformation correponds to if we choose a more 
conventional coordinate system to the fiber torus,
let us denote by
$u$ is the conventional flat coordinate on the fiber torus, so that the
torus is described through the identification 
$u\equiv u+1\equiv u+\tau$. It can be shown that
$u$ is related to $(x,y)$ through the relation 
$du = K dx/y$ where $K$ is a constant. 
Thus $(x,y)\to (x,-y)$ corresponds to $u\to -u$, {\it i.e.}
reversing the orientation of both the circles on the fiber torus.
This is nothing but an SL(2,Z) transformation with matrix 
$\pmatrix{-1 & \cr & -1}$.
Thus we see that as we move around any of the points $z=z_m$ on a the
base once, we make an SL(2,Z) transformation with
this matrix.
But now recall that the SL(2,Z) transformation with the matrix
$\pmatrix{-1 & \cr & -1} $
can be identified to the transformation
$ (-1)^{F_L}\cdot\Omega$ as discussed in section \ref{s21}. 
Thus as we go around the point $z=z_m$ we make a global symmetry
transformation by $(-1)^{F_L}\cdot\Omega$. 
This shows that the F-theory on K3 at this particular point in
the moduli space can be identified to
\begin{center}
Type IIB on $T^2/\II_2\cdot(-1)^{F_L}\cdot\Omega$
\end{center}
We have encountered this theory earlier in section \ref{srecov}.
As discussed there, by making
$R\to (1/R)$ duality transformation on both coordinates
of $T^2$, we can relate this theory 
type IIB on $T^2/\Omega$. But
since type IIB modded out by $\Omega$ is type I theory, we see
that this model can be identified to
type I on $T^2$. Finally, using
type I - SO(32) heterotic duality in ten dimension, we can relate
this model to heterotic string theory on $T^2$.
This finally establishes the equivalence between
\begin{center}
(F-theory on K3) and (Heterotic on $T^2$)
\end{center}

Similar strategy has been used to establish many other dualities
involving F-theory compactification\cite{OTHERF}.

As in the case of type IIA string theory and M-theory, F-theory
on a singular manifold can also develop enhanced non-abelian
gauge symmetry\cite{VMORF,VAFSIX,DASMUKT}. In this case the extra
massless states required for the symmetry enhancement come from
open string states lying on the base\cite{JOHAN,GABZWI}.

The F-theory type compactification can be generalized in the
following way.
One starts with type II string theory compactified on
$T^n$ with a duality group $G$, and compactifies it further on
a base $\BB$ with the monodromy on the base being a subgroup of
$G$. Such compactifications have been discussed in \cite{KUVA}.

\sectiono{Microscopic Derivation of Black Hole Entropy}
\label{sblack}

One of the major stumbling blocks to our understanding of nature
has been the apparent
incompatibility between quantum mechanics and general relativity.
Three of the four known forces of nature $-$ strong, weak and
electromagnetic 
interactions $-$ are very well explained by quantum field fheory.
The current model of elementary particle physics $-$ the 
standard model $-$ has
explained most of the observed phenomena involving these three
interactions.
However, this is not so for gravity.

There are many
problems of quantizing gravity using quantum field theory. 
First of all, 
gravity is not perturbatively renormalizable {\it i.e.}
the usual rules in quantum field theory for extracting finite answers
for all physical quantities from infinite answers at the intermediate
stages of calculation, are not valid for gravity.
As mentioned in section \ref{spert},
this problem is automatically solved in string theory.
However,
during the last twenty five years a more serious objection to the 
compatibility of general relativity and quantum mechanics has
been raised\cite{HAWRAD,BHENT}. 
This is the problem to which we now turn. The
discussion in this section will follow closely
refs.\cite{SENOLD,STVA,CALMAL,MALSUSS,DASMAT,MALSTR}.

\subsection{Problem with black holes in quantum mechanics}

The starting point is the existence of a class of classical
solutions in
general relativity (possibly coupled to other fields)
known as black holes.
Classically black holes are completely black.
In other words,
objects can fall into the black hole, but nothing can ever come out of
a black hole.
(In more technical terms, one says that black holes have event
horizons.) Black holes in general relativity
also satisfy a classical
no hair theorem which states that
a black hole solution is completely characterized by 
its mass, angular momentum, and gauge charges.
Thus all other information (quantum numbers) carried by an object falling
into the black hole is lost for ever.

However due to the work of Bekenstein, Hawking and others during the last
twenty five years it has become clear 
that once quantum effects are taken into
account, this picture of black hole undergoes dramatic modification.
In particular two things happen.
\begin{itemize}
\item Black holes emit thermal radiation at temperature\cite{HAWRAD}:
\be \label{eb1}
T = {\kappa\over 2\pi}\, ,
\ee
where $\kappa$ is the surface gravity of the black hole (the
acceleration due to gravity felt by a static observer at the
event horizon).
\item
Black holes carry entropy\cite{BHENT}:
\be \label{eb2}
S={1\over 4G_N} A\, .
\ee
where $A$ is the area of the event horizon and $G_N$ is the Newton's
constant.
\end{itemize}
Black holes satisfy the usual laws of thermodynamics in terms of
these variables. In particular the first law of black hole
thermodynamics states that
\be \label{eb3}
dM = TdS\, .
\ee
This relates the change in the mass $M$ of a black hole to its change
in entropy $S$ and the Hawking temperature $T$.
The second law of black hole thermodynamics states that
\be \label{eb4}
dS\ge 0\, ,
\ee
{\it i.e.} the sum of the entropy of the black hole and the usual
thermodynamic entropy of its surroundings increases with time.
In the presence of U(1) gauge charges we can also define chemical
potential 
associated with each gauge charge. In the presence of these charges
the laws of thermodynamics are modified in the usual manner.

It is
this thermodynamic description of the black hole that causes an
apparent conflict with quantum mechanics. This is best
illustrated by considering the following thought experiment.
\begin{itemize}
\item
Consider a black hole formed out of the collapse of a pure state.
We can imagine a spherical shell of matter described by an s-wave
state collapsing to form a black hole.
\item
It will then emit thermal radiation and at the end evaporate
completely.
If the outgoing radiation is really thermal, then the final state
is a mixed state.
\end{itemize}
Thus the 
net result of this two step process is the evolution of a 
pure state to a mixed state, 
in conflict with the rules of quantum mechanics.

At this stage, it is useful to
compare this with the phenomenon of thermal radiation from a
star (or any other hot object).
If an object in a pure quantum state is thrown into a star, it
comes out as thermal radiation, so 
why doesn't this contradict quantum mechanics?
The answer to this is that
the thermal description of the radiation from a star is a result of
averaging over the microstates of the star.
We could, in principle, start from a pure quantum state of the star, and
give a microscopic description of the radiation coming out of the star.
In this description a pure state evolves to a pure state.
In other words, although on average the star emits thermal radiation, 
the radiation coming out of the star has subtle dependence on what goes
into the star, and a detailed analysis of this radiation can be used to
completely reconstruct the initial state.

Let us now come back to black holes.
Why can't the same reasoning be used for black holes to resolve
the apparent conflict with quantum mechanics?
The reason is that there is no similar microscopic description of the 
radiation from a black hole in conventional
semiclassical gravity $-$ the approximation that is used in
demonstrating that black holes emit thermal radiation.
This is
related to the problem that there is no understanding
of the black hole entropy in terms of counting of microstates.
In other words, the entropy formula \refb{eb2} is derived purely
in analogy with thermodynamics, but not as the logarithm of
density of states as in statistical mechanics. Thus there is no
possibility of giving a quantum mechanical description of Hawking 
radiation by studying the evolution of individual microstates of
the black hole.

Since string theory claims to be a consistent quantum theory of gravity,
it should be able to explain black hole entropy and Hawking radiation
in terms of conventional quantum mechanics.
We shall now discuss some of the attempts to explain black hole 
thermodynamics in string theory.

\subsection{Black holes as elementary string states}

As we have discussed earlier (see, for example, eq.\refb{eheto})
the
spectrum of an elementary string contains an infinite tower of
states. Since the Schwarzschild radius (the radius
of the event horizon) of a black hole is proportional to its mass,
for sufficiently large mass the Schwarzschild radius of a black hole
will be larger than the string scale. In that case an elementary string
state of the same mass will lie inside its Schwarzschild radius,
and  become a black hole\cite{THOOFT}.
This opens up the possibility of a statistical description of black
hole entropy as follows\cite{SUSS}.
For a given mass $M$, the microscopic entropy $S_{micro}$ can be 
defined to be the logarithm of the number of elementary string states at
that mass level.
We can then compare this with the Bekenstein-Hawking entropy $S_{BH}$ 
of the black hole with the same mass. If the two expressions
agree, then we can give a statistical interpretation of black
hole entropy by attributing it to the degeneracy of string states.
Unfortunately this attempt fails, since one finds that
\be \label{eb5}
S_{micro}\propto M, \qquad  S_{BH}\propto M^2\, .
\ee
The above formulae hold for chargeless black holes, but a similar
discrepancy is present even for black holes carrying electric
charge.
This seems to be a severe blow to the attempt at giving a microscopic 
description of Bekenstein-Hawking entropy in string theory.
But one should keep in mind that
in the region of parameter space where the elementary string state
becomes 
a black hole, there are strong coupling effects, and  hence
the mass of an elementary string state can get 
renormalized\cite{SUSS}.
Thus the parameter
$M$ that appears in the computation of $S_{BH}$ and the $M$ that
appears in the computation of $S_{micro}$ may not be the same,
but may be related by a renormalization factor. 

Various attempts
have been made to get out of this difficulty for Schwarschild
black holes\cite{SUSS,HORPOL,SFET,KALYAN}, 
but we shall not discuss them here.
Instead, we shall try to get out of
this impasse by working with states for which there is no
mass renormalization, namely the BPS states\cite{SENOLD}.
Since the degeneracy of a BPS state does not change as we change
the coupling (at least for theories with $\ge 16$ supersymmetry
charges) we can proceed as follows. First we
compute the degeneracy of BPS states at weak coupling where the
microscopic description of the state is reliable. Then
we increase the string coupling constant
to a sufficiently large value 
where the state becomes a black hole. For this black hole,
$S_{micro}$ should be given by the logarithm of the degeneracy
computed at weak coupling.
This leads to the following strategy for comparing $S_{micro}$
and $S_{BH}$:

\begin{enumerate}

\item Identify the BPS states among elementary 
string states and calculate their degeneracy.
This gives $S_{micro}$.
\item Identify BPS black holes (also known as
extremal black holes) with the same quantum
numbers and find $S_{BH}$ by computing the area of the event horizon.
\item Compare the two expressions.

\end{enumerate}

We shall illustrate this with the help of a
specific model $-$ heterotic string theory 
compactified on $T^6$\cite{SENOLD}.
As discussed above eq.\refb{eheto},
this theory has two classes of U(1) charges, $\vec k_L$ and $\vec k_R$.
{}From eqs.\refb{eheto}, \refb{ehett} we see that among the
elementary string states,
BPS states are those satisfying\footnote{When we regard
elementary strings as classical black hole solutions, the field
$\lambda_2$ varies as a function of the radial distance from the
origin. Due to this fact we are specifically using the notation
$\langle\lambda_2\rangle$ to denote the asymptotic value of
$\lambda_2$, {\it i.e.} the expectation value of $\lambda_2$ in
the vacuum.}
\be \label{eb6}
m^2 = (2\vec k_R^2/\langle \lambda_2 \rangle)\, , \qquad
N_L ={1\over 4} \langle\lambda_2\rangle \Big(m^2 - 
{2\vec k_L^2\over \langle
\lambda_2\rangle}\Big)+1\, .
\ee
The degeneracy of these states can be calculated from
eq.\refb{ehetf}. For large $N_L$, this gives
\be \label{eb7}
d(m, k_L, \langle \lambda_2\rangle) \sim
\exp(4\pi \sqrt{N_L}) \sim
\exp\Bigg({2\pi \sqrt{\langle\lambda_2\rangle}} 
\sqrt{m^2-{2\vec k_L^2\over \langle\lambda_2\rangle}}\Bigg).
\ee
This gives
\be \label{eb8}
S_{micro} \simeq {2 \pi \sqrt{\langle\lambda_2\rangle}} 
\sqrt{m^2-{2\vec k_L^2\over \langle\lambda_2\rangle}}\, .
\ee

Our next task is to calculate the Bekenstein-Hawking entropy
$S_{BH}$ for black holes carrying the same quantum numbers. 
It turns out that the black holes carrying these quantum numbers
have vanishing area of the event horizon, and hence $S_{BH}=0$.
Thus again we seem to have run into a contradiction!

However, again there is a subtlety.
In constructing the black hole solution, one uses the low energy
effective field theory which is valid only when the curvature is
much smaller than the string scale. But since this black
hole has vanishing area of the event horizon, it follows that the
solution actually has a curvature singularity at the
horizon.\footnote{One might wonder whether such a solution can 
be called a black
hole at all, but the reason that they are called black holes is
that they can be obtained as a limit of black hole solutions with
non-singular horizons with finite area. As the mass of the black
hole approaches the Bogomol'nyi bound \refb{eb6}, we get
the singular black hole.} Thus we would expect that the solution
near the horizon will be modified by the higher derivative terms
in the string effective action, and hence might give a different
answer for the area of the event horizon.

In order to estimate what the modified area will be, one needs to
understand what kind of corrections we must include in the
effective action. Typically in string theory there are two types
of correction $-$ the ones due to string world-sheet effects (the
higher derivative terms),
and the ones due to the string loop effects. In the present case
when one examines the classical black hole solution one finds
that the field $\lambda_2$ approaches $\infty$ as we approach the
event horizon. Thus the string coupling $\lambda_2^{-1/2}$
vanishes in this region
and we do not expect the string loop corrections to be
significant. This leaves us with the string world-sheet corrections.
Although we cannot explicitly calculate the effect
of these corrections, we can use a scaling argument to
determine the form of these corrections up to an overall
numerical factor. The argument goes as follows. If we use the
string metric 
(as opposed to the canonical metric) to describe the
black hole solution, then one finds that with a suitable choice
of coordinate system the solution near the origin becomes
completely independent of all parameters $m$, $\vec k_L$ and
$\langle \lambda_2\rangle$, except for an additive factor of 
\be \label{eb9}
-\ln\Bigg({ \sqrt{\langle\lambda_2\rangle}} 
\sqrt{m^2-{2\vec k_L^2\over \langle\lambda_2\rangle}}\Bigg)\, ,
\ee
in the expression for the dilaton. This additive constant does not
affect the string world-sheet lagrangian. Thus whatever be the
effect of the corrections due to the string world-sheet effects,
these corrections are universal, and do not depend on any
parameters. Thus after taking into account the string world-sheet
corrections, the area of the event horizon, as measured in
the string metric, will be a universal numerical constant,
independent of all external parameters. However since the
Bekenstein-Hawking entropy is to be identified with the area of
the event horizon measured in the {\it canonical metric}, which differs
from the string metric by a  factor of $e^{-\Phi}$, and since 
$\Phi$ has an additive factor \refb{eb9} that depends on the
parameters, the modified Bekenstein-Hawking entropy will be given
by, 
\be \label{eb10}
S_{BH} = {C \sqrt{\langle\lambda_2\rangle}} 
\sqrt{m^2-{2\vec k_L^2\over \langle\lambda_2\rangle}}\, ,
\ee
where $C$ is an unknown numerical constant. This is in 
complete agreement with the answer for $S_{micro}$ given in
\refb{eb8} if we choose
\be \label{eb11}
C = 2\pi\, .
\ee
Note, in particular, that $S_{micro}$ and $S_{BH}$ have the
same functional dependence on $m$, $\vec k_L$ and
$\langle\lambda_2\rangle$. Considering the fact that these are
all dimensionless parameters (we are working in units $\hbar=1,
c=1, \alpha'=1$) this agreement is impressive. This calculation
can also be extended to other toroidal compactification of
heterotic string theory\cite{PEET}, and to non-toroidal compactification
of heterotic and type II string theories\cite{NTBL}.

\subsection{Black holes and D-branes}

Although the result described in the previous
subsection is encouraging, we would like to do better
and compare $S_{BH}$ and $S_{micro}$ without encountering any
undetermined numerical factor.
The strategy is to 
try to identify black hole solutions which are
\begin{itemize}
\item BPS states, and
\item have non-vanishing area of the event horizon even without 
stringy corrections. 
\end{itemize}
It turns out that there are indeed such black holes present in the theory,
but they do not carry the same quantum numbers as elementary string states.
Instead they carry the same quantum numbers as a configuration of
D-branes.
Thus in order to calculate the microscopic entropy we need to calculate
the degeneracy of this D-brane configuration. 
As has already been discussed earlier, the dynamics of collective
coordinates of D-branes is given by the massless open string
states propagating on the D-branes. Thus we can explicitly
determine the Hamiltonian describing this dynamics and calculate
the degeneracy of states to calculate $S_{micro}$. This is
precisely what is done.

Thus our strategy is as follows:
\begin{enumerate}
\item
Identify a BPS black hole with non-vanishing area of the event horizon
and calculate $S_{BH}$ from this area.
\item
Identify the D-brane configuration carrying the same quantum numbers as
this black hole and calculate $S_{micro}$ by computing the degeneracy of
these states.
\item
Compare the two answers.
\end{enumerate}
The analysis is simplest in five dimensions, so we shall
concentrate on this case\cite{STVA,MALREV}.
We focus on type IIB string theory on $T^5$.
The D-brane configuration that we consider has
\begin{itemize}
\item $Q_5$ D-5-branes wrapped on $T^5$,
\item
$Q_1$ D-1-branes wrapped on one of the circles $S^1$ of $T^5$,
and
\item
$-n$ units of momentum along $S^1$. If $R$ denotes the radius of
$S^1$, this corresponds to a momentum of $-n/R$.
\end{itemize}
The counting of states for this D-brane system can be done as
follows. The world-volume theory
of a system of $Q_5$ parallel D-5 branes is a supersymmeric
$U(Q_5)$ gauge theory in $(5+1)$ dimensions, obtained by the
dimensional reduction of N=1 supersymmetric $U(Q_5)$ gauge theory
in (9+1) dimensions. It can be shown that a single D-1 brane
inside $Q_5$ coincident D-5 branes can be identified to a single
instanton in this $U(Q_5)$ gauge
theory\cite{DOUGLAS}.\footnote{An instanton is a classical
solution in Yang-Mills theory in four euclidean dimensions. Thus
in the (5+1) dimensional gauge theory it represents a solution
that is independent of time and one spatial directions, {\it
i.e.} a static string.} Thus a system of $Q_1$ parallel D-1 branes
inside $Q_5$ coincident D-5 branes can be described as a system
of $Q_1$ instantons in the $U(Q_5)$ gauge theory. The moduli of
this $Q_1$ instanton solution act as collective coordinates of
this system. These moduli span a $4Q_1Q_5$ dimensional hyper-kahler
manifold. As a result the low energy dynamics of a system of
$Q_1$ D-1 branes inside $Q_5$ D-5 branes is described by a (1+1)
dimensional supersymmetric $\sigma$-model with this instanton
moduli space as the target space. Since this space is
hyper-kahler, the corresponding supersymmetric $\sigma$-model is
conformally invariant\cite{ALVFRM}. Furthermore, since the
moduli space has dimension $4Q_1Q_5$, the central charge is given
by:
\be \label{ebe1}
c ={3\over 2} \cdot 4Q_1Q_5 = 6 Q_1Q_5\, ,
\ee
taking into account the contribution of 1 from each scalar and
(1/2) from each fermion.
The BPS states in this theory correspond to states with 
$L_0=0$. $L_0-\bar L_0$ represents the total number of
momentum units carried by the system along $S^1$; for this system
this is equal to $-n$. For large $n$ the degeneracy of such states
can be computed\cite{CARDY}. The answer is
\be \label{ebe2}
d(Q_1,Q_5,n) \sim \exp\Big( 2\pi \sqrt{cn\over 6}\Big) =
\exp(2\pi\sqrt{Q_1Q_5n})\, .
\ee

Since this argument is somewhat abstract, we shall now
give a simplified description of the counting of
states of this system\cite{CALMAL,DASMATFR,MALSUSS,HASWAD}. 
First consider the case when there is one D-5 brane and $Q_1$ D-1
branes inside the D-5 brane (so that the D-1 branes are free to
move inside the D5-brane but not free to leave the D-5 brane). 
Now, since $Q_1$ D-1 branes, each wrapped once around
$S^1$, has the same charge as a single D-1 brane wrapped $Q_1$
times, we must also include this configuration in our counting of
states. It turns out that the contribution to the total
degeneracy is dominated by the later configuration, so we can
restrict our attention to this configuration. Now consider the
case when there are $Q_5$ D-5 branes instead of just one. Again
the dominant contribution comes from the
configuration where instead of $Q_5$ D-5
branes each wrapped once around $S^1\times T^4$, we have a single
D-5 brane wrapped $Q_5$ times. Thus we have a configuration
of a single D1 brane of length $2\pi R Q_1$ and a single
D-5 brane of length $2\pi R Q_5$ along the direction of $S^1$. 
We shall call these branes
long D-1 and long D-5 branes respectively. Now take $Q_1$ 
and $Q_5$ to be 
relatively prime. In that case, when we go around the long D-1 brane
once by travelling $Q_1$ times around the circle $S^1$, we do not
come back to the same point on the long D-5 brane. On the other
hand, if we go around the long D-5 brane once by travelling $Q_5$
times around $S^1$, we do not come back to the same point on the
long D-1 brane. In fact, we
need to go around the long D-1 brane $Q_5$ times in order to come
back to the same point on the long D-5 brane and the long D-1
brane. This amounts to
going around $S^1$ $Q_1Q_5$ times. Thus to an open string stretched 
between the D-1 brane and the D-5 brane the
configuration will appear to be
that of a single D-string of length $2\pi R Q_1
Q_5$, which is free to move in the four transverse directions
inside a D-5 brane\cite{MALSUSS}. This gives four bosonic collective
coordinates $X^i$. Due to supersymmetry, this system will also have
four Majorana fermions $\lambda^i$ moving on the D-string world-sheet.

For each of these four bosonic and fermionic coordinates there are left
moving modes as well as right moving modes on the D-1 brane.
A quantum of the $m$th left (right) moving mode carries $-m$ ($m$) units
of momentum along $S^1$, with each unit of momentum now being equal
to $1/(RQ_1 Q_5)$. We need a state with a total of $-nQ_1Q_5$
units of momentum.  It turns out that in order to saturate the
BPS mass formula, we need to concentrate on states containing
only quanta of left moving modes and no quanta of right moving modes.
If there are $N^i_m$ quanta of the $m$th left moving mode of $X^i$, 
and $n^i_m$ quanta of the $m$th left moving mode of $\lambda^i$,
then in order to get $-Q_1Q_5n$ units of momentum along $S^1$ we need:
\be \label{eb12}
Q_1Q_5n = \sum_{i=1}^4 \sum_{m=1}^\infty m(N_m^i + n_m^i)\, .
\ee
The degeneracy $d(Q_1, Q_5, n)$ is the 
number of ways we can choose integers $N_m^i$ and $n_m^i$ satisfying the 
above relation. This can be computed using standard procedure,
and the answer is
\be \label{eb13}
d(Q_1,Q_5,n)\sim \exp(2\pi\sqrt{Q_1Q_5n})\, ,
\ee
which is the same as \refb{ebe2}. This gives,
\be \label{eb14}
S_{micro}=\ln d \simeq 2\pi\sqrt{Q_1Q_5n}\, .
\ee

We now need to compare this result with the Bekenstein-Hawking
entropy of the black hole solution of the low energy effective
field theory carrying the same set of charges and the same mass.
In the normalization convention of eqs.\refb{edil}, \refb{egmunu}
and \refb{eiia}
the five dimensional Newton's constant is given by,
\be \label{efnewton}
G_N^5 = {\pi \over 4 }\, .
\ee
Instead of writing down the full black hole
solution, we shall only write
down the canonical metric for this black hole in the five
non-compact directions, since this is what is required to compute
the area of the event horizon, and hence $S_{BH}$.
The metric is\cite{MALREV,PEETREV}\footnote{Our ten dimensional
metric, in which $R$ and $V$ are measured, differs from that of
\cite{MALREV,PEETREV} by a factor of $g^{-1/2}$, and our five
dimensional metric differs from that of \cite{MALREV,PEETREV} by
a factor of $(RV)^{2/3}g^{-1/2}$.}
\be \label{eb15}
ds^2 = - \lambda^{-2/3} dt^2 
+ \lambda^{1/3} [ dr^2 + r^2 d\Omega_3^2]\, ,
\ee
where,
\be \label{eb16}
\lambda = (1 + r_1^2/r^2) (1 + r_5^2/r^2) (1 + r_n^2/r^2)\, ,
\ee
\be \label{eb17}
r_1^2 = (RV)^{2/3}g^{-1/2} Q_1/V, \qquad r_5^2 = 
(RV)^{2/3} g^{1/2}Q_5, \quad r_n^2 = (RV)^{2/3} n/R^2
V\, .
\ee
$d\Omega_3$ denotes line element on a unit three sphere,
$(2\pi)^4 V$ is the volume of $T^4$ and
$R$ is the radius of $S^1$, both measured in the ten dimensional
canonical metric, and
$g(\equiv e^{\langle \Phi^{(10)}\rangle/2}$) is the 
string coupling constant {\it in ten dimension}. 
Here $T^4$ denotes the
subspace of the full $T^5$ that does not include the special
$S^1$ on which the D-1 branes are wrapped. The event horizon is
located at $r=0$, and the area $A$ of the even horizon can be easily
computed from eqs.\refb{eb15}, \refb{eb16} to be 
$2\pi^2 r_1r_5r_n$. 
This gives,
\be \label{eb18}
S_{BH} = {A\over 4G_N^5} =
(2\pi) \sqrt{Q_1Q_5n}\, .
\ee
This is in exact agreement with
$S_{micro}$ computed in eq.\refb{eb14}. 

Similar agreement between
the Bekenstein-Hawking entropy and the microscopic entropy has
been demonstrated for black
holes carrying angular momentum\cite{ANGMOM}, and also for black
holes in four dimensions\cite{FOURDBH}.
The analysis can also be easily extended to
compactification of type II string theory on $K3\times S^1$
with sixteen supersymmetry charges. 
More non-trivial case is the extension to black holes in theories
with eight supercharges. This has been done in many cases, and in
every case that has been studied, the Bekenstein-Hawking
entropy of the BPS saturated black hole agrees with the
microscopic entropy computed from analyzing the dynamics of
the brane configuration\cite{NEQTBH}. An alternate approach to
calculating $S_{micro}$ has been advocated in \cite{ALTER} that
also gives answer in agreement with $S_{BH}$.

Given the success of the D-brane dynamics in giving a microscopic
description of the entropy of an extremal black hole, one might
wonder if a similar analysis can be carried out for black holes
which are not extremal, but are nearly extremal\cite{CALMAL,NONEX}.
For the D-brane
configuration such states are obtained by relaxing the
requirement that there is no quanta of right-moving modes.
However, we restrict the  number of such excitations so that the
interaction between the left- and the right-moving modes can be
neglected; this is known as the dilute gas approximation. 
Let the left moving modes carry a total 
momentum $-N_L/(Q_1Q_5R)$ along $S^1$ and the
right moving modes carry a total 
momentum $N_R/(Q_1Q_5R)$ along $S^1$.
Then we require
\be \label{eb19}
N_L-N_R=Q_1Q_5n\, .
\ee
The degeneracy of states is obtained by computing the number of
ways these momenta can be distributed between different left- and
the right-moving modes. The microscopic entropy, computed this way, is
given by
\be \label{eb20}
S_{micro} = 2\pi (\sqrt{N_L}+\sqrt{N_R})\, .
\ee
Since for this configuration the mass is no longer given by the
BPS formula, 
the black hole solution also gets modified and gives a new $S_{BH}$.
The answer turns out to be
\be \label{eb21}
S_{BH} = 2\pi (\sqrt{N_L}+\sqrt{N_R})\, ,
\ee
again in perfect agreement with \refb{eb20}. In the expression for
$S_{BH}$ the combination $N_L-N_R$ enters through the dependence
of the solution on the various charges, whereas the combination
$N_L+N_R$ enters through the dependence of the solution on the
mass of the black hole which is now an independent parameter.

{\it A priori} we should not have expected such an agreement
between $S_{BH}$ and $S_{micro}$ for these black holes,
since for non-BPS states one did not expect any
non-renormalization theorem to hold and there was no reason why the
two answers computed in different domains of validity should
agree. An explanation for this agreement was provided later with
the help of a new non-renormalization theorem\cite{MALNON}.

Having found agreement between $S_{BH}$ and $S_{micro}$, one can now
ask if we can also reproduce Hawking radiation from these black holes
from the dynamics of D-branes\cite{DMW,DASMAT,MALSTR}. 
It turns out that extremal black holes
have zero temperature and hence do not Hawking radiate.
This is consistent with the fact that in the microscopic
description, BPS states are stable
and hence cannot decay into other states.
But non-extremal black holes do Hawking radiate and we can try to 
compare this radiation with the radiation due to the decay of a non-BPS
D-brane.
Computation of the Hawking radiation rate from the near extremal black
hole can be done by standard technique. One subtlety comes from
the fact that although the black hole horizon gives out thermal
radiation, there is a frequency dependent filtering of this
radiation as it passes through the black hole background and
reaches the asymptotic observer. This effect is known as the grey body
factor, and can be computed by
knowing the background fields associated with the specific black
hole solution under consideration. For the specific non-extremal
black holes that we are considering, the net Hawking radiation
of a specific class of scalar particles,
as seen by an asymptotic observer is given
by\cite{MALSTR},\footnote{Again in comparing this expression to those
in refs.\cite{MALREV,PEETREV} we need to take into account the
rescaling of the metric.}
\ben \label{eb22}
\Gamma_H &=& 2 \pi^2 (RV)^{4/3} {Q_1Q_5\over V} 
{\pi k_0\over 2} {d^4k \over (2\pi)^4}
{1\over e^{k_0\over 2T_L}-1} \, \,{1\over e^{k_0\over 2T_R}-1} 
\een
where
\ben
(T_R) &=& {(RV)^{-1/3} \over \pi R   Q_1 Q_5}\sqrt{N_R} \nonumber
\\
(T_L) &=& {(RV)^{-1/3} \over \pi R   Q_1 Q_5}\sqrt{N_L} \, ,
\een
and $d^4k$ denotes the four dimensional phase-space in this
$(4+1)$ dimensional theory.

In the D-brane description, the radiation of these specific scalar
particles is due to the annihilation of the
left and right moving modes on the D-1 brane.
This decay rate can be calculated by using standard string theoretic 
technique, and the answer is\cite{DASMAT},
\ben
\Gamma_{micro} &=& 2 \pi^2 (RV)^{4/3} {Q_1Q_5\over V} 
{\pi k_0\over 2} {d^4k \over (2\pi)^4}
{1\over e^{k_0\over 2T_L}-1} \, \,{1\over e^{k_0\over 2T_R}-1}\, .
\een
Again we see that this is 
in exact agreement with $\Gamma_{H}$!

Thus we see that
at least for a class of black holes we now have a concrete microscopic
derivation of  
the Bekenstein-Hawking entropy and the phenomenon of Hawking radiation.
The description is completely quantum mechanical. 
The only caveat is that these microscopic calculations are done
in regions of parameter space where the system is not a black hole,
and then, 
with the help of non-renormalization theorems,
the answer is continued to the region where the system is
a black hole. 
If we could understand in more detail how this continuation works, then
perhaps we shall be able to show that there is really
no conflict between Hawking radiation and quantum mechanics, 
in the same sense that thermal radiation from a star is not in conflict
with quantum mechanics.

\sectiono{Matrix Theory} \label{smatrix}

In section \ref{smtheory} we postulated the existence of an
eleven dimensional theory, known as M-theory, with the following
two properties:
\begin{itemize}
\item
The low energy limit of M-theory is eleven dimensional N=1
supergravity.
\item
M-theory compactified on a circle is dual to type IIA string
theory. As the radius of this circle goes to infinity, the type
IIA coupling constant also goes to infinity.
\end{itemize}
{}From this we can define M-theory as the strong coupling limit
of type IIA string theory. However this definition does not give
us any clue as to how to systematically compute S-matrix elements
in M-theory, since the type IIA string theory is defined only by
the rules for its perturbation expansion in the coupling
constant. Thus finding a non-perturbative definition of M-theory
is of importance. One might try various approaches:
\begin{enumerate}
\item M-theory is the N=1 supergravity theory in D=11: This
proposal by itself does not make sense since this supergravity
theory is not renormalizable and hence is plagued by the usual
ultra-violet divergences. One might argue that there is some
intrinsic regularization that accompanies this supergravity
theory, but till we find such a regularisation, describing
M-theory as the supergravity theory remains an empty statement.
\item M-theory is a string theory: One might imagine that
M-theory can be made finite by regarding it as the low energy
limit of some string theory in the same way that the various
supergravity theories in ten dimensions are made finite. However
nobody has been able to find any such string theory so far.
There is also another compelling reason to believe that M-theory
is not a string theory. With the sole exception of type I string
theory, every other string theory has the property that the
corresponding low energy supergravity theory contains the
fundamental string as a supersymmetric
soliton. N=1 supergravity in eleven 
dimensions does not contain any such soliton solution.
\item M-theory is a theory of membranes or five-branes: The
eleven dimensional supergravity theory does contain membrane and
five-brane like soliton solutions\cite{DUFFREV}. 
Thus one might argue that
M-theory should be formulated as a theory of membranes or
five-branes. The difficulty with this proposal is
that unlike string theory, which is tractable due to the infinite
dimensional conformal symmetry on its world-sheet, the
world-volume theory of membranes or five-branes have no such
infinite dimensional symmetry and hence are extremely difficult
to handle. However, the Matrix theory that we are about to
discuss does in
some sense regard M-theory as a theory of
membranes\cite{MEMDIF}. 
\end{enumerate}
These difficulties in formulating M-theory has led to a radically
new way of thinking about M-theory, and in fact string theories
in general. This proposal, known as Matrix
theory\cite{BFSS,SUSSDL},
is based on describing the theory in terms of its Hamiltonian in
the discrete light cone quantization (DLCQ). In this section we
shall give a brief overview of this formulation. 
The contents of this section will form a
small fraction of the material covered in
\cite{BANKSREV,SUSSREV}. 

\subsection{Discrete light-cone quantization (DLCQ)}

We shall illustrate the procedure of DLCQ in the context of a
scalar field theory. Let us begin with a free scalar field theory
in $d+1$ dimension. The action of the system can be
expressed as 
\be \label{em1}
S = \int d x^+ dx^- d^{d-1} \vec x_{\perp} (\p_+\phi \p_-\phi
-{1\over 2} \sum_{i=1}^{d-1}\p_i\phi \p_i\phi)\, ,
\ee
where 
\be \label{em2}
x^\pm = {1\over\sqrt 2} (x^0\pm x^d)\, , \qquad \p_\pm = {\p\over
\p x^\pm}\, ,
\ee
and $\vec x_\perp \equiv (x^1, \ldots x^{d-1})$ denotes the
transverse coordinates. We shall now canonically
quantize this theory by
regarding $x^+$ as time. Thus the momentum conjugate to $\phi$ is
given by,
\be \label{em3}
\pi = {\delta S\over \delta(\p_+\phi(x))} = \p_-\phi\, .
\ee
Note that the relationship between $\phi$ and $\pi$ given in
eq.\refb{em3} does not involve a `time' derivative, reflecting
the fact that the original lagrangian is linear in $\p_+\phi$. As
a result eq.\refb{em3} represents a constraint. The
canonical equal time commutation relations can be found by using the
standard formalism for quantization of constrained system.
However, we can simplify this problem by going to the Fourier
transformed variables defined through the decomposition: 
\ben \label{em6}
\phi(x^+,x^-,\vec x_\perp) &=& \int_0^\infty {dk_-\over
\sqrt{4\pi k_-}} [a(x^+, k_-, \vec x_\perp) e^{-ik_- x^-} +
a^*(x^+, k_-, \vec x_\perp) e^{ik_-x^-}] \nonumber \\
\een
The action \refb{em1} now takes the form:
\be \label{eme1}
S = i\int dx^+ \int_0^\infty dk^- \int d^{d-1}
\vec x_\perp [a^*
(x^+,k_-,\vec x_\perp) \p_+ a(x^+, k_-, \vec x_\perp)
+ \ldots]\, ,
\ee
where $\ldots$ denote terms without any $x^+$ derivatives. Thus
if we now regard $a$ as the coordinate, its canonically conjugate
momentum is given by,
\be \label{eme2}
{\delta S\over \delta (\p_+ a)} = i a^*
\ee
Upon going to the quantum theory, $a^*$ should be regarded as
the hermitian conjugate of the operator $a$.
This gives the following equal time commutation rules:
\ben \label{em8}
&& [a(x^+,k_-,\vec x_\perp), a(x^+,l_-,\vec y_\perp)] = 0
= [a^\dagger(x^+,k_-,\vec x_\perp), a^\dagger(x^+,l_-,\vec
y_\perp)]\, \nonumber \\
&& [a(x^+,k_-,\vec x_\perp), a^\dagger(x^+,l_-,\vec y_\perp)] = 
\delta(k_--l_-) \delta(\vec x_\perp-\vec y_\perp)\, .
\een
{}From this we see that $a^\dagger$ and $a$ behave respectively
as creation and annihilation operators for particles carrying
momentum $k_-$ located at the transverse location $\vec x_\perp$.
Note that the argument $k_-$ of $a$ and $a^\dagger$ extends over
positive values only. This can be traced to the fact that
\refb{em3} is a constraint equation, hence only half of the
degrees of freedom of $\phi$ are true coordinates, the other half
being momenta.

We shall now compactify the light-like direction $x^-$ on a circle
of radius $L$. This of course is not a physically meaningful
system since it has closed light-like curves and hence violates
causality, but we simply use it as an infrared regulator and at
the end take the $L\to\infty$ limit to recover physically
meaningful answers for various processes. In this case the
momentum $k_-$ is quantized as:
\be \label{em9}
k_- = {n\over L}\, .
\ee
Eqs.\refb{em6}  and\refb{eme1} are now modified as:
\ben \label{em10}
\phi(x^+,x^-,\vec x_\perp) &=&  a_0 (x^+,\vec x_\perp) 
+ {1\over \sqrt{4\pi}} \sum_{n=1}^\infty {1\over \sqrt n}
\Big[a_n(x^+, \vec x_\perp) e^{-in x^-/L} \nonumber \\
&& + a_n^*(x^+, \vec x_\perp) e^{inx^-/L}\Big] 
\een
\be \label{eme3}
S = i\int d x^+ \int d^{d-1}\vec x_\perp
\sum_{n=1}^\infty (a_n^* \p_+ a_n + \ldots)\, ,
\ee
This shows that $i a_n^*$ is the momentum conjugate to $a_n$. In
quantum theory $a_n^*$ will be represented by the hermitian
conjugate $a_n^\dagger$ of the operator $a_n$. This gives the
following commutation relations:
\ben \label{ecomm}
&& [a_n(x^+, \vec x_\perp), a_m(x^+,\vec y_\perp)] = 0
= [a_n^\dagger(x^+, \vec x_\perp), a_m^\dagger(x^+,\vec
y_\perp)]\, , \nonumber \\
&& 
[a_n(x^+, \vec x_\perp), a_m^\dagger(x^+,\vec y_\perp)] =
\delta_{mn} \delta(\vec x_\perp - \vec y_\perp)\, .
\een
Note that the action
does not contain any term containing $x^+$ derivative of $a_0$.
Thus we can interprete it as the Lagrange multiplier field and
integrate it out. The hamiltonian computed from the action
\refb{em1} takes the form:
\be \label{em4}
H = \sum_{n=1}^\infty {L\over 2n} \int d^{d-1} \vec
x_\perp \p_i a_n^\dagger(x^+, k_-, \vec x_\perp) \p_i a_n(x^+,
k_-, \vec x_\perp)\, .
\ee
We can define the ground state of this system as the state
annihilated by all the $a_n$'s. The $p$-particle state is created
by acting with $p$ of the $a^\dagger$'s on this ground state.

Let us now consider the effect of introducing interactions in
the original action \refb{em1}. This
will add new terms to the hamiltonian.
In particular, since the interaction
terms in the lagrangian
will involve $a_0$, integrating out $a_0$ will in general
produce infinite series of additional terms in $H$ even if the
lagrangian itself contains only a finite number of
terms.\footnote{If the theory contains moduli fields which have
flat potential and hence can acquire arbitrary vacuum expectation
value, then the process of integtrating out the zero modes of
these fields requires choosing a definite set of vacuum
expectation values of these fields. This corresponds to choosing
a specific point in the moduli space. At different points of the
moduli space we shall get different DLCQ theory. Thus the DLCQ
formalism is not independent of the choice of background.}
But as long as the
interactions are invariant under translation, all the terms in
the Hamiltonian will conserve $k_-$. Let us now focus on a sector
with $k_-=N/L$. This sector contains $N$-particle states with
each particle carrying momentum $1/L$, but it may also contain
states with less than $N$ particles, with some particles carrying
momentum larger than $1/L$. However, this sector does not cantain
states with more than $N$ particles, since all particles in this
description carry strictly positive $k_-$ in units of $1/L$. This
suggests that the dynamics in this sector should be describable
by an $N$-body {\it quantum mechanical hamiltonian}
$\HH_N^{DLCQ}$.
(One should distinguish this from the hamiltonian $H$ of the
second quantized theory.)
$\HH_N^{DLCQ}$, by definition, describes the complete dynamics in
the $k_-=N/L$ sector. Thus for example it should correctly
reproduce the spectrum, as well as all scattering amplitudes in
this sector. In particular,
in this description a single particle state carrying $N$ units of
momentum should appear as a bound state of $N$ particles in the
spectrum of $\HH_N^{DLCQ}$. Thus $\HH_N^{DLCQ}$ 
must possess an appropriate
bound state of this form. Similarly it must contain in its
spectrum appropriate bound states representing 
$p$ particle state carrying total momentum $N/L$ for all $p<N$,
and for all possible distribution of the minus component of the
total momentum $N/L$ among
these $p$ particles.

Given the action of an interacting quantum field theory, it
should in principle be possible to construct $\HH_N^{DLCQ}$.
Conversely, if we are given $\HH_N^{DLCQ}$ for all $N$, we can,
in principle, completely reconstruct the spectrum and the
S-matrix elements of the theory. For example, calculation of
a specific $n$ particle $\rightarrow$ $m$ particle process 
will involve the following steps:
\begin{enumerate}
\item
Calculate the scattering amplitude for a set of $n$ states
carrying momenta $k^i_-=n_i/L$ going to a 
set of $m$ states carrying
momenta $l^i_-=m_i/L$ with 
\be \label{esm1}
\sum_{i=1}^n n_i = \sum_{i=1}^m m_i =N\, .
\ee
\item Now take the limit $N\to\infty$, $L\to\infty$, $n_i\to\infty$
and $m_i\to\infty$ keeping fixed
\be \label{esm2}
N/L, \qquad k^i_-=n_i/L, \qquad l^i_-=m_i/L\, .
\ee
\end{enumerate}
In this limit we shall reproduce the physical scattering
amplitude involving $(m+n)$ external legs. We are of course
implicitly assuming that there is an unambiguous procedure for
taking the large $N$ limit.

Given that $\HH_N^{DLCQ}$ defines a theory, we can now try to
define M-theory by specifying $\HH_N^{DLCQ}$ for M-theory. This
is what is done in the Matrix-theory approach to M-theory.

\subsection{DLCQ of M-theory}

Although DLCQ is used for giving a non-perturbative definition of
M-theory, it can in principle be used to describe any theory. We
shall first give a general recipe for constructing $\HH_N^{DLCQ}$
for any (compactified) string theory or M-theory, and then
specialize to the case of M-theory\cite{SMATR,SEIMATR}. 
Let $\TH$ be
such a (compactified) string or M-theory. Typically $\TH$ has one
mass parameter $M$, which can be taken to be the Planck mass for
M-theory, and $(\sqrt{\alpha'})^{-1}$
for string theory, and a set of
dimensionless parameters, {\it e.g.} string coupling constant,
the dimensions of the compact manifold measured in 
units of $M^{-1}$,
etc. We shall label all these dimensionless parameters by $\{\vec
y\}$. (From now on we shall display all factors of the mass
parameter explicitly, and not work in the $\alpha'=1$ unit, as
we have been doing till now.)
Let us denote by $\HH_N^{DLCQ}(M,L,\{\vec y\})$ the DLCQ
hamiltonian for this system. $L$ as usual denotes the radius of
the compact light-cone direction. We now propose the following
recipe for constructing $\HH_N^{DLCQ}$\cite{SMATR,SEIMATR}.
\begin{enumerate}
\item Consider the same theory $\TH$ with the same values of the
dimensionless parameters $\{\vec y\}$, but a different value $m$ of the
mass parameter. Let us compactify this theory on a
space-like circle $S^1$ of radius $R$. We shall call this theory
the auxiliary theory.
\item When $R$ is small, the Kaluza-Klein modes carrying momentum
along $S^1$ are heavy, and one would expect that the dynamics of
this system will be described by a non-relativistic quantum
mechanical hamiltonian. Let us focus on the sector with 
$N$-particles
each carrying 
momentum $1/R$ along $S^1$, and let us denote the $N$-body
hamiltonian describing the dynamics of this system
by $\HH_N^{KK}(m,R,\{\vec y\})$. (We subtract off the rest mass
energy $N/R$ of these particles in defining $\HH_N^{KK}$.)
\item $\HH_N^{DLCQ}$ is constructed from $\HH_N^{KK}$ by taking
the limit
\be \label{emm1}
\HH_N^{DLCQ}(M,L,\{\vec y\}) = \lim_{R\to
0}\HH_N^{KK}(m=M\sqrt{L/R}, R, \{\vec y\})\, .
\ee
\end{enumerate}
We shall first explore some of the consequences of this recipe. 
Later we shall discuss how this recipe might be `derived'.

First of all, note that since the duality transformations in
$\TH$ (before compactification on $S^1$) leave the momenta along
the non-compact directions unchanged, it leaves the sector with
$N$ units of momentum along $S^1$ invariant after we compactify $\TH$
on $S^1$. As a result $\HH_N^{KK}$, and hence
$\HH_N^{DLCQ}$ defined in \refb{emm1} is
expected to possess the full set of duality symmetries of 
$\TH$\cite{GRT}-\cite{OBER}.

Second, note that although this recipe gives $\HH_N^{DLCQ}$ for
any theory $\TH$, it is particularly useful for (compactified)
M-theory, since in the $R\to 0$ limit M-theory on a circle of
radius $R$ gets mapped to {\it weakly coupled} type IIA string
theory as discussed in section \ref{smtheory}. As discussed
there, states carrying momenta along $S^1$ correspond to D0
branes in the type IIA theory. Thus the recipe given above relates
$\HH_N^{DLCQ}$ to a specific weak coupling
limit of the hamiltonian describing
$N$ D0 branes in type IIA string theory. We shall later construct
this hamiltonian explicitly in some cases.

One might feel that the assertion made above is a bit too
simplified, since D0-branes represent only a subset of
particles in M-theory carrying momenta along $S^1$ $-$ the eleven
dimensional graviton and its supersymmetric partners. For
example, if $\TH$ corresponds to M-theory compactified on $T^2$,
then $\TH$ has solitonic states corresponding to the membrane of
M-theory wrapped around $T^2$. These states are distinct from the
supergravitons. Thus one would expect that in constructing
$\HH_N^{KK}$ (and hence $\HH_N^{DLCQ}$) for this theory one needs
to add to the D0-brane hamiltonian new degrees of freedom which
are capable of describing these wrapped membrane
states carrying momentum along $S^1$. However, due to a truly 
marvellous
property of the D0-brane system, this is not necessary. It turns
out that the D0-brane hamiltonian automatically contains the
required degrees of freedom that gives rise to these new states.
We shall explicitly see an example of this later.

Let us now apply this recipe to construct $\HH_N^{DLCQ}$ for
M-theory on $T^n$. Let $M_p$ be the Planck mass of the M-theory,
$L_i$ be the radii of the $n$ circles\footnote{For
simplicity we are assuming that the torus is made of product of
$n$ circles, without any background field.} 
which make up $T^n$
and $L$ be the radius
of the light-like circle. Let $m_p$ be the Planck mass of the
auxiliary M-theory,
$R_i$ be the radii of the circles that make up the
$n$-dimensional torus in this auxiliary M-theory, and $R$ be the
radius of the extra $S^1$ on which this auxiliary M-theory on
$T^n$ is further compactified. Then from
\refb{emm1} we get the following relation between the parameters
of the two theories:
\be \label{emm2}
m_p = M_p\sqrt{L/R}, \qquad m_p R_i = M_p L_i, 
\ee
where the second equation reflects that the dimensionless
parameters obtained by taking the product of the Planck mass and
the radii of the compact directions must be the same in the two
theories. 
We now map this to the hamiltonian of a set of D0 branes in type
IIA string theory by identifying M-theory on $S^1$ of radius $R$
with type IIA
string theory according to the rules given in section
\ref{smtheory}. 
Let $g_S (\equiv e^{\Phi/2})$ and 
$m_S(\equiv (\alpha')^{-1/2})$ denote the 
coupling constant and the string mass
in this type IIA theory. Then the first of eq.\refb{e32} gives
\be \label{exx1}
m_p R= g_S^{2/3}\, ,
\ee
where we have explicitly put in the eleven dimensional Planck
mass that was set to unity in \refb{e32}.
Another relation between $m_p$, $m_S$ and $g_S$ comes from
restoring the appropriate factors of $m_p$, $m_S$ and $g_S$ in
\refb{esugra} and \refb{eiia}. In particular
\refb{esugra} contains a multiplicative
factor of $(m_p)^{9}$, whereas
\refb{eiia} contains a multiplicative
factor of $(\alpha')^{-4}g_S^{-2}=m_S^8
g_S^{-2}$ (analog of eq.\refb{e8a} for the heterotic
theory). Thus the equality between the two actions upon
compactification of the eleven dimensional theory on $S^1$
requires that:
\be \label{exx2}
m_p^9 R = m_S^8 g_S^{-2}\, .
\ee
Inverting the relations \refb{exx1} and \refb{exx2} we get,
\be \label{emm6}
m_S = m_p^{3/2} R^{1/2}, \qquad g_S= (m_p R)^{3/2}\, .
\ee
If
$\HH_N^{D0}$ denotes the hamiltonian for $N$ D0 branes (with the
rest mass subtracted) in this theory, then, using 
eqs.\refb{emm1}, \refb{emm2}
and \refb{emm6} we get,
\ben \label{emm7}
&& \HH_N^{DLCQ}(M_p,L,\{R_i\}) \nonumber \\
&=& \lim_{R\to 0}
\HH_N^{D0}(m_S=M_p^{3/2}L^{3/4}R^{-1/4}, g_S=M_p^{3/2}(LR)^{3/4},
R_i=R^{1/2}L^{-1/2}L_i)\, . \nonumber \\
\een
This gives sensible answer for all $n$ up to 5 
and does not give sensible
answer for $n\ge 6$\cite{SMATR,SEIMATR}. 
Here we shall only discuss two special
cases, $n=0$ and $n=2$.

First we consider the case $n=0$. This corresponds to eleven 
dimensional M-theory.
As seen from \refb{emm7}, as $R\to 0$, $g_S\to 0$ and $m_S\to
\infty$. Thus we can use low energy and weak coupling
approximation of type IIA string theory. This limit has been
studied in detail in \cite{DKPS}. The action governing the
dynamics of the  D0 brane system in this limit is given by the
dimensional reduction of $N=1$ supersymmetric U(N) gauge theory
in ten dimensions:
\ben \label{emm8}
S &\sim& m_S^{-3} g_S^{-1} \int dt \Big[ \sum_{m=1}^9
Tr(\p_t \Phi^m \p_t \Phi^m) - \sum_{m<n =1}^9
Tr([\Phi^m,\Phi^n]^2)\Big]\nonumber \\
&& + \hbox{fermionic terms} \nonumber \\
&=& M_p^{-6} L^{-3} \int dt \Big[ \sum_{m=1}^9
Tr(\p_t \Phi^m \p_t \Phi^m) - \sum_{m<n =1}^9
Tr([\Phi^m,\Phi^n]^2)\Big] \nonumber \\
&& + \hbox{fermionic terms} \nonumber \\
\een
where $\Phi^m$ are $(N\times N)$ hermitian matrices.
In going from the first to the second line we have used
eq.\refb{emm7}. This gives a well-defined hamiltonian for DLCQ
M-theory. The flat direction in the potential corresponds to a
configuration where all the $\Phi^m$'s are simultaneously
diagonalized. The $N$ eigenvalues of $\Phi^m$ represent the
$m$'th coordinate of the $N$ different D0 branes.

Let us now consider M-theory on $T^2$. From eq.\refb{emm7} we see
that in the $R\to 0$ limit, 
the radii $R_i$ vanish. We can remedy this
problem by giving a different description of the same system by
making an $R_i\to (1/m_S^2R_i)$ duality transformation on both
circles. This converts 
the original type IIA theory to type IIA
theory on a dual torus $\wt T^2$ and
the system of $N$ D0 branes to a system of
$N$ D-2 branes wrapped on 
$\wt T^2$.\footnote{Quite generally one can show that an 
$R_i\to 1/(m_S^2R_i)$ duality transformation converts a
Dirichlet boundary condition to Neumann boundary condition and
vice versa\cite{POLREV}. Thus this duality transformation converts
a D0 brane with Dirichlet boundary conditions along $T^2$ into
a D2 brane wrapped on $\wt T^2$ which has Neumann boundary condition
along $\wt T^2$.} 
The new theory has parameters:
\ben \label{emm10}
&& \wt g_S =
g_S/(m_S^2 R_1 R_2) = M_p^{-3/2} L^{1/4} R^{1/4} L_1^{-1}
L_2^{-1}, \nonumber \\
&& \wt m_S = m_S = M_p^{3/2} L^{3/4} R^{-1/4}, \qquad 
\wt R_i = R_i^{-1} m_S^{-2} = M_p^{-3} L^{-1}
L_i^{-1}\, . \nonumber \\
\een
In the $R\to 0$ limit
this again gives a theory at weak coupling and large string mass.
Furthermore the new radii $\wt R_i$ are finite. The dynamics of
wrapped D-2 branes in this limit is described by a (2+1)
dimensional $N=8$
supersymmetric U(N) Yang-Mills theory compactified on $\wt T^2$.
This theory has
seven scalars $\Phi^m$ $(1\le m\le 7$) in the
adjoint representation of the gauge group. The bosonic part of
the lagrangian is given by:
\ben \label{emm11}
S &\sim& \wt m_S^{-1} \wt g_S^{-1} \int dt \int_0^{\wt R_1}
dx^8 \int_0^{\wt R_2} dx^9 
\Big[Tr (F_{\mu\nu} F^{\mu\nu}) \nonumber \\
&& + 
Tr(D_\mu \Phi^m D^\mu \Phi^m) - \sum_{m<n =1}^7
Tr([\Phi^m,\Phi^n]^2)\Big] \nonumber \\
&=& L^{-1} L_1 L_2 \int dt \int_0^{\wt R_1}
dx^8 \int_0^{\wt R_2} dx^9
\Big[Tr (F_{\mu\nu} F^{\mu\nu}) \nonumber \\
&& + 
Tr(D_\mu\Phi^m D^\mu\Phi^m) - \sum_{m<n =1}^7
Tr([\Phi^m,\Phi^n]^2)\Big]\, , \nonumber \\
\een
where $x^\mu$ ($\mu =0,8,9$) denote coordinates along the D-2
brane world-volume, and the $N$ eigenvalues of $\Phi_m$
represent the $m$th transverse coordinate of the $N$ different
D-2 branes.

Let us now address the problem alluded to earlier, namely that
M-theory on $T^2$ contains solitonic states in the form of
wrapped membranes. Thus the complete $\HH_N^{DLCQ}$ must contain
these states as well. In the auxiliary type IIA theory, these
wrapped membranes correspond to D2-branes wrapped on $T^2$. Upon
T-duality in both circles, these become D0 branes moving on the
dual torus $\wt T^2$. Thus the relevant question is, does the system
described in \refb{emm11} automatically contain these states, or
do we need to add new degrees of freedom in this system so as to
be able to describe these states? It turns out that the D0-brane
charge in this dual theory simply corresponds to the flux
of the U(1) component of the magnetic field through $\wt T^2$.
Thus a state with $k$ D0-branes (which correspond to a membrane
wrapped $k$ times on the original torus) can be described by a
specific excitation of the system \refb{emm11} carrying $k$ units
of magnetic flux through $\wt T^2$. There is no need to add new
degrees of freedom.

Finally let us give a `derivation' of the recipe described at the
beginning of this section following \cite{SEIMATR}. First of all,
we note that in the auxiliary theory $\TH$ on $S^1$,
if we multiply all the masses by some constant
$\lambda$, and simultaneously multiply all the lengths by
$\lambda^{-1}$, then the hamiltonian gets multiplied by $\lambda$
due to purely dimensional reasons.
This gives the following identity:
\be \label{emm12}
\HH_N^{KK}(m, R, \{\vec y\})
=\lambda^{-1} \HH_N^{KK}(\lambda m, \lambda^{-1} R, \{\vec y\})
\, .
\ee
where the first and the second arguments denote
respectively the overall mass scale and
the radius of $S^1$ as usual. Let us now
choose:
\be \label{emm12a}
r=\sqrt{RL}, \qquad m = M\sqrt{L/R}, \qquad \lambda = \sqrt{R/L} = (r/L) .
\ee
Substituting this in \refb{emm12} we get,
\be \label{emm12b}
\HH_N^{KK}(M\sqrt{L/R}, R, \{\vec y\})
= {L\over r} \HH_N^{KK}(M, r, \{\vec y\})\, .
\ee
Thus the recipe \refb{emm1} can now be rewritten as
\be \label{emm13}
\HH_N^{DLCQ}(M,L,\{\vec y\}) = \lim_{r\to 0} {L\over r}
\HH_N^{KK}(M,r,\{\vec y\})\, .
\ee
It is this form of the identity that we shall attempt to prove.

The basic idea behind this proof is to regard the light-like circle
as an infinitely boosted space-like circle of zero
radius\cite{BFOUR,HELPOL}. Let us start with theory $\TH$
compactified on a space-like circle $S^1$ of radius $r$. If $x$
and $t$ denote the coordinate along $S^1$ and the time coordinate
respectively, then we have an identification:
\be \label{emm14}
\pmatrix{x\cr t} \equiv \pmatrix{x\cr t} + 2\pi \pmatrix{r\cr
0}\, .
\ee
Let us now define new coordinates $(x',t')$ and
$x^{\prime\pm}$ as follows:
\be \label{emm15}
\pmatrix{x'\cr t'} = \pmatrix{ x\cosh\alpha - t\sinh\alpha\cr
t\cosh\alpha - x\sinh \alpha}\, ,
\ee
\be \label{emm17}
x^{\prime \pm} ={1\over \sqrt 2} (t' \pm x')\, .
\ee
In this coordinate system eq.\refb{emm14} takes the form:
\be \label{emm18}
\pmatrix{x^{\prime +}\cr x^{\prime -}} \equiv 
\pmatrix{x^{\prime +}\cr x^{\prime -}} 
- \sqrt 2\pi r
\pmatrix{-e^{-\alpha} \cr e^\alpha}\, .
\ee
Now consider the limit $r\to 0$, $\alpha\to\infty$ keeping fixed
\be \label{emm19}
L \equiv {r\over \sqrt 2} e^\alpha.
\ee
In this limit eq.\refb{emm18} reduces to
\be \label{emm20}
\pmatrix{x^{\prime +}\cr x^{\prime -}} \equiv
\pmatrix{x^{\prime +}\cr x^{\prime -}} 
- 2\pi 
\pmatrix{0 \cr L}\, .
\ee
This is equivalent to compactifying $x^{\prime -}$ on a circle of
radius $L$.

Under this map, a system carrying momentum $N/R$ along $S^1$ gets
mapped to a system carrying total momentum $k^{\prime -}=N/L$
along the $x^{\prime -}$ direction. Thus it is not surprising
that there is a relation between the Hamiltonian describing the
two systems. To find the precise relation between these two
hamiltonians, we need to study the relation between the usual
time coordinate $t$ of the original theory and the light-cone
time $x^{\prime +}$ of the boosted theory. From eqs.\refb{emm15},
\refb{emm17} it follows that:
\be \label{emm22}
{\p\over \p x^{\prime +}} = {1\over \sqrt 2} e^\alpha
\Big({\p\over \p t} + {\p\over \p x}\Big)\, .
\ee
Since the quantum operators which generate $i(\p/\p x^{\prime
+})$, $i(\p/\p t)$ and $i(\p/\p x)$ are $\HH_N^{DLCQ}$,
$\HH_N^{KK}+M_N$ and $-N/r$ respectively, with $M_N$ being the
rest mass of the $N$ Kaluza-Klein modes in $\TH$ on $S^1$,
we see from eq.\refb{emm19}, \refb{emm22}
that in the $r\to 0$ limit with $L$
fixed,
\be \label{emm23}
\HH_N^{DLCQ}(M,L,\{\vec y\}) = {L\over r} (\HH_N^{KK}(M,r,\{\vec
y\}) + M_N - {N\over r}) = 
{L\over r} \HH_N^{KK}(M,r,\{\vec
y\})\, ,
\ee
since $M_N=N/r$. This reproduces \refb{emm13}.

If we recall that $\HH_N^{DLCQ}$ is supposed to describe the
theory $\TH$, whereas $\HH_N^{KK}$ describes
theory $\TH$ compactified on a small circle, then by the above
argument, quite generally we can reconstruct a
theory by knowing its behaviour when compactified on a small
circle. This seems counterintuitive, so let us examine the steps
leading to this conclusion. They may be summarized 
as follows:
\begin{enumerate}
\item We start with a small circle.
\item We convert this to an almost light-like circle of
finite radius via a
large boost.
\item We then take the limit where the radius of this light-like
circle goes to infinity.
\end{enumerate}

As we can see, the key point in this proof is the assumption that
a light-like circle can be considered as a space-like
circle of zero radius in the limit of infinite boost. Of course,
this may be taken as a definition of the light-like circle.
However,
we are interested in a definition in which the radius of the
light-like circle acts as an infra-red regulator in the
uncompactified theory, so that in the end by taking $L\to\infty$
limit we recover the amplitudes in the uncompactified theory.
Clearly there is a possibility that these two definitions do not
match\cite{HELPOL}. 
Indeed there are explicit computations which show that
these two definitions do not always match for finite 
$N$\cite{DOS,GGR,DDM,DO,DIRA}, although they do match for some specific
terms in the supergraviton scattering 
amplitudes\cite{BFSS,BECKO,BFOUR}. (Note that the
`proof' given above did not involve taking the $N\to\infty$ limit.) 
It has
been suggested\cite{BANKSREV,SUSSREV} that this problem might go
away in the $N\to\infty$ limit, but there is no compelling
argument as of now in favour of this. This of course does not
mean that Matrix theory is wrong, it is just that we do not know
for sure if it is right, and even if it is right,
we do not quite know why it is right. Perhaps the arguments of
ref.\cite{SEIMATR} together with supersymmetry
non-renormalization theorems and properties of the large $N$
limit can be combined to constitute such a proof.

{\bf Acknowledgement}: I wish to thank S. Panda for useful
comments and suggestions.


\begin{thebibliography}{999}


\bibitem{REVB}
A. Sen, Int. J. Mod. Phys. {\bf A9} (1994) 3707 [hep-th/9402002].

\bibitem{DUFKHREV}
M. Duff, R. Khuri and J. Lu, Phys. Rep. {\bf 259} (1995) 213
[hep-th/9412184].

\bibitem{SCHREV}
J. Schwarz, Nucl. Phys. Proc. Suppl. {\bf 55B} 1 [hep-th/9607201].

\bibitem{SJPREV}
S. Chaudhuri, C. Johnson and J. Polchinski, hep-th/9602052.

\bibitem{POLREV}
J. Polchinski, Rev. Mod. Phys. {\bf 68} (1996) 1245 [hep-th/9607050]. 

\bibitem{POLREVT}
J. Polchinski, hep-th/9611050.

\bibitem{TOWNREV}
P. Townsend, hep-th/9612121; gr-qc/9707012; hep-th/9712004.

\bibitem{DUFFREV}
M. Duff, hep-th/9611203.

\bibitem{DOUGREV}
M. Douglas, hep-th/9610041.

\bibitem{LERCREV}
W. Lerche, hep-th/9611190; hep-th/9710246.

\bibitem{FOLO}
S. Forste and J. Louis, hep-th/9612192.

\bibitem{VAFREV}
C. Vafa, hep-th/9702201.

\bibitem{KLRE}
A. Klemm, hep-th/9705131.

\bibitem{KIRE}
E. Kiritsis, hep-th/9708130.

\bibitem{DELO}
B. de Wit and J. Louis, hep-th/9801132.

\bibitem{MARIO}
M. Trigiante, hep-th/9801144.

\bibitem{YOUM}
D. Youm, hep-th/9710046.

\bibitem{MALREV}
J. Maldecena, hep-th/9705078.

\bibitem{PEETREV}
A. Peet, hep-th/9712253.

\bibitem{BILALREV}
A. Bilal, hep-th/9710136.

\bibitem{BANKSREV}
T. Banks, hep-th/9710231.

\bibitem{SUSSREV}
D. Bigatti and L. Susskind, hep-th/9712072.

\bibitem{TAYLOR}
W. Taylor, hep-th/9801182.

\bibitem{REVE}
S. Mukhi, hep-ph/9710470. 

\bibitem{GSW}
M. Green, J. Schwarz and E. Witten,  Superstring Theory vol. 1
and 2, Cambridge University Press (1986); \\
D. Lust and S. Theisen, Lectures on String Theory, Springer
(1989); \\
J. Polchinski, hep-th/9411028.

\bibitem{WITTEND}
E. Witten, Nucl. Phys. {\bf B443} (1995) 85 [hep-th/9503124].

\bibitem{DABH}
A. Dabholkar, Phys. Lett. {\bf B357} (1995) 307 [hep-th/9506160].

\bibitem{HULLOPEN}
C. Hull, Phys. Lett. {\bf B357} (1995) 545 [hep-th/9506194]. 

\bibitem{POLCWIT}
J. Polchinski and E. Witten, Nucl. Phys. {\bf B460} (1996) 525
[hep-th/9510169].

\bibitem{HULLTOWN}
C. Hull and P. Townsend, Nucl. Phys. {\bf B438} (1995) 109
[hep-th/9410167].

\bibitem{DUFFSS}
M. Duff, Nucl. Phys. {\bf B442} (1995) 47 [hep-th/9501030]; \\
M. Duff and R. Khuri, Nucl. Phys. {\bf B411} (1994) 473
[hep-th/9305142].

\bibitem{SSSD}
A. Sen, Nucl. Phys. {\bf B450} (1995) 103 [hep-th/9504027].

\bibitem{HARSTRSSD}
J. Harvey and A. Strominger, Nucl. Phys. {\bf B449} (1995) 535
[hep-th/9504047].

\bibitem{GIVETC}
A. Giveon, M. Porrati and E. Rabinovici, Phys. Rep. {\bf 244}
(1994) 77 [hep-th/9401139].

\bibitem{FREREV}
P. Fre, Nucl. Phys. {\bf B [Proc. Sup.] 45B,C} (1996) 59
[hep-th/9512043] and references therein.

\bibitem{WITOLI}
E. Witten and D. Olive, Phys. Lett. {\bf B78} (1978) 97.


\bibitem{DEROO}
M. de Roo, Nucl. Phys. {\bf B255} (1985) 515.

\bibitem{FERR}
S. Ferrara, C. Kounnas and M. Porrati, Phys. Lett. {\bf B181}
(1986) 263.

\bibitem{TEREN}
M. Terentev, Sov. J. Nucl. Phys. {\bf 49} (1989) 713.

\bibitem{HASS}
S.F. Hassan and A. Sen, Nucl. Phys. {\bf B375} (1992) 103 
[hep-th/9109038].

\bibitem{MAHSCH}
J. Maharana and J. Schwarz, Nucl. Phys. {\bf B390} (1993) 3
[hep-th/9207016].

\bibitem{TSW}
A. Shapere, S. Trivedi and F. Wilczek, Mod. Phys. Lett. {\bf A6}
(1991) 2677.

\bibitem{SONE}
A. Sen, Nucl. Phys. {\bf B404} (1993) 109 [hep-th/9207053].

\bibitem{FILQ}
A. Font, L. Ibanez, D. Lust and F. Quevedo, Phys. Lett. {\bf
B249} (1990) 35.

\bibitem{REY}
S.J. Rey, Phys. Rev. {\bf D43} (1991) 526.

\bibitem{SCHONE}
J. Schwarz, hep-th/9209125.

\bibitem{STWO}
A. Sen, Phys. Lett. {\bf B303} (1993) 22 [hep-th/9209016].

\bibitem{STHREE}
A. Sen, Mod. Phys. Lett. {\bf A8} (1993) 2023 [hep-th/9303057].

\bibitem{SSCH}
J. Schwarz and A. Sen, Nucl. Phys. {\bf B411} (1994) 35
[hep-th/9304154]; Phys.
Lett. {\bf B312} (1993) 105 [hep-th/9305185].


\bibitem{SEIOLD}
N. Seiberg, Nucl. Phys. {\bf B303} (1988) 286.

\bibitem{IIBSUGRA}
M. Green and J. Schwarz, Phys. Lett. {\bf 122B} (1983) 143; \\
J. Schwarz and P. West, Phys. Lett. {\bf 126B} (1983) 301; \\
J. Schwarz, Nucl. Phys. {\bf B226} (1983) 269; \\
P. Howe and P. West, Nucl. Phys. {\bf B238} (1984) 181.


\bibitem{JULIA}
E. Cremmer, in `Unification of Fundamental Particle
Interactions', Plenum (1980); \\
B. Julia, in `Superspace and Supergravity', Cambridge Univ. Press
(1981).

\bibitem{FERRKOU}
S. Ferrara and C. Kounnas, Nucl. Phys. {\bf B328} (1989) 406.

\bibitem{CHL}
S. Chaudhuri, G. Hockney and J. Lykken, Phys. Rev. Lett. {\bf 75}
(1995) 2264 [hep-th/9505054].

\bibitem{CP}
S. Chaudhuri and J. Polchinski, Phys. Rev. {\bf D52} (1995) 
7168 [hep-th/9506048]. 


\bibitem{MONOLI}
C. Montonen and D. Olive, Phys. Lett. {\bf B72} (1977) 117.

\bibitem{OSBORN}
H. Osborn, Phys. Lett. {\bf B83} (1979) 321.

\bibitem{SBPS}
A. Sen, Phys. Lett. {\bf B329} (1994) 217 [hep-th/9402032].

\bibitem{AH}
M. Atiyah and N. Hitchin, Phys. Lett. {\bf 107A} (1985) 21; Phil.
Trans. Roy. Soc. Lond. {\bf A315} (1985) 459; Geometry and
Dynamics of Magnetic Monopoles, Cambridge Univ. Press.

\bibitem{GIBMAN}
G. Gibbons and N. Manton, Nucl. Phys. {\bf B274} (1986) 183.

\bibitem{GAUNT}
J. Gauntlett, Nucl. Phys. {\bf B400} (1993) 103 [hep-th/9205008];
Nucl. Phys. {\bf B411} (1994) 443 [hep-th/9305068].


\bibitem{BLUM}
J. Blum, Phys. Lett. {\bf B333} (1994) 92 [hep-th/9401133].

\bibitem{MANSCH}
N. Manton and B. Schroers, Annals Phys. {\bf 225} (1993) 290.

\bibitem{GIBRUB}
G. Gibbons and P. Ruback, Comm. Math. Phys. {\bf 115} (1988) 267.

\bibitem{SEG}
G. Segal and A. Selby,    Comm. Math. Phys. {\bf 177} (1996) 775.

\bibitem{PORR}
M. Porrati, Phys. Lett. {\bf B377} (1996) 67 [hep-th/9505187].

\bibitem{GAUHAR}
J. Gauntlett and J. Harvey, [hep-th/9407111].

\bibitem{SCHSLT}
J. Schwarz, Phys. Lett. {\bf B360} (1995) 13 [hep-th/9508143].

\bibitem{WITTDB}
E. Witten, Nucl. Phys. {\bf B460} (1996) 335 [hep-th/9510135].

\bibitem{DASMATFR}
S. Das and S. Mathur, Phys. Lett. {\bf B375} (1996) 103
[hep-th/9601152].

\bibitem{DABHAR}
A. Dabholkar and J. Harvey, Phys. Rev. Lett. {\bf 63} (1989) 
478; \\
A. Dabholkar, G. Gibbons, J. Harvey and F. Ruiz, Nucl. Phys. {\bf
B340} (1990) 33.

\bibitem{POLD}
J. Polchinski, Phys. Rev. Lett. {\bf 75} (1995) 4724
[hep-th/9510017].

\bibitem{DOLD}
J. Dai, R. Leigh and J. Polchinski, Mod. Phys. Lett. {\bf A4}
(1989) 2073; \\
R. Leigh, Mod. Phys. Lett. {\bf A4} (1989) 2767; \\
J. Polchinski, Phys. Rev. {\bf D50} (1994) 6041 [hep-th/9407031].

\bibitem{BSVO}
M. Bershadsky, V. Sadov and C. Vafa, Nucl. Phys. {\bf B463}
(1996) 398 [hep-th/9510225].

\bibitem{SUD}
A. Sen, Phys. Rev. {\bf D53} (1996) 2874 [hep-th/9711026].

\bibitem{VAFUD}
C. Vafa, Nucl. Phys. {\bf B469} (1996) 415 [hep-th/9511088];
Nucl. Phys. {\bf B463} (1996) 435 [hep-th/9512078].

\bibitem{SETHSTTWO}
S. Sethi and M. Stern, Phys. Lett. {\bf B398} (1997) 47
[hep-th/9607145].

\bibitem{BSV}
M. Bershadsky, V. Sadov and C. Vafa, Nucl. Phys. {\bf B463}
(1996) 420 [hep-th/9511222].

\bibitem{WITTSM}
E. Witten, Nucl. Phys. {\bf B460} (1996) 541 [hep-th/9511030].

\bibitem{PORRH}
M. Porrati, Phys. Lett. {\bf B387} (1996) 492 [hep-th/9607082].

\bibitem{BLU}
J. Blum, Nucl. Phys. {\bf B506} (1997) 223 [hep-th/9705030].

\bibitem{UNITY}
A. Sen, hep-th/9609176.

\bibitem{STHD}
A. Sen, Nucl. Phys. {\bf B434} (1995) 179 [hep-th/9408083];
Nucl. Phys. {\bf B447} (1995) 62 [hep-th/9503057].

\bibitem{VAFWITFI}
C. Vafa and E. Witten, hep-th/9507050.

\bibitem{DUORBI}
A. Sen, Nucl. Phys. {\bf B474} (1996) 361 [hep-th/9604070].

\bibitem{KACVAF}
S. Kachru and C. Vafa, Nucl. Phys. {\bf B450} ((1995) 69
[hep-th/9605105].

\bibitem{FHSV}
S. Ferrara, J. Harvey, A. Strominger and C. Vafa, Phys. Lett.
{\bf B361} (1995) 59 [hep-th/9505162].

\bibitem{LERCH}
A. Klemm, W. Lerche and P. Mayr, Phys. Lett. {\bf B357} (1995)
313 [hep-th/9506112].

\bibitem{ALDFON}
G. Aldazabal, A. Font, L. Ibanez and F. Quevedo, Nucl. Phys. {\bf
B461} (1996) 537 [hep-th/9510093].

\bibitem{SCHLYN}
B. Hunt and R. Schimmrigk, Phys. Lett. {\bf B381} (1996) 427
[hep-th/9512138]; \\
B. Hunt, M. Lynker and R. Schimmtigk, hep-th/9609082.

\bibitem{DECOUP}
B. de Wit, P. Lauwers and A. van Proeyen, Nucl. Phys. {\bf B255}
(1985) 569.

\bibitem{KAPL}
V. Kaplunovsky, J. Louis and S. Theisen, Phys. Lett. {\bf B357}
(1995) 71 [hep-th/9506110].

\bibitem{NARA}
I. Antoniadis, E. Gava, K. Narain and T. Taylor, Nucl. Phys. {\bf
B455} (1995) 109 [hep-th/9507115].

\bibitem{NONPER}
S. Kachru, A. Klemm, W. Lerche, P. Mayr and C. Vafa, Nucl. Phys.
{\bf B459} (1996) 537 [hep-th/9508155]; \\
A. Klemm, W. Lerche, P. Mayr, C. Vafa and N. Warner, Nucl. Phys.
{\bf B477} (1996) 746 [hep-th/9604034].

\bibitem{SEIWIT}
N. Seiberg and E. Witten, Nucl. Phys. {\bf B426} (1994) 19
[hep-th/9407087];
Nucl. Phys. {\bf B431} (1994) 
484 [hep-th/9408099].

\bibitem{TOWNM}
P. Townsend, Phys. Lett. {\bf 350B} (1995) 184 [hep-th/9501068]. 

\bibitem{IIAM}
F. Giani and M. Pernici, Phys. Rev. {\bf D30} (1984) 325; \\
I. Campbell and P. West, Nucl. Phys. {\bf B243} (1984) 112; \\
M. Huq and M. Namazie, Class. Quant. Grav. {\bf 2} (1985) 293. 

\bibitem{SETHSTEO}
S. Sethi and M. Stern, hep-th/9705046.

\bibitem{SMARG}
A. Sen, Phys. Rev. {\bf D54} (1996) 2964 [hep-th/9510229].

\bibitem{SCHM}
J. Schwarz, Phys. Lett. {\bf B367} (1996) 97 [hep-th/9510086].

\bibitem{ASPINM}
P. Aspinwall, Nucl. Phys. Proc. Suppl. {\bf 46} (1996) 30
[hep-th/9508154].

\bibitem{HORWIT}
P. Horava and E. Witten, Nucl. Phys. {\bf B460} (1996) 506
[hep-th/9510209]; Nucl. Phys. {\bf B475} (1996) 94
[hep-th/9603142].

\bibitem{DASMUK}
K. Dasgupta and S. Mukhi, Nucl. Phys. {\bf B465} (1996)
399 [hep-th/9512196].

\bibitem{WITTOR}
E. Witten, Nucl. Phys. {\bf B463} (1996) 383 [hep-th/9512219].

\bibitem{MORBI}
A. Sen, Mod. Phys. Lett. {\bf A11} (1996) 1339 [hep-th/9603113].

\bibitem{KAPEE}
E. Caceres, V. Kaplunovsky and M. Mandelberg, Nucl. Phys. {\bf
B493} (1997) 73 [hep-th/9606036].

\bibitem{VAFAF}
C. Vafa, Nucl. Phys. {\bf B469} (1996) 403 [hep-th/9602022].

\bibitem{VMORF}
D. Morrison and C. Vafa, Nucl. Phys. {\bf B473} (1996) 74
[hep-th/9602114]; Nucl. Phys. {\bf B476} (1996) 437 [hep-th/9603161].

\bibitem{STRCOS}
B. Greene, A. Shapere, C. Vafa and S.T. Yau, Nucl. Phys. {\bf
B337} (1990) 1.

\bibitem{FMW}
R. Friedman, J. Morgan and E. Witten, Comm. Math. Phys. {\bf 187} (1997)
679 [hep-th/9701162].

\bibitem{BERET}
M. Bershadsky, A. Johansen, T. Pantev and V. Sadov, Nucl. Phys.
{\bf B505} (1997) 165 [hep-th/9701165].

\bibitem{FTHEORY}
A. Sen, Nucl. Phys. {\bf B475} (1996) 562 [hep-th/9605150].

\bibitem{ORIENT}
A. Sagnotti, `Open Strings and their Symmetry Groups', Talk at
Cargese Summer Inst., 1987; \\
G. Pradisi and A. Sagnotti, Phys. Lett. {\bf B216} (1989) 59; \\
M. Bianchi, G. Pradisi and A. Sagnotti, Nucl. Phys. {\bf B376}
(1992) 365; \\
P. Horava, Nucl. Phys. {\bf B327} (1989) 461; Phys. Lett. {\bf
B231} (1989) 251.

\bibitem{GIMPOL}
E. Gimon and J. Polchinski, Phys. Rev. {\bf D54} (1996) 1667
[hep-th/9601038].

\bibitem{OPENREF}
A. Chamseddine, Phys. Rev. {\bf D24} (1981) 3065; \\
E. Bergshoeff, M. de Roo, B. de Wit and P. van Niewenhuizen,
Nucl. Phys. {\bf B195} (1982) 97; \\
E. Bergshoeff, M. de Roo and B. de Wit, Nucl. Phys. {\bf B217}
(1983) 143; \\
G. Chapline and N. Manton, Phys. Lett. {\bf 120B} (1983) 105.

\bibitem{HYPER}
H. Nicolai, Phys. Lett. {\bf B276} (1992) 333.

\bibitem{MONMOD}
E. Weinberg, Phys. Rev. {\bf D20} (1979) 936; \\
E. Corrigan and P. Goddard, Comm. Math. Phys. {\bf 80} (1981)
575; \\
C. Taubes, Comm. Math. Phys. {\bf 91} (1983) 235.

\bibitem{MANTON}
N. Manton, 
Phys. Lett. {\bf 110B} (1982) 54.

\bibitem{INDEX}
C. Callias, Comm. Math. Phys. {\bf 62} (1978) 213.

\bibitem{NARAIN}
K. Narain, Phys. Lett. {\bf 169B} (1986) 41; \\
K. Narain, H. Sarmadi and E. Witten, Nucl. Phys. {\bf B279}
(1987) 369.

\bibitem{DASMAT}
S. Das and S. Mathur, Nucl. Phys. {\bf B478} (1996) 561
[hep-th/9606185].

\bibitem{BANDIX}
T. Banks, L. Dixon, D. Friedan and E. Martinec, Nucl. Phys. {\bf
B299} (1988) 613; \\
T. Banks and L. Dixon, Nucl. Phys. {\bf B307} (1988) 93.

\bibitem{MIRREV}
B. Greene, hep-th/9702155 and references therein.

\bibitem{SPL}
B. de Wit and A. van Proeyen, Nucl. Phys. {\bf B245} (1984) 89;
\\
E. Cremmer, C. Kounnas, A. van Proeyen, J. Derendinger, S.
Ferrara, B. de Wit and L. Girardello, Nucl. Phys. {\bf B250}
(1985) 385; \\
S. Ferrara and A. Strominger, in `Strings 89', World Scientific
(1989); \\
A. Strominger, Comm. Math. Phys. {\bf 133} (1990) 163.

\bibitem{HYPERM}
J. Bagger and E. Witten, Nucl. Phys. {\bf B222} (1983) 1; \\
K. Galicki, Comm. Math. Phys. {\bf 108} (1987) 117.

\bibitem{CRJL}
E. Cremmer and B. Julia, Nucl. Phys. {\bf B159} (1979) 141.

\bibitem{HAWRAD}
S. Hawking, Nature {\bf 248} (1974) 30; Comm. Math. Phys. {\bf
43} (1975) 199.

\bibitem{BHENT}
J. Bekenstein, Lett. Nuov. Cim. {\bf 4} (1972) 737; Phys. Rev.
{\bf D7} (1973) 2333; Phys. Rev. {\bf D9} (1974) 3192; \\
G. Gibbons and S. Hawking, Phys. Rev. {\bf D15} (1977) 2752..

\bibitem{THOOFT}
G. 't Hooft, Nucl. Phys. {\bf B335} (1990) 138.

\bibitem{SUSS}
L. Susskind, hep-th/9309145; \\
L. Susskind and J. Uglam, Phys. Rev. {\bf D50} (1994) 2700
[hep-th/9401070]; \\
J. Russo and L. Susskind, Nucl. Phys. {\bf B437} (1995) 611
[hep-th/9405117].

\bibitem{HORPOL}
G. Horowitz and J. Polchinski, Phys. Rev. {\bf D55} (1997) 6189
[hep-th/9612146].

\bibitem{SFET}
K. Sfetsos and K. Skenderis, hep-th/9711138.

\bibitem{SENOLD}
A. Sen, Mod. Phys. Lett. {\bf A10} (1995) 2081 [hep-th/9504147].

\bibitem{PEET}
A. Peet, Nucl. Phys. {\bf B456} (1995) 732 [hep-th/9506200].

\bibitem{NTBL}
A. Sen, hep-th/9712150.

\bibitem{STVA}
A. Strominger and C. Vafa, Phys. Lett. {\bf B379} (1996) 99
[hep-th/9601029].

\bibitem{CALMAL}
C. Callan and J. Maldacena, Nucl. Phys. {\bf B472} (1996) 591
[hep-th/9602043].

\bibitem{MALSUSS}
J. Maldacena and L. Susskind, Nucl. Phys. {\bf B475} (1996) 679
[hep-th/9604042].

\bibitem{NONEX}
G. Horowitz and A. Strominger, Phys. Rev. Lett. {\bf 77} (1996) 2368
[hep-th/9602051].

\bibitem{MALNON}
J. Maldacena, Phys. Rev. {\bf D55} (1997) 7645 [hep-th/9611125].

\bibitem{DMW}
A. Dhar, G. Mandal and S. Wadia, Phys. Lett. {\bf B388} (1996) 51
[hep-th/9605234].

\bibitem{MALSTR}
J. Maldacena and A. Strominger, Phys. Rev. {\bf D55} 
(1997) 861 [hep-th/9609026].

\bibitem{NEQTBH}
S. Ferrara, R. Kallosh and A. Strominger, Phys. Rev. {\bf D52}
(1995) 5412 [hep-th/9508072]; \\
A. Strominger, Phys. Lett. {\bf B383} (1996) 39 [hep-th/9602111];
\\
S. Ferrara and R. Kallosh, Phys. Rev. {\bf D54} (1996) 1514
[hep-th/9602136]; Phys. Rev. {\bf D54} (1996) 1525
[hep-th/9603090]; \\
M. Shmakova, Phys. Rev. {\bf D56} (1997) 540 [hep-th/9612076]; \\
K. Behrndt, G. Lopez Cardoso, B. de Wit, R. Kallosh, D. Lust and
T. Mohaupt, Nucl. Phys. {\bf B488} (1997) 236 [hep-th/9610105];
\\
S. Rey, Nucl. Phys. {\bf B508} (1997) 569 [hep-th/9610157]; \\
D. Kaplan, D. Lowe, J. Maldacena and A. Strominger, Phys. Rev.
{\bf D55} (1997) 4898 [hep-th/9609204]; \\
K. Behrndt and T. Mohaupt, Phys. Rev. {\bf D56} (1997) 2206
[hep-th/9611140]; \\
J. Maldacena, Phys. Lett. {\bf B403} (1997) 20 [hep-th/9611163];
\\
A. Chou, R. Kallosh, J. Rahmfeld, S. Rey, M. Shmakova and W.
Wong, Nucl. Phys. {\bf B508} (1997) 147 [hep-th/9704142]; \\
J. Maldacena, A. Strominger and E. Witten, hep-th/9711053; \\
C. Vafa, hep-th/9711067.



\bibitem{ANGMOM}
J. Bekenridge, R. Myers, A. Peet and C. Vafa, Phys. Lett. {\bf
B391} (1996) 93 [hep-th/9602065]; \\
J. Bekenridge, D. Lowe, R. Myers, A. Peet, A. Strominger
and C. Vafa, Phys. Lett. {\bf B381} (1996) 423 [hep-th/9603078].

\bibitem{FOURDBH}
C. Johnson, R. Khuri and R. Myers, Phys. Lett. {\bf B378} (1996)
78 [hep-th/9603061]; \\
J. Maldacena and A. Strominger, Phys. Rev. Lett. {\bf 77} (1996)
428 [hep-th/9603060]; \\
G. Horowitz, D. Lowe and J. Maldacena, Phys. Rev. Lett. {\bf 77}
(1996) 430 [hep-th/9603195].

\bibitem{HIDDEN}
M. Gaillard and B. Zumino, Nucl. Phys. {\bf B193} (1981) 221; \\
S. Ferrara, J. Scherk and B. Zumino, 
Nucl. Phys. {\bf B121} (1977) 393; \\
E. Cremmer, J. Scherk and S. Ferrara, Phys. Lett. {\bf 74B}
(1978) 61; \\
B. de Wit, Nucl. Phys. {\bf B158} (1979) 189; \\
B. de Wit and H. Nicolai, Nucl. Phys. {\bf B208} (1982) 323; \\
E. Cremmer and B. Julia, Phys. Lett. {\bf 80B} (1978) 48; \\
B. de Wit and A. van Proeyen, Nucl. Phys. {\bf B245} (1984) 89.

\bibitem{HETEROTIC}
D. Gross, J. Harvey, E. Martinec and R. Rohm, Phys. Rev. Lett.
{\bf 54} (1985) 502; Nucl. Phys. {\bf B256} (1985) 253; Nucl.
Phys. {\bf B267} (1986) 75.

\bibitem{TYPEI}
M. Green and J. Schwarz, Phys. Lett. {\bf 109B} (1982) 444;
Phys. Lett. {\bf 149B} (1984) 117;
Phys. Lett. {\bf 151B} (1985) 21; Nucl. Phys. {\bf B255} (1985)
93.

\bibitem{ALTER}
F. Larsen and F. Wilczek, Phys. Lett. {\bf B375} (1996) 37
[hep-th/9511064]; Nucl. Phys. {\bf B475} (1996) 627
[hep-th/9604134]; Nucl. Phys. {\bf B488} (1997) 261
[hep-th/9609084]; \\
M. Cvetic and A. Tseytlin, Phys. Rev. {\bf D53} (1996) 5619
[hep-th/9512031]; \\
A. Tseytlin, Mod. Phys. Lett. {\bf A11} (1996) 689
[hep-th/9601177]; Nucl. Phys. {\bf B477} (1996) 431
[hep-th/9605091].

\bibitem{HTS}
C. Hull and P. Townsend, Nucl. Phys. {\bf B451} (1995) 525
[hep-th/9505073].

\bibitem{STROMSM}
A. Strominger,   Nucl. Phys. {\bf B451} (1995) 96 
[hep-th/9504090]; \\
B. Greene, D. Morrison and A. Strominger, Nucl. Phys. {\bf B451}
(1995) 109 [hep-th/9504145].

\bibitem{DOUGENH}
M. Douglas and G. Moore, hep-th/9603167; \\
J. Polchinski,   Phys. Rev. {\bf B55} (1997) 6423 [hep-th/9606165]; \\
C. Johnson and R. Myers, Phys. Rev. {\bf D55} (1997) 6382
[hep-th/9610140]; \\
M. Douglas, [hep-th/9612126] published in JHEP electronics
journal; \\
D. Diaconescu, M. Douglas and J. Gomis, [hep-th/9712230].

\bibitem{WITTENTEN}
E. Witten, [hep-th/9507121].

\bibitem{DOUGLAS}
M. Douglas, [hep-th/9512077]; [hep-th/9604198].

\bibitem{ALVFRM}
L. Alvarez-Gaume, D. Freedman and S. Mukhi, Ann. Phys. (NY) {\bf 134}
(1981) 85;  \\
L. Alvarez-Gaume and D. Freedman, Comm. Math. Phys. {\bf 80} (1981)
443.

\bibitem{HASWAD}
S. Hassan and S. Wadia,   [hep-th/9712213].

\bibitem{MEMDIF}
E. Bergshoeff, E. Sezgin and P. Townsend, Phys. Lett. {\bf B189}
(1987) 75; Ann. Phys. (NY) {\bf 185} (1988) 330; \\
M. Duff, Class. Quant. Grav. {\bf 5} (1988) 189; \\
B. de Wit, M. Luscher and H. Nicolai, Nucl. Phys. {\bf B320}
(1989) 135; \\
B. de Wit, J. Hoppe and H. Nicolai, Nucl. Phys. {\bf B305} (1988)
545.

\bibitem{BFSS}
T. Banks, W. Fischler, S. Shenker and L. Susskind, Phys. Rev.
{\bf D55} (1997) 5112 [hep-th/9610043].

\bibitem{SUSSDL}
L. Susskind, [hep-th/9704080].

\bibitem{SMATR}
A. Sen, hep-th/9709220, to appear in Adv. Theor. Math. Phys.

\bibitem{SEIMATR}
N. Seiberg, Phys. Rev. Lett. {\bf 79} (1997) 3577 [hep-th/9710009].

\bibitem{GRT}
O. Ganor, S. Ramgoolam and W. Taylor, Nucl. Phys. {\bf B492}
(1997) 191 [hep-th/9611202]. 

\bibitem{SUSSDU}
L. Susskind, [hep-th/9611164].

\bibitem{ROZ} 
M. Rozali, Phys. Lett. {\bf B400} (1997) 260 [hep-th/9702136].

\bibitem{BRS}
M. Berkooz, M. Rozali and N. Seiberg, Phys. Lett. {\bf B408}
(1997) 105 [hep-th/9704089].

\bibitem{SEIB}
N. Seiberg, Phys. Lett. {\bf B408} (1997) 98 [hep-th/9705221].

\bibitem{GOVIND}
S. Govindarajan, Phys. Rev. {\bf D56} (1997) 5276
[hep-th/9705113].

\bibitem{BERROZ}
M. Berkooz and M. Rozali, [hep-th/9705175].

\bibitem{BERK}
M. Berkooz, [hep-th/9712012].

\bibitem{BLA}
M. Blau and M. O'Loughlin, [hep-th/9712047].

\bibitem{OBER}
N. Obers, B. Pioline and E. Rabinovici, [hep-th/9712084].

\bibitem{DKPS}
M. Douglas, D. Kabat, S. Poulio and S. Shenker,  Nucl. Phys. {\bf
B485} (1997) 85 [hep-th/9608024].

\bibitem{BFOUR}
K. Becker, M. Becker, J. Polchinski and A. Tseytlin, Phys. Rev.
{\bf D56} (1997) 3174 [hep-th/9706072].

\bibitem{HELPOL}
S. Hellerman and J. Polchinski, [hep-th/9711037].

\bibitem{DOS}
M. Douglas, H. Ooguri and S. Shenker, Phys. Lett. {\bf B402}
(1997) 36 [hep-th/9702203].

\bibitem{GGR}
O. Ganor, R. Gopakumar and S. Ramgoolam, hep-th/9705188.

\bibitem{DDM}
J. David, A. Dhar and G. Mandal, hep-th/9707132.

\bibitem{DO}
M. Douglas and H. Ooguri, [hep-th/9710178].

\bibitem{DIRA}
M. Dine and A. Rajaraman, [hep-th/9710174].

\bibitem{BECKO}
K. Becker and M. Becker, Nucl. Phys. {\bf B506} (1997) 48
[hep-th/9705091].

\bibitem{RAMOND}
P. Ramond,  Phys. Rev. {\bf D3} (1971) 2415.

\bibitem{NEVSCH}
A. Neveu and J. Schwarz, Nucl. Phys. {\bf B31} (1971) 86; Phys.
Rev. {\bf D4} (1971) 1109.

\bibitem{GSO}
F. Gliozzi, J. Scherk and D. Olive, Phys. Lett. {\bf 65B} (1976)
282; Nucl. Phys. {\bf B122} (1977) 253.

\bibitem{KTDUAL}
P. Aspinwall and D. Morrison, [hep-th/9404151]; \\
P. Aspinwall, [hep-th/9611137] and references therein.

\bibitem{ANOM}
M. Duff and R. Minasian, Nucl. Phys. {\bf B436} (1995) 507
[hep-th/9406198]; \\
C. Vafa and E. Witten, Nucl. Phys. {\bf B447} (1995) 261
[hep-th/9505053]; \\
M. Duff, J. Liu and R. Minasian, Nucl. Phys. {\bf B452} (1995)
261 [hep-th/9506126]; \\
M. Duff, R. Minasian and E. Witten, Nucl. Phys. {\bf B463} (1996)
435 [hep-th/9601036]; \\
M. Berkooz, R. Leigh, J. Polchinski, J. Schwarz, N. Seiberg and
E. Witten, Nucl. Phys. {\bf B475} (1996) 115 [hep-th/9605184].

\bibitem{KALYAN}
S. Das, S. Kalyanarama, P. Ramadevi and S. Mathur,
[hep-th/9711003].

\bibitem{CAND}
P. Candelas, G. Horowitz, A. Strominger and E. Witten, Nucl.
Phys. {\bf B258} (1985) 46.

\bibitem{OTHERF}
E. Gimon and C. Johnson, Nucl. Phys. {\bf B479} (1996) 285
[hep-th/9606176]; \\
J. Blum and A. Zaffaroni, Phys. Lett. {\bf B387} (1996)
71 [hep-th/9607019]; \\
A. Dabholkar and J. Park, Phys. Lett. {\bf B394} (1997)
302 [hep-th/9607041]; \\
R. Gopakumar and S. Mukhi, Nucl. Phys. {\bf B479} (1996) 260
[hep-th/9607057]; \\
J. Park, [hep-th/9611119]; \\
A. Sen, Nucl. Phys. {\bf B489} (1997) 139
[hep-th/9611186]; Nucl. Phys. {\bf B498} (1997) 135;
[hep-th/9702061]; 
Phys. Rev. {\bf D55} (1997) 7345;
[hep-th/9702165]; \\
O. Aharony, J. Sonneschein, S. Yankielowicz and S. Theisen, Nucl.
Phys. {\bf B493} (1997) 177 [hep-th/9611222].

\bibitem{VAFSIX}
M. Bershadsky, K. Intrilligator, S. Kachru, D. Morrison, V. Sadov
and C. Vafa, Nucl. Phys. {\bf B481} (1996) 215
[hep-th/9605200]; \\
P. Aspinwall and M. Gross, Phys. Lett. {\bf B387}
(1996) 735 [hep-th/9605131].

\bibitem{DASMUKT}
K. Dasgupta and S. Mukhi, Phys. Lett. {\bf B385} (1996) 125
[hep-th/9606044].

\bibitem{JOHAN}
A. Johansen, Phys. Lett. {\bf B395} (1997) 36 [hep-th/9608186].

\bibitem{GABZWI}
M. Gaberdiel and B. Zwiebach, [hep-th/9709013]; \\
M. Gaberdiel, T. Hauer and B. Zwiebach, [hep-th/9801205].

\bibitem{WITTBR}
E. Witten, Nucl. Phys. {\bf B500} (1997) 3 [hep-th/9703166].

\bibitem{ENH}
A. Sen, [hep-th/9707123] published in JHEP electronic journal.

\bibitem{KUVA}
A. Kumar and C. Vafa, Phys. Lett. {\bf B396} (1997) 85
[hep-th/9611007].

\bibitem{WITTSH}
E. Witten, Nucl. Phys. {\bf B471} (1996) 135 [hep-th/9602070].

\bibitem{MCY}
A. Cadavid, A. Ceresole, R. D'Auria and S. Ferrara,
[hep-th/9506144]; \\
I. Antoniadis, S. Ferrara and T. Taylor, Nucl. Phys. {\bf B460}
(1996) 489; \\
S. Ferrara, R. Khuri and R. Minasian, [hep-th/9602102]; \\
E. Witten, Nucl. Phys. {\bf B471} (1996) 195 [hep-th/9603150].

\bibitem{DUFFNREV}
M. Duff, B. Nilsson and C. Pope, Phys. Rep. {\bf 130} (1986) 1.

\bibitem{CARDY}
J. Cardy, Nucl. Phys. {\bf B270} (1986) 186.

\end{thebibliography}
\end{document}